\documentclass[12pt]{article}
\usepackage{epsf,a4,slashed,latexsym}
\usepackage[small,bf]{caption}

\newcommand{\sixth}{\mbox{\small $\frac{1}{6}$}}         % 1/6
\newcommand{\half}{\mbox{\small $\frac{1}{2}$}}          % 1/2
\newcommand{\quart}{\mbox{\small $\frac{1}{4}$}}         % 1/4
\newcommand{\third}{\mbox{\small $\frac{1}{3}$}}         % 1/3
\newcommand{\threehalf}{\mbox{\small $\frac{3}{2}$}}     % 3/2
\newcommand{\twelfth}{\mbox{\small $\frac{1}{12}$}}      % 1/12
\newcommand{\msbar}{\mbox{\tiny $\overline{MS}$}}        % Msbar
\newcommand{\mom}{\mbox{\tiny $M\!O\!M$}}                % MOM
\newcommand{\ripmom}{\mbox{\tiny $R\!I^\prime\!\!-\!\!M\!O\!M$}} % RIp-MOM
\newcommand{\rgi}{\mbox{\tiny $R\!G\!I$}}                % RGI
\newcommand{\ti}{\mbox{\tiny $T\!I$}}                    % TI
\newcommand{\lat}{\mbox{\tiny $L\!A\!T$}}                % LAT
\newcommand{\plaq}{\Box}                                 % PLAQ
\newcommand{\born}{\mbox{\tiny $B\!O\!R\!N$}}            % LAT
\newcommand{\NS}{\mbox{\tiny $N\!S$}}                    % NS
\newcommand{\NR}{\mbox{\tiny $N\!R$}}                    % Nr

\def\lsim{\mathrel{\rlap{\lower4pt\hbox{\hskip1pt$\sim$}}
    \raise1pt\hbox{$<$}}}                % less than or approx. symbol
\def\gsim{\mathrel{\rlap{\lower4pt\hbox{\hskip1pt$\sim$}}
    \raise1pt\hbox{$>$}}}                % greater than or approx. symbol
\begin{document}

\title{
\vspace{-2.5cm}
\flushleft{\normalsize DESY 04-194} \\
\vspace{-0.35cm}
{\normalsize Edinburgh 2004/24} \\
\vspace{-0.35cm}
{\normalsize LTH 638} \\
\vspace{-0.35cm}
{\normalsize LU-ITP 2004/039} \\
\vspace{-0.35cm}
{\normalsize October 2004} \\
\vspace{0.5cm}
\centering{\Large \bf A lattice determination of moments of 
                      unpolarised nucleon structure functions
                      using improved Wilson fermions}}

\author{\large M. G\"ockeler$^{1,2}$, R. Horsley$^3$, D. Pleiter$^4$, \\
               P.~E.~L. Rakow$^5$ and G. Schierholz$^{4,6}$ \\[1em]
         -- QCDSF Collaboration -- \\[1em]
        \small
           $^1$ Institut f\"ur Theoretische Physik,
                Universit\"at Leipzig, \\[-0.5em]
        \small
                D-04109 Leipzig, Germany \\[0.25em]
        \small
           $^2$ Institut f\"ur Theoretische Physik,
                Universit\"at Regensburg, \\[-0.5em]
        \small
                D-93040 Regensburg, Germany \\[0.25em]
        \small
           $^3$ School of Physics, University of Edinburgh, \\[-0.5em]
        \small 
                Edinburgh EH9 3JZ, UK \\[0.25em]
        \small
           $^4$ John von Neumann-Institut f\"ur Computing NIC, \\[-0.5em]
        \small
                Deutsches Elektronen-Synchrotron DESY, \\[-0.5em]
        \small
                D-15738 Zeuthen, Germany \\[0.25em]
        \small
           $^5$ Department of Mathematical Science, University of Liverpool,
                                                            \\[-0.5em]
        \small
                Liverpool L69 3BX, UK \\[0.25em]
        \small
           $^6$ Deutsches Elektronen-Synchrotron DESY, \\[-0.5em] 
        \small
                D-22603 Hamburg, Germany}

%\date{\today}
\date{October 5, 2004}

\maketitle

% ----------------------------------------------------------------------

\begin{abstract}
Within the framework of quenched lattice QCD and using $O(a)$ improved
Wilson fermions and non-perturbative renormalisation, a high stat\-istics
computation of low moments of the unpolarised nucleon structure functions
is given. Particular attention is paid to the chiral and continuum
extrapolations.
% We find that the lowest moment is higher
%than the phenomenological value and this situation also persists
%to a lesser degree for the next two higher moments.
\end{abstract}

\clearpage

% ----------------------------------------------------------------------

\section{Introduction}
\label{introduction}

The results of a lattice simulation of Quantum Chromodynamics (QCD)
give in principle a direct probe of certain low energy aspects
of the theory, such as hadronic masses and matrix elements.
This is at present the only way of getting these quantities from QCD,
without additional model-dependent assumptions.
A useful theoretical tool in conjunction with QCD
and deep inelastic scattering (or DIS) experiments is the
operator product expansion, OPE.
At leading twist the OPE relates moments
of an experimentally measured structure function,
generically denoted by $F$, to certain matrix elements
$v_n$ where
\begin{equation}
   \int_0^1 dx x^{n-2} F^{\NS}(x,Q^2) = f
         E_{F;v_n}^{\cal S}\left( {M^2\over Q^2}, g^{\cal S}(M) \right)
         v_{n}^{\cal S}(g^{\cal S}(M)) \,.
\label{moment_def}
\end{equation}
$F$ is a function of two variables -- $Q^2$, the space-like momentum
transfer to the nucleon and $x$, the Bjorken variable
($f$ is a normalisation factor).
$v_n$ are the nucleon matrix elements of certain operators
and $E$ are the associated Wilson coefficients. These are perturbatively
known at high energies where the coupling constant $g$ becomes small
and are found in a specified scheme ${\cal S}$ at scale $M$.
Usually, of course, we take ${\cal S} \equiv \overline{MS}$
at scale $M \equiv \mu \sim$ few GeV. We also assume that $Q^2$ is
large enough, so that higher twist terms ie $O(1/Q^2)$ terms are negligible.

As will be discussed later
(section \ref{determining_me}), lattice computations are presently
restricted to determining non-singlet, NS, nucleon structure functions
\begin{equation}
   F^{\NS}(x,Q^2) \equiv F^p(x,Q^2) - F^n(x,Q^2) \,,
\end{equation}
ie the difference between proton, $p$, and neutron, $n$, results.
Note in particular that this means that nucleon matrix elements
of gluonic operators have cancelled.

In this article we shall only be concerned with unpolarised
structure functions. The same matrix elements $v_n$ contribute
to the scattering of charged leptons and of neutrinos, but the weights
$f$ are different in the two cases.
Thus for charged lepton--nucleon DIS, $lN \to l X$,
which is mediated by a photon, we have $F^{\NS} \to F_2^{\gamma;\NS}$
with $n = 2, 4, \ldots$ and $f^\gamma = 2(e_u^2 - e_d^2)/2 = 1/3$.
For neutrino--nucleon charged weak current interactions
for example $\nu N \to lX$, ($l^+ N \to \overline{\nu}X$) or
$lN \to \nu X$, ($\overline{\nu} N \to l^+X$) which are mediated
by $W^+$, $W^-$ respectively, then we have $F^{\NS} \to F_2^{W^{\pm};\NS}$
and $f^{W^+} = 2(-1)$, $f^{W^-} = 2(+1)$ (neglecting the CKM mixing matrix)
with $n = 3, 5, \ldots$. (Alternatively setting
$F^{\NS} \to 2x F_1^{\NS}$ in all cases one has
the same matrix elements and $f$s as for $F_2^{\NS}$ in
eq.~(\ref{moment_def}), but different Wilson coefficients. The additional
$F_3$ structure functions, occuring because of parity non-conservation
also obey eq.~(\ref{moment_def}) with $F^{\NS} \to xF_3^{W^\pm;\NS}$ and
again $f^{W^\pm} = \mp 2$ with $n = 2, 4, \ldots$.)
Similar expressions hold for the neutral currents, but with more
complicated expressions for the $f$s.

In all cases the relevant matrix elements are given by
first defining the sequence of quark bilinear forms
\begin{eqnarray}
 {\cal O}^{\mu_1\cdots\mu_n}_q
   &=& i^{n-1} \overline{q}\gamma^{\mu_1}
           \stackrel{\leftrightarrow}{{D}^{\mu_2}} \cdots
           \stackrel{\leftrightarrow}{{D}^{\mu_n}}q, \qquad q = u, d \,,
\label{Doperators}
\end{eqnarray}
where $\stackrel{\leftrightarrow}{D} = \half ( \stackrel{\rightarrow}{D} -
\stackrel{\leftarrow}{D} )$. Symmetrising the indices and removing
traces, gives the Lorentz decomposition of the proton
(ie nucleon, N) matrix element of%
\footnote{The nucleon states are normalised with the convention
$\langle N(\vec{p}\,) | N(\vec{p}^{\,\prime} ) \rangle
= (2\pi)^3 2 E_{\vec{p}} \delta (\vec{p} - \vec{p}^{\,\prime} )$.}
\begin{equation}
 \langle N(\vec{p}\,) | \left[ {\cal O}^{ \{ \mu_1\cdots\mu_n \} }_q -
                    \mbox{\rm Tr} 
                      \right] | N(\vec{p}\,) \rangle^{\cal S}
 \equiv 2 v^{(q){\cal S}}_n  [p^{\mu_1} \cdots p^{\mu_n} - \mbox{\rm Tr}] \,.
\label{minkowski_vn}
\end{equation}
For example we have for $n = 2$,
\begin{eqnarray}
   \lefteqn{\hspace*{-1.00in}\langle N(\vec{p}\,) | i \overline{q} 
                     \left[
        \half [ \gamma^{\mu_1} \stackrel{\leftrightarrow}{{D}^{\mu_2}} +
                     \gamma^{\mu_2} \stackrel{\leftrightarrow}{{D}^{\mu_1}}
              ] - 
        \quart \stackrel{\leftrightarrow}{\slashed{D}} \eta^{\mu_1\mu_2}
                     \right] q | N(\vec{p}\,) \rangle^{\cal S} }
   & &
                                                      \nonumber \\
   &=& 2 v^{(q){\cal S}}_2 \left[ 
                        p^{\mu_1} p^{\mu_2} - \quart m_N^2 \eta^{\mu_1\mu_2}
                          \right] \,,
\end{eqnarray}
and more complicated expressions hold for higher moments.
Finally, the non-singlet, NS, matrix element is defined as
\begin{equation}
   v^{\cal S}_{n} \equiv v^{(u){\cal S}}_n - v^{(d){\cal S}}_n \,.
\end{equation}   

In this article, we shall compute $v_2$,
$v_3$ and $v_4$ in the quenched approximation
($n_f = 0$), by finding the appropriate
matrix elements in eq.~(\ref{minkowski_vn}).
As will be seen most effort will be spent on $v_2$, as this is
technically less complicated than the higher moments, and also
numerically the lattice results are more precise.
The lattice approach discretises Euclidean space-time,
with lattice spacing $a$, in the path integral
and simulates the resulting high-dimensional integral 
for the partition function using Monte Carlo techniques.
Matrix elements can then be obtained from suitable
ratios of correlation functions, \cite{martinelli89a,gockeler95a}.
Note that the lattice programme is rather like an experiment:
careful account must be taken of error estimations and extrapolations.
There are three limits to consider:
\begin{itemize}
   \item The spatial box size $L_S$ must be large enough so that finite
         size effects are small.
         Currently sizes of $ \gsim 2\,\mbox{fm}$
         seem large enough (the nucleon diameter is about $1.5\,\mbox{fm}$)%
         \footnote{Additionally all our current lattices have
         $m_{ps}L_S \gsim 4$ where $m_{ps}$ is the pseudoscalar mass.}.
         This situation is probably more favourable for
         quenched QCD (where one drops the fermion determinant in
         the path integral, see section \ref{lattice}), as due to 
         the suppression of the pion cloud, we would expect the
         radius of the nucleon to be somewhat smaller.
         This indeed seems to be the case, see eg \cite{gockeler01a}.
   \item The continuum limit, $a \to 0$. We use $O(a)$ improved Wilson
         fermions (where the discretisation effects of the action
         and matrix elements have been arranged
         to be $O(a^2)$). For unimproved Wilson fermions,
         or where one has not succeeded in entirely $O(a)$ improving
         the matrix element we should extrapolate in $a$ rather
         than $a^2$.
   \item The chiral limit, when the quark mass approaches zero.
         There has recently been much activity on deriving formulae
         for this limit,
         \cite{thomas00a,detmold01a,chen01a,chen01b,arndt01a,chen01c}.
         However while most of these results are valid around the
         physical pion mass on the lattice,
         it is difficult to calculate quark propagators
         at quark masses much below the strange quark mass.
         Thus the use of these formulae is not straightforward.
\end{itemize}
In addition the lattice matrix element must also be renormalised.

Previous lattice studies gave discrepancies to the phenomenological
results, especially for $v_2$. In this work we want to try to narrow
down the sources for this difference. In particular we shall present
here non-perturbative, NP, results for the renormalisation
constants (and as many previous studies used results based on
perturbation theory compare with these other results). We also consider
$O(a)$ improvement and operator mixing to enable a reliable
continuum extrapolation to be performed.

Compared to our previous work \cite{gockeler95a} we have improved
our techniques in several respects:
\begin{itemize}
   \item We employ non-perturbatively improved Wilson fermions instead of
         unimproved Wilson fermions. This should reduce cut-off effects.
   \item Modified operators are used for $v_3$, $v_4$, which improves
         the numerical signal.
   \item In \cite{gockeler95a} we had simulations for a single lattice
         spacing only. Here we shall present results for three different
         values of the lattice spacing so that we can monitor lattice
         artefacts.
   \item As mentioned before, the 1-loop perturbative renormalisation
         factors of \cite{gockeler95a} have been replaced by
         non-perturbatively calculated renormalisation constants.
         In addition we shall pay close attention to possible mixing
         problems of the operators involved.
   \item We have increased the number of quark masses at each value of the
         lattice spacing in order to improve the chiral extrapolation.
\end{itemize}

The organisation of this paper is as follows. In section~\ref{continuum}
various continuum results for the $\beta$ and $\gamma$ functions and
for the Wilson coefficients in the $\overline{MS}$ scheme are collated
and renormalisation group invariant quantities are introduced, while in 
section~\ref{experiment_theory} some NS phenomenological
results are discussed to compare later with the lattice results.
Section~\ref{lattice} describes our lattice techniques:
choice of operators, operator mixing problems, $O(a)$ improvement
and gives the bare, ie unrenormalised results.
Section~\ref{renormalisation} discusses and compares various
renormalisation results: one-loop perturbation theory and a
tadpole improvement, together with the $\rm{RI}^\prime - \rm{MOM}$
non-perturbative method. In section~\ref{chiral_extraps} we discuss
the results and give continuum and chiral extrapolations. Finally in
section~\ref{conclusions} we present our conclusions.

% --------------------------------------------------------------------------

\section{Continuum QCD results}
\label{continuum}

In this section we shall consider the RHS of eq.~(\ref{moment_def}).
Much of the functional form is already known: the lattice input
is reduced to the computation of a single number (for each moment).

The running of the coupling constant as the scale $M$
is changed is controlled by the $\beta$ function.
This is defined as
\begin{eqnarray}
   \beta^{\cal S}  \left(g^{\cal S}(M) \right)
        &\equiv&   \left. {\partial g^{\cal S}(M) \over
                           \partial \log M }\right|_{bare}
                                          \nonumber  \\
        &=&        - b_0(g^{\cal S})^3 - b_1(g^{\cal S})^5
                          - b_2^{\cal S}(g^{\cal S})^7 
                          - b_3^{\cal S}(g^{\cal S})^9 - \ldots \,,
\label{beta_def}
\end{eqnarray}
with the bare parameters being held constant.
This function is given perturbatively as a power series expansion
in the coupling constant. The first two coefficients in the expansion
are universal (ie scheme independent). In the $\overline{MS}$ 
scheme where $({\cal S}, M) = (\overline{MS}, \mu)$,
the expansion is now known to four loops \cite{ritbergen97a,vermaseren97a}.
The three-loop result for quenched QCD is
\begin{equation}
   b_0 = {11\over (4\pi)^2} \,,    \qquad
   b_1 = {102\over (4\pi)^4} \,,   \qquad
   b_2^{\msbar} = { 1\over (4\pi)^6}\left[ {2857\over 2} \right] \,.
\end{equation}
We may immediately integrate eq.~(\ref{beta_def}) to obtain
the solution,
\begin{equation}
   M = \Lambda^{\cal S} \exp{\left[{1\over 2b_0 g^{\cal S}(M)^2}\right]} 
         \left[b_0 g^{\cal S}(M)^2 \right]^{b_1\over 2b_0^2}
         \exp{\left\{ \int_0^{g^{\cal S}(M)} \!\! d\xi
          \left[ {1 \over \beta^{\cal S}(\xi)} +
                 {1\over b_0 \xi^3} - {b_1\over b_0^2\xi} \right]\right\} }\,,
\label{lambda_def}
\end{equation}
where $\Lambda^{\cal S}$ is an integration constant.

While eq.~(\ref{moment_def}) is the conventional definition of the 
moment of a structure function, for us it is more convenient to re-write
it using renormalisation group invariant (or RGI) functions.
The bare operator (or matrix element) must first be renormalised
\begin{equation}
   {\cal O}^{\cal S}(M) 
    = Z_{{\cal O}; bare}^{\phantom{{\cal O};}{\cal S}}(M) {\cal O}_{bare} \,,
\end{equation}   
giving $\gamma$, the anomalous dimension of the operator,
\begin{eqnarray}
   \gamma^{\cal S}_{\cal O} \left(g^{\cal S}(M) \right)
          &\equiv& \left. {\partial \log 
              Z_{{\cal O}}^{\phantom{{\cal O}}\cal S}(M) \over
                           \partial \log M }\right|_{bare}
                                          \nonumber \\
          &=&      d_{{\cal O};0}(g^{\cal S})^2 + 
                   d_{{\cal O};1}^{\cal S}(g^{\cal S})^4 +
                   d_{{\cal O};2}^{\cal S}(g^{\cal S})^6 + \ldots \,.
\label{gamma_fun_def}
\end{eqnarray}
(The first coefficient is again scheme independent.)
One may also change the scale and/or scheme
$({\cal S}, M) \to ({\cal S}^\prime , M^\prime)$ for the operator by
\begin{equation}
   {\cal O}^{{\cal S}^\prime} ( M^\prime )
        = Z_{{\cal O}; {\cal S}}^{\phantom{{\cal O},} {\cal S}^{\prime}}
                (  M^\prime , M )
                {\cal O}^{\cal S} ( M ) \,.
\label{Z_O_Sp_S_def} 
\end{equation}
This leads to two anomalous dimension equations,
obtained by either differentiating with respect to $M^{\prime}$ or $M$.
Integrating these equations gives
\begin{equation}
   Z_{{\cal O}; {\cal S}}^{\phantom{{\cal O};} {\cal S}^{\prime}}
                                                           ( M^\prime , M )
      = { Z_{{\cal O}; bare}^{\phantom{{\cal O};} {\cal S}^\prime} (M^\prime)
          \over
          Z_{{\cal O}; bare}^{\phantom{{\cal O};} {\cal S}} (M) }
      \equiv
        { \Delta Z_{\cal O}^{\cal S}(M) \over 
                    \Delta Z_{\cal O}^{{\cal S}^\prime}(M^\prime)} \,,
\label{Z_O_Sp_S_res}
\end{equation}
where we have defined
\begin{equation}
   [\Delta Z_{\cal O}^{\cal S}(M)]^{-1} = 
          \left[ 2b_0 g^{\cal S}(M)^2 \right]^{- {d_{{\cal O};0}\over 2b_0}}
          \exp{\left\{ \int_0^{g^{\cal S}(M)} d\xi
          \left[ {\gamma^{\cal S}_{\cal O}(\xi)
                             \over \beta^{\cal S}(\xi)} +
                 {d_{{\cal O};0}\over b_0 \xi} \right] \right\} } \,.
\label{deltaZ_def}
\end{equation}
From eqs.~(\ref{Z_O_Sp_S_def}) and (\ref{Z_O_Sp_S_res}),
we see that we can define a RGI operator by
\begin{equation}
   {\cal O}^{\rgi} = \Delta Z_{\cal O}^{\cal S}(M) {\cal O}^{\cal S}(M)
                   \equiv Z^{\rgi}_{\cal O} {\cal O}_{bare} \,.
\label{Zrgi_def}
\end{equation}
Then obviously ${\cal O}^{\rgi}$ is independent of the scale and
scheme. The $\gamma$ function thus controls how the matrix element changes 
as the scale $M$ is varied.
Note also that the normalisation of ${\cal O}^{\rgi}$ depends on the
convention chosen for $\Delta Z_{\cal O}^{\cal S}(M)$,
here given in eq.~(\ref{deltaZ_def}).

As the LHS of eq.~(\ref{moment_def}) is a physical
quantity, the RGI form for the Wilson coefficient is given by
\begin{equation}
   E_{F;v_n}^{\rgi}(Q^2)
      = \Delta Z_{v_n}^{\cal S}(M)^{-1} E_{F;v_n}^{\cal S}
                   \left( {M^2\over Q^2}, g^{\cal S}(M) \right) \,.
\end{equation}
It is convenient to choose $M^2 \equiv Q^2$, as then
\begin{equation}
   E_{F;v_n}^{\rgi}(Q^2) 
      = \Delta Z_{v_n}^{\cal S}(Q)^{-1}
               E_{F;v_n}^{\cal S}(1,g^{\cal S}(Q)) \,,
\label{Ergi_M=Q}
\end{equation}
and $E_{F;v_n}^{\cal S}(1,g^{\cal S}(Q))$ has no large numbers in it,
so that a perturbative power series in $g^{\cal S}$ becomes tenable.
In two schemes ${\cal S}$ and ${\cal S}^\prime$ from eq.~(\ref{Ergi_M=Q})
we have
\begin{equation}
   { E_{F;v_n}^{\cal S}(1,g^{\cal S}(Q)) \over
             E_{F;v_n}^{{\cal S}^\prime}(1,g^{{\cal S}^\prime}(Q)) }
       = { \Delta Z_{v_n}^{\cal S}(Q) \over
                 \Delta Z_{v_n}^{{\cal S}^\prime}(Q) }
         \to 1 \qquad \mbox{as} \quad Q^2 \to \infty \,,
\end{equation}
because in this limit $g^{{\cal S}^\prime} = g^{\cal S} + \ldots \to 0$.
Hence $E_{F;n}^{\cal S}(1,0)$ is independent of the scheme.
With our convention for $f$ this is $1$, so that
\begin{equation}
   E_{F;v_n}^{\cal S} ( 1, g^{\cal S} )
          = 1 + e^{\cal S}_{F;n;0} (g^{\cal S})^2
              + e^{\cal S}_{F;n;1} (g^{\cal S})^4
              + \ldots \,.
\label{wilson_coeff_expan}
\end{equation}

\begin{figure}[t]
   \hspace*{1.0in}
   \epsfxsize=11.00cm \epsfbox{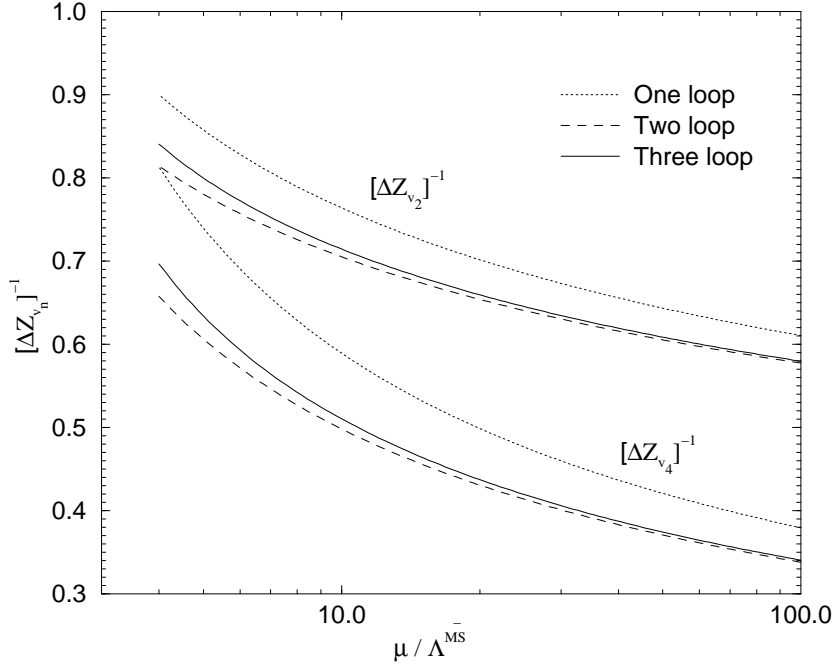}
   \caption{The one, two and three loop results
            for $[\Delta Z^{\msbar}_{v_2}(\mu)]^{-1}$
            and $[\Delta Z^{\msbar}_{v_4}(\mu)]^{-1}$ for quenched
            QCD versus $\mu/\Lambda^{\msbar}$.}
\label{fig_Del_Z_v2+v4_inv_nf0_lat01}
\end{figure}

Practically we shall here only consider the $n=2$, $3$ and $4$ moments.
For these moments we have, for quenched QCD (ie $n_f=0$)
\cite{bardeen78a,vanneerven91a,larin93a,neerven99a},
\begin{eqnarray}
   e_{F_2;n;0}^{\msbar} &=& {1 \over (4\pi)^2}
                                            \left\{ {4\over 9}; \quad
                                                   {29\over 9};\quad
                                                   {91\over 15}
                                            \right\}
                                  \qquad\qquad\qquad n = \{2;3;4\} \,,
                                                     \\
   e_{F_2;n;1}^{\msbar} &=& { 1\over (4\pi)^4}
                                           \left\{
                             {363604\over 3645}
                            -{1024\over 15}\zeta(3); \quad
                             62.74;                 
                                           \right.
                                           \nonumber \\
                          & & \hspace*{2.0in}      \left.
                             {1112778271 \over 3645000} -
                             {1220 \over 9}\zeta(3)
                                            \right\} \,,
                                           \nonumber
\end{eqnarray}
for $n = 2$, $3$ and $4$ respectively
($\zeta(3) = 1.20206\ldots$). The operator has anomalous
dimensions given by, \cite{gonzalez79a,larin93b,retey00a},
\begin{eqnarray}
   d_{v_n;0}        &=& - {1 \over (4\pi)^2} \left\{ {64\over 9};  \quad
                                                   {100\over 9}; \quad
                                                   {628\over 45}
                                            \right\}
                                  \qquad\qquad\qquad n = \{2;3;4\} \,,
                                           \nonumber \\
 d_{v_n;1}^{\msbar} &=& - { 1\over (4\pi)^4} \left\{ {23488\over 243}; \quad
                                                   {34450\over 243}; \quad
                                                   {5241914\over 30375}
                                            \right\} \,,
\label{gammaN=2_res}
                                                     \\
 d_{v_n;2}^{\msbar} &=& - { 1\over (4\pi)^6} \left\{ {11028416\over 6561}
                                             + {2560\over 81}\zeta(3); \quad
                                             {64486199\over 26244}
                                             + {2200 \over 81} \zeta(3);
                                            \right.
                                           \nonumber \\
                   & & \hspace*{2.0in}      \left.
                                            {245787905651 \over 82012500}
                                             + {11512 \over 405}\zeta(3)
                                            \right\} \,,
                                           \nonumber
\end{eqnarray}
again for $n = 2$, $3$ and $4$, respectively.
Solving first eq.~(\ref{lambda_def}) for $g^{\msbar}$ and then
using eq.~(\ref{deltaZ_def}) gives the
results for $[\Delta Z_{v_n}^{\msbar}(\mu)]^{-1}$
shown in Fig.~\ref{fig_Del_Z_v2+v4_inv_nf0_lat01}.
Note that by loop expansion, we mean using the $\beta$ and $\gamma$
function result to the appropriate order; we do not expand
eqs.~(\ref{lambda_def}), (\ref{deltaZ_def}) any further, but solve
them numerically.

\begin{figure}[t]
   \hspace*{1.0in}
   \epsfxsize=11.00cm \epsfbox{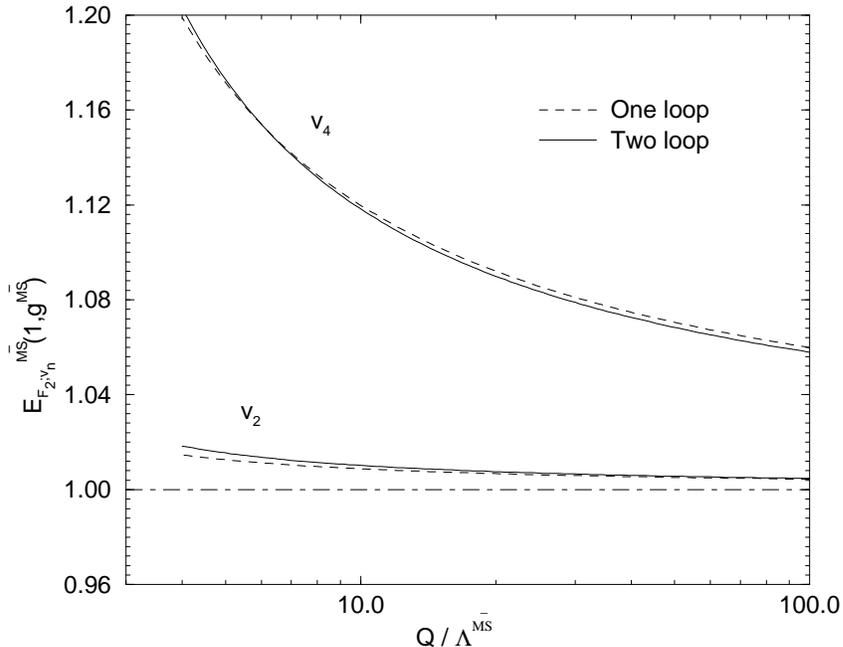}
   \caption{The one and two loop results for $E_{F_2;v_n}^{\msbar}$,
            $n=2$, $4$ for quenched QCD versus $Q/\Lambda^{\msbar}$.
            The two loop results at $2\,\mbox{GeV}$ are 1.011(1),
            1.130(3) for $n = 2$, $4$ respectively
            where the error is a reflection of the error in
            $\Lambda^{\msbar} r_0$.}
\label{fig_wc_v2+v4_nf0_msbar_lat01}
\end{figure}

To determine a result in $\mbox{GeV}$, we shall use the $r_0$ scale here,
\cite{sommer94a}. From \cite{capitani98a, booth01a}
we take $\Lambda^{\msbar} r_0 =  0.602(48)$ for quenched QCD
and together with the scale choice
$r_0 = 0.5\,\mbox{fm} \equiv 1/(394.6\mbox{MeV})$ this gives
for an energy of $Q \equiv \mu = 2\,\mbox{GeV}$ for example,
$\mu/\Lambda^{\msbar} \sim 8.4$. The Wilson coefficient,
$E_{F_2;v_n}^{\msbar}$ can also be found and is shown in
Fig.~\ref{fig_wc_v2+v4_nf0_msbar_lat01} for $n=2$, $4$.
To obtain $E^{\rgi}_{F_2;v_n}$ from eq.~(\ref{Ergi_M=Q}) we must
simply multiply the results from Fig.~\ref{fig_Del_Z_v2+v4_inv_nf0_lat01}
with those of Fig.~\ref{fig_wc_v2+v4_nf0_msbar_lat01}. From this latter
figure we see that the change from the tree level result for the $n=2$
moment in the Wilson coefficient is at most $\sim 1\%$ for
$Q \sim 2\,\mbox{GeV}$ and is practically negligible.
This is not so for the higher moments, when the Wilson coefficient
deviates significantly from one. Useful values for
$\Delta Z^{\msbar}_{v_n}$ (relevant for the forthcoming lattice results)
are given in Table~\ref{table_v2sbar_values} in Appendix~\ref{table_v2sbar}.

% --------------------------------------------------------------------------

\section{Phenomenology and experimental data}
\label{experiment_theory}

Ideally we would like to make a direct comparison between
the theoretical and experimental result, by re-writing
eq.~(\ref{moment_def}) as,
\begin{equation}
   \int_0^1 dx x^{n-2} F^{\NS}(x,Q^2) = f
         E_{F;{v_n}}^{\rgi}(Q^2) v_{n}^{\rgi} \,.
\label{rgi_mom_vn}
\end{equation}
The RHS of this equation has a clean separation between a
number $v_{n}^{\rgi}$, which can only be obtained
using a non-perturbative method (eg the lattice approach)
and a function, $E_{F;v_n}^{\rgi}(Q^2)$,
which describes all the momentum behaviour of the moment.

More conventional (and practical) however is to use
parton densities. Usually phenomenological fits using parton
densities are obtained from global fits (such as MRST, \cite{martin02a}
and CTEQ, \cite{pumplin02a}) to the data.
In this section we shall compare whether taking moments
of the structure function gives the same answer as taking moments
of the parton density. This could also help in estimating the
error in the phenomenological fit. Parton densities $q$, $\overline{q}$
are implicitly defined by
\begin{equation}
   \int_0^1 dx x^{n-1} \left[q^{\cal S}(x, Q) +
                                  (-1)^n \overline{q}^{\cal S}(x,Q)\right]
      = v^{(q){\cal S}}_{n}(Q) \,.
\label{parton_density_def}
\end{equation}
We may relate the structure function to the parton density
via a convolution. Defining similar but separate Mellin transformations 
for even and odd $n$ by
\begin{equation}
   h_n = \int_0^1 dx x^{n-1} h(x) 
                          \quad n = \mbox{even/odd} \,,
                          \qquad
   h(x) = \int_{c - i\infty}^{c + i\infty}
                    { dn \over 2\pi i} x^{-n} h_n  \,, 
\end{equation}
(where in the inverse transformation, $n$ in $h_n$ is analytically
continued from even/odd integer $n$ values to complex numbers and
$c$ is chosen so that all singularities lie to the left of the
line $n = c$) then gives,
\begin{eqnarray}
   F^{\NS}(x,Q^2) &=& f x \int_0^1 {dy \over y} 
                            E^{\cal S}\left( x/y, Q \right)
                                                      \nonumber \\
                  & & \hspace*{0.15in}
                      \times \left[ (u^{\cal S}(y, Q) - d^{\cal S}(y, Q) ) +
                                   \eta (\overline{u}^{\cal S}(y, Q) -
                                        \overline{d}^{\cal S}(y, Q) )
                            \right] \,,
\label{Ftoq}
\end{eqnarray}
where $\eta = +1$ for even $n$ (ie for $F_2^{\gamma;\NS}$)
and $\eta = -1$ for odd $n$ (ie for $F_2^{W^{\pm};\NS}$).
To lowest order in the coupling constant we get from
eq.~(\ref{wilson_coeff_expan}),
$E^{\cal S}(z,Q) = \delta(1-z) + O(g^{\cal S})^2$ and so%
\footnote{There are many references to the relationship between
structure functions and parton densities. See for example 
\cite{londergan98a}.}
\begin{equation}
   F^{\NS}(x,Q^2) = f x \left[ (u^{\cal S}(x, Q) - d^{\cal S}(x, Q) ) +
                               \eta (\overline{u}^{\cal S}(x, Q) -
                                     \overline{d}^{\cal S}(x, Q) )
                        \right] + \ldots \,.
\label{F_parton_density_leading}
\end{equation}
The parton densities are usually determined from global fits
to the data, with an assumed functional form, typically for MRST
results like
\begin{equation}
   xq^{\msbar}(x, Q_0) = 
               A_q x^{\lambda_q} (1 - x )^{\eta_q}
                  \left( 1 + \epsilon_q \sqrt{x} + \gamma_q x \right) \,,
\label{parton_density_fit}
\end{equation}
with parameters $A_q$, $\lambda_q$, $\eta_q$, $\epsilon_q$ and
$\gamma_q$ at some given reference scale $Q_0$.
For the MRST results given here we use the fit MRST200E, \cite{martin01a}
together with the error analysis of \cite{martin02a} at
a scale of $Q_0^2 = 4\,\mbox{GeV}^2$. As a comparison we also consider
the CTEQ fit CTEQ61M, with errors calculated from \cite{pumplin02a}.
(Practically, in both cases, we use the
parton distribution calculator \cite{durhampdf} to compute
the moments in eq.~(\ref{parton_density_def}).)

\begin{figure}[t]
   \hspace*{0.50in}
   \epsfxsize=11.00cm 
       \epsfbox{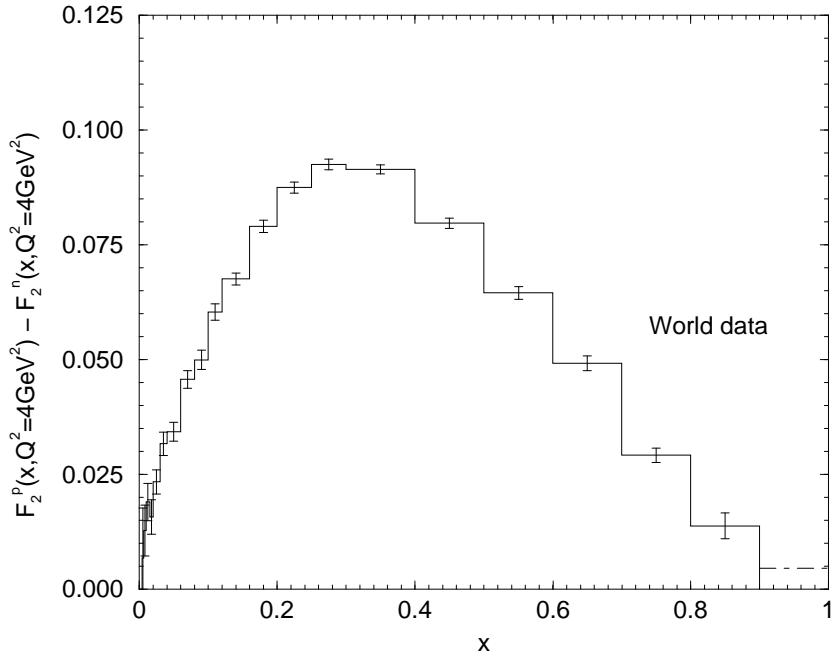}
   \caption{World experimental data for $F_2^{\gamma;\NS}(x,Q_0^2)$,
            \cite{bruell95a}, at $Q_0^2=4\,\mbox{GeV}^2$
            in the form of bins, plotted against $x$ using a
            linear scale. Errors in the bins are also shown.
            The dot-dashed line is a rough estimate, obtained
            by a linear extrapolation of the last bin to nought
            (at $x=1$).}
\label{fig_f2p-f2n_pnq4.f2np_nmcslbc_noht_world}
\end{figure}

Let us now briefly consider some lepton--nucleon DIS
experimental results. While $F^{\gamma;p}_2(x,Q^2)$ is well known,
experiments with deuterium to find $F^{\gamma;n}_2(x,Q^2)$
are much more difficult, due to target nuclear effects.
We shall use here the results from \cite{arneodo94a} which 
employ both proton and neutron (deuterium) targets in the same
experiment, which thus minimizes systematic errors. 
In \cite{bruell95a} this has been combined with
the world data, \cite{amaudruz92a}, and is shown in
Fig.~\ref{fig_f2p-f2n_pnq4.f2np_nmcslbc_noht_world}
in the form of a series of bins at different
$x$ values. (Naively, if there were no QCD interactions,
the parton model would give a delta-function distribution at 
$x = 1/3$. This distribution has been considerably washed
out here though.) There is a paucity of data for larger $x$.
However $F_2^{\gamma;\NS}$ is dropping rapidly to zero,
so any error here will not affect the low moments.
As shown in the figure, we have simply made a linear
extrapolation to $x = 1$, to estimate this region.
To find the moments for eq.~(\ref{moment_def}), we simply need
to find the area under the bins weighted with the appropriate
$x$-moment, ie
\begin{eqnarray}
   { 1 \over f} \int_0^1 dx x^{n-2} F_2^{\gamma;\NS}(x, Q_0^2)
      &\approx&
          {1 \over f} \sum_{bins,b} {1 \over n-1}
               \left( x_{b+1}^{n-1} - x_b^{n-1} \right)
                          F_2^{\gamma;\NS}(x_b, Q_0^2)
                                            \nonumber \\
      &\approx& \left\{ \begin{array}{ll}
                           0.164(4)(1)     & n = 2  \\
                           0.0289(10)(10)  & n = 4  \\
                        \end{array}
                \right. \,,
\end{eqnarray}
where the first error is from the data and the second error is the
effect of dropping the estimated last bin. (As expected, we see
that higher moments are more sensitive to this bin.)

In Table~\ref{table_expt_moments} we give estimates of
\begin{table}[t]
   \begin{center}
      \begin{tabular}{||c||l|l|l||}
         \hline
         \hline
                            & `World'    & MRST       & CTEQ       \\
         \hline
         \hline
      $v_{2}^{\msbar}(Q_0)$ & 0.161(4)   & 0.157(9)   & 0.155(17)  \\
      $v_{2}^{\rgi}$        & 0.226(14)  & 0.220(18)  & 0.217(27)  \\
         \hline
         \hline
      $v_{3}^{\msbar}(Q_0)$ & \multicolumn{1}{c|} - 
                                         & 0.0565(26) & 0.0551(51) \\
      $v_{3}^{\rgi}$        &  \multicolumn{1}{c|} - 
                                         & 0.0972(95) & 0.095(12)  \\
         \hline
         \hline
      $v_{4}^{\msbar}(Q_0)$ & 0.0241(13) & 0.0231(11) & 0.0230(23) \\
      $v_{4}^{\rgi}$        & 0.0480(58) & 0.0460(54) & 0.0458(67) \\
         \hline
         \hline
      \end{tabular}
   \end{center}  
\caption{Values of $v_{n}^{\msbar}(Q_0)$ at $Q_0 = 2\,\mbox{GeV}$.
         The Wilson coefficients (needed for `World') have been
         computed from eq.~(\ref{wilson_coeff_expan}), using
         $\Lambda^{\msbar}|_{n_f=4} = 250(50)\mbox{MeV}$, giving
         $E_{F_2;v_2} = 1.018(3)$, $E_{F_2;v_4} = 1.200(30)$, 
         where the error is a reflection of the error in
         $\Lambda^{\msbar} r_0$. Similarly, to convert to the RGI
         form, $[\Delta Z^{\msbar}_{v_n}(2\,\mbox{GeV})]^{-1}|_{n_f=4}$
         from eq.~(\ref{Zrgi_def}) again uses $\Lambda^{\msbar}_{n_f=4}$
         giving $0.713(40)$, $0.581(50)$ and $0.502(54)$ for
         $n = 2$, $3$ and $4$ respectively.}
\label{table_expt_moments}
\end{table}
$v_{n}^{\msbar}(Q_0)$, $v_{n}^{\rgi}(Q_0)$, using estimates
of the Wilson coefficients and
$[\Delta Z^{\msbar}_{v_n}(2\,\mbox{GeV})]^{-1}$  given in the figure
caption. (These numbers are similar to the quenched results,
as can be seen from the caption of
Fig.~\ref{fig_wc_v2+v4_nf0_msbar_lat01} and
Table~\ref{table_v2sbar_values}.) We find that there is good
agreement between the two methods, with the lowest moment from MRST
or CTEQ being slightly smaller than the experimental result.
Thus these global fits describe the (low) moment data well%
\footnote{Note that a recent analysis, \cite{blumlein04a}, gives 
similar results of $v_2^{\msbar}(Q_0) \approx 0.159$ and
$v_4^{\msbar}(Q_0) \approx 0.0245$.}.
In future for definiteness we use the MRST results.

% --------------------------------------------------------------------------

\section{The Lattice Approach}
\label{lattice}

Euclidean lattice operators%
\footnote{Our Euclidean conventions are described in \cite{best97a}.}
are defined by
\begin{equation}
   {\cal O}^{\Gamma}_{q;\mu_1 \cdots \mu_n}
       = \overline{q}\Gamma_{\mu_1 \cdots \mu_i} 
         \stackrel{\leftrightarrow}{D}_{\mu_{i+1}} \cdots
         \stackrel{\leftrightarrow}{D}_{\mu_n} q \,,
\label{gen_euclid_op}
\end{equation}
where $q$ is taken to be either a $u$ or $d$ quark
and $\Gamma$ is an arbitrary product of Dirac gamma matrices.
(The $q$ index will be usually suppressed.)
We have used the lattice definitions
\begin{eqnarray}
   \stackrel{\rightarrow}{D}_\mu q(x)
          &=& {1 \over 2a} \Big[
               U_\mu (x) q(x + a \hat{\mu}) -
               U_\mu^\dagger (x - a \hat{\mu}) q (x - a \hat{\mu})
                           \Big] \,,
                                            \nonumber \\
   \overline{q}(x) \stackrel{\leftarrow}{D}_\mu 
          &=& {1 \over 2a} \Big[ 
               \overline{q}(x + a \hat{\mu}) U_\mu^\dagger (x) - 
               \overline{q}(x - a \hat{\mu}) U_\mu (x - a \hat{\mu})
                           \Big] \,,
\label{Ddefs}
\end{eqnarray}
and $\stackrel{\leftrightarrow}{D} =
\half ( \stackrel{\rightarrow}{D} - \stackrel{\leftarrow}{D} )$.
For the operators corresponding to eq.~(\ref{Doperators}),
we shall only need $\Gamma = \gamma$. However for
the discussion on mixing under renormalisation,
we shall also use $\Gamma = \gamma\gamma_5$ and $\sigma\gamma_5$.

% ----------------------------------------------------------------------

\subsection{Choice of lattice operators}

We now take the simplifying choices of two momenta
$\vec{p} = \vec{0}$ or $\vec{p} = (p_1, 0, 0 ) \equiv \vec{p}_1$
with $p_1$ being the lowest non-zero momentum possible on a periodic lattice
ie $p_1 = 2\pi /L_S$ where the number of sites in each spatial direction
is $L_S/a$. We take our lattice operators as%
\footnote{${\cal O}_{v_{2a}}$, ${\cal O}_{v_{2b}}$ are the same
operators as we used previously in \cite{gockeler95a}, while
there are some modifications to ${\cal O}_{v_3}$ and ${\cal O}_{v_4}$.
For ${\cal O}_{v_3}$ we have effectively made the proper rotation
(ie preserving $\gamma_5$) of $4 \leftrightarrow 1$ (and
$2 \leftrightarrow 3$) of the operator in \cite{gockeler95a}.
This means that the measured ratio in
eq.~(\ref{R_def_practical}) is $-ip_1E_{\vec{p}_1} v_3$ rather than
$- p_1^2 v_3$ and hence the signal is better by a factor
$O(E_{\vec{p}_1}/p_1)$. For ${\cal O}_{v_4}$ we have symmetrised over
the $2$, $3$ indices of the operator given in \cite{gockeler95a},
which makes the measurement of the ratio for the new operator numerically
a little less noisy.},
\begin{eqnarray}
   {\cal O}_{v_{2a}} &=& {\cal O}^{\gamma}_{\{14\}} \,,
                                                    \nonumber \\
   {\cal O}_{v_{2b}} &=& {\cal O}^{\gamma}_{\{44\}} 
                          - \third \left( {\cal O}^{\gamma}_{\{11\}}
                                         +{\cal O}^{\gamma}_{\{22\}}
                                         +{\cal O}^{\gamma}_{\{33\}} \right)
                                                    \,,
                                                    \nonumber \\
   {\cal O}_{v_3}    &=& {\cal O}^{\gamma}_{\{441\}} - 
                          \half \left( {\cal O}^{\gamma}_{\{221\}} +
                                       {\cal O}^{\gamma}_{\{331\}} \right)
                                                    \,,
                                                    \nonumber \\
   {\cal O}_{v_4}    &=&  {\cal O}^{\gamma}_{\{1144\}} 
                         + {\cal O}^{\gamma}_{\{2233\}}
                         - \half
                         \left(   {\cal O}^{\gamma}_{\{1133\}}
                                + {\cal O}^{\gamma}_{\{1122\}}
                                + {\cal O}^{\gamma}_{\{2244\}}
                                + {\cal O}^{\gamma}_{\{3344\}} 
                         \right) \,.
\label{lattice_ops}
\end{eqnarray}
Of course, there are other possibilities. However these will all
require non-zero momenta in two directions or suffer from more severe
mixing problems. As we shall see, even a non-zero momentum
in one direction leads to a strong degradation
of the signal and with two non-zero momenta very
little signal is observed, \cite{brower96a}.

The transformation properties under the hypercubic group $H(4)$
are given in Table~\ref{table_H4_transformation}, \cite{gockeler96a}.
\begin{table}[t]
   \begin{center}
      \begin{tabular}{||c|c||}
         \hline
         \hline
         \multicolumn{1}{||c}{Operator}   &
         \multicolumn{1}{|c||}{$(\tau_k^{(l)},{\cal C})$} \\
         \hline
         \hline
         ${\cal O}_{v_{2a}}$  & $(\tau^{(6)}_3,+)$      \\
         ${\cal O}_{v_{2b}}$  & $(\tau^{(3)}_1,+)$      \\
         \hline
         ${\cal O}_{v_3}$     & $(\tau^{(8)}_1,-)$      \\
         ${\cal O}_{v_4}$     & $(\tau^{(2)}_1,+)$      \\
         \hline
         \hline
      \end{tabular}
   \end{center}
\caption{Transformation of the various operators under the 
         hypercubic group $H(4)$, \protect\cite{gockeler96a},
         where $l$ represents the dimension of the representation
         $\tau_k^{(l)}$, $k$ labels different representations of the
         same dimension and ${\cal C}$ is the charge conjugation
         parity of the operator.}
 \label{table_H4_transformation}
\end{table}
Note in particular that the `off-diagonal' (${\cal O}_{v_{2a}}$)
and `diagonal' operators (${\cal O}_{v_{2b}}$) for $v_2$ belong
to different representations, in distinction to the continuum operator.
Thus we expect that although these operators should have different
lattice artefacts and renormalisation factors, in the continuum limit
both should lead to the same result -- potentially a useful check.

% ----------------------------------------------------------------------

\subsection{Mixing of lattice operators}
\label{operator_mixing}

A given operator of engineering dimension $d_{\cal O}$ can mix with
operators (with the same quantum numbers) of lower dimension,
the same dimension or higher dimension. 
$O(a)$ improvement involves mixing with one dimension higher operators 
(irrelevant operators) where the choice of
improvement coefficients is conventionally treated separately
to mixing with operators of dimension $\le d_{\cal O}$ (relevant operators).
We shall follow this practice here.

% ----------------------------------------------------------------------

\subsubsection{Operator mixing with additional relevant operators}
\label{op_mising_rel}

While for ${\cal O}_{v_2}$ there is no mixing
with relevant operators, unfortunately for ${\cal O}_{v_3}$
and ${\cal O}_{v_4}$ there are other relevant operators transforming
identically under $H(4)$ which can thus mix with the original operator,
\cite{gockeler96a}. Specifically we have%
\footnote{Using the operators given in \cite{gockeler95a} we would have
\begin{eqnarray}
   {\cal O}_{v_3}^{m_1} =
     {\cal O}^{\gamma}_{\langle\langle 411 \rangle\rangle} -
     \half \left(
       {\cal O}^{\gamma}_{\langle\langle 422 \rangle\rangle} +
       {\cal O}^{\gamma}_{\langle\langle 433 \rangle\rangle}
            \right) \,, \qquad
  {\cal O}_{v_3}^{m_2} = 2 {\cal O}^{\gamma\gamma_5}_{|| 213 ||} \,,
                                             \nonumber
\end{eqnarray}
and
         \begin{eqnarray}
            {\cal O}_{v_4}^{m_1} \hspace*{-0.075in}&=&\hspace*{-0.075in}
                 - {\cal O}^{\gamma}_{1144} - {\cal O}^{\gamma}_{4114}
                 - {\cal O}^{\gamma}_{1441} - {\cal O}^{\gamma}_{4411}
                 + 2 {\cal O}^{\gamma}_{1414} + 2 {\cal O}^{\gamma}_{4141}
                                             \nonumber \\ 
                                \hspace*{-0.075in}& &\hspace*{-0.075in}
                 - {\cal O}^{\gamma}_{2233} - {\cal O}^{\gamma}_{3223}
                 - {\cal O}^{\gamma}_{2332} - {\cal O}^{\gamma}_{3322}
                 + 2 {\cal O}^{\gamma}_{2323} + 2 {\cal O}^{\gamma}_{3232}
                                             \nonumber \\ 
                               \hspace*{-0.075in}& &\hspace*{-0.075in}
                 + {\cal O}^{\gamma}_{1133} + {\cal O}^{\gamma}_{3113}
                 + {\cal O}^{\gamma}_{1331} + {\cal O}^{\gamma}_{3311}
                 - 2 {\cal O}^{\gamma}_{1313} - 2 {\cal O}^{\gamma}_{3131}
                                             \nonumber \\ 
                                \hspace*{-0.075in}& &\hspace*{-0.075in}
                 + {\cal O}^{\gamma}_{2244} + {\cal O}^{\gamma}_{4224}
                 + {\cal O}^{\gamma}_{2442} + {\cal O}^{\gamma}_{4422}
                 - 2 {\cal O}^{\gamma}_{2424} - 2 {\cal O}^{\gamma}_{4242}
                                             \,,
                                             \nonumber \\
            {\cal O}_{v_4}^{m_2} \hspace*{-0.075in}&=&\hspace*{-0.075in}
                 + {\cal O}^{\gamma\gamma_5}_{1234}
                 - {\cal O}^{\gamma\gamma_5}_{3214}
                 - {\cal O}^{\gamma\gamma_5}_{1432}
                 + {\cal O}^{\gamma\gamma_5}_{3412}
                 + {\cal O}^{\gamma\gamma_5}_{2143}
                 - {\cal O}^{\gamma\gamma_5}_{4123}
                 - {\cal O}^{\gamma\gamma_5}_{2341}
                 + {\cal O}^{\gamma\gamma_5}_{4321}
                                             \nonumber \\
                                \hspace*{-0.075in}& &\hspace*{-0.075in}
                 + {\cal O}^{\gamma\gamma_5}_{1243} 
                 - {\cal O}^{\gamma\gamma_5}_{4213}
                 - {\cal O}^{\gamma\gamma_5}_{1342}
                 + {\cal O}^{\gamma\gamma_5}_{4312}
                 + {\cal O}^{\gamma\gamma_5}_{2134}
                 - {\cal O}^{\gamma\gamma_5}_{3124}
                 - {\cal O}^{\gamma\gamma_5}_{2431}
                 + {\cal O}^{\gamma\gamma_5}_{3421} \,,
                                             \nonumber \\
            {\cal O}_{v_4}^{m_3} \hspace*{-0.075in}&=&\hspace*{-0.075in} 
              i \left (
    {\cal O}^{\sigma\gamma_5}_{2314} - {\cal O}^{\sigma\gamma_5}_{2341} 
  + {\cal O}^{\sigma\gamma_5}_{1423} - {\cal O}^{\sigma\gamma_5}_{1432} 
  + {\cal O}^{\sigma\gamma_5}_{2413} - {\cal O}^{\sigma\gamma_5}_{2431} 
  + {\cal O}^{\sigma\gamma_5}_{1324} - {\cal O}^{\sigma\gamma_5}_{1342}
              \right ) \,.
                                             \nonumber
         \end{eqnarray}
}
\begin{itemize}
   \item Operators mixing with ${\cal O}_{v_3}$:
         \begin{eqnarray}
            {\cal O}_{v_3}^{m_1} &=&
                {\cal O}^{\gamma}_{\langle\langle 144 \rangle\rangle} -
                \half \left(
                  {\cal O}^{\gamma}_{\langle\langle 133 \rangle\rangle} +
                  {\cal O}^{\gamma}_{\langle\langle 122 \rangle\rangle}
                      \right) \,,
                                             \nonumber \\
            {\cal O}_{v_3}^{m_2} &=&
                  {\cal O}^{\gamma\gamma_5}_{|| 432 ||}
                  + 3 {\cal O}^{\gamma\gamma_5}_{| 432 |} 
                  = 2 {\cal O}^{\gamma\gamma_5}_{|| 342 ||} \,,
            \label{Ov3mi}
         \end{eqnarray}
         where using the notation of \cite{gockeler96a} we have defined:
         \begin{eqnarray}
            {\cal O}^\Gamma_{| \nu_1 \nu_2 \nu_3 |} &=&
                 {\cal O}^\Gamma_{\nu_1 \nu_2 \nu_3 } 
               - {\cal O}^\Gamma_{\nu_1 \nu_3 \nu_2 }
               - {\cal O}^\Gamma_{\nu_3 \nu_1 \nu_2 } 
               + {\cal O}^\Gamma_{\nu_3 \nu_2 \nu_1 } \,,
                                             \nonumber \\
            {\cal O}^\Gamma_{||\nu_1 \nu_2 \nu_3||} &=&
                 {\cal O}^\Gamma_{\nu_1 \nu_2 \nu_3 } 
               - {\cal O}^\Gamma_{\nu_1 \nu_3 \nu_2 }
               + {\cal O}^\Gamma_{\nu_3 \nu_1 \nu_2 } 
               - {\cal O}^\Gamma_{\nu_3 \nu_2 \nu_1 }
               - 2 {\cal O}^\Gamma_{\nu_2 \nu_3 \nu_1 } 
               + 2 {\cal O}^\Gamma_{\nu_2 \nu_1 \nu_3 } \,,
                                             \nonumber \\
            {\cal O}^\Gamma_{\langle \langle \nu_1 \nu_2 \nu_3 
                             \rangle \rangle} &=&
                 {\cal O}^\Gamma_{\nu_1 \nu_2 \nu_3 } 
               + {\cal O}^\Gamma_{\nu_1 \nu_3 \nu_2 }
               - {\cal O}^\Gamma_{\nu_3 \nu_1 \nu_2 }
               - {\cal O}^\Gamma_{\nu_3 \nu_2 \nu_1 } \,.
         \end{eqnarray}
   \item Operators mixing with ${\cal O}_{v_4}$:
         \begin{eqnarray}
            {\cal O}_{v_4}^{m_1} \hspace*{-0.075in}&=&\hspace*{-0.075in}
              - {\cal O}^{\gamma}_{1144} - {\cal O}^{\gamma}_{4114}
              - {\cal O}^{\gamma}_{1441} - {\cal O}^{\gamma}_{4411}
              + 2 {\cal O}^{\gamma}_{1414} + 2 {\cal O}^{\gamma}_{4141}
                                             \nonumber \\ 
                           \hspace*{-0.075in}& &\hspace*{-0.075in}
              - {\cal O}^{\gamma}_{2233} - {\cal O}^{\gamma}_{3223}
              - {\cal O}^{\gamma}_{2332} - {\cal O}^{\gamma}_{3322}
              + 2 {\cal O}^{\gamma}_{2323} + 2 {\cal O}^{\gamma}_{3232}
         \label{Ov4mi}
                                             \nonumber \\ 
                               \hspace*{-0.075in}& &\hspace*{-0.075in}
              + \half \left(
              + {\cal O}^{\gamma}_{1133} + {\cal O}^{\gamma}_{3113}
              + {\cal O}^{\gamma}_{1331} + {\cal O}^{\gamma}_{3311}
              - 2 {\cal O}^{\gamma}_{1313} - 2 {\cal O}^{\gamma}_{3131}
                \right.
                                             \nonumber \\ 
                                \hspace*{-0.075in}& &\hspace*{-0.075in}
                \left. \hspace*{0.30in}
              + {\cal O}^{\gamma}_{1122} + {\cal O}^{\gamma}_{2112}
              + {\cal O}^{\gamma}_{1221} + {\cal O}^{\gamma}_{2211}
              - 2 {\cal O}^{\gamma}_{1212} - 2 {\cal O}^{\gamma}_{2121}
                \right.
                                             \nonumber \\ 
                                \hspace*{-0.075in}& &\hspace*{-0.075in}
                \left. \hspace*{0.30in}
              + {\cal O}^{\gamma}_{2244} + {\cal O}^{\gamma}_{4224}
              + {\cal O}^{\gamma}_{2442} + {\cal O}^{\gamma}_{4422}
              - 2 {\cal O}^{\gamma}_{2424} - 2 {\cal O}^{\gamma}_{4242}
                \right.
                                             \nonumber \\ 
                                \hspace*{-0.075in}& &\hspace*{-0.075in}
                \left. \hspace*{0.30in}
              + {\cal O}^{\gamma}_{3344} + {\cal O}^{\gamma}_{4334}
              + {\cal O}^{\gamma}_{3443} + {\cal O}^{\gamma}_{4433}
              - 2 {\cal O}^{\gamma}_{3434} - 2 {\cal O}^{\gamma}_{4343}
                \right) \,,
                                             \nonumber \\
            {\cal O}_{v_4}^{m_2} \hspace*{-0.075in}&=&\hspace*{-0.075in}
                \half \left(
              + {\cal O}^{\gamma\gamma_5}_{1234}
              - {\cal O}^{\gamma\gamma_5}_{3214}
              - {\cal O}^{\gamma\gamma_5}_{1432}
              + {\cal O}^{\gamma\gamma_5}_{3412}
              + {\cal O}^{\gamma\gamma_5}_{2143}
              - {\cal O}^{\gamma\gamma_5}_{4123}
              - {\cal O}^{\gamma\gamma_5}_{2341}
              + {\cal O}^{\gamma\gamma_5}_{4321}
                \right.
                                             \nonumber \\ 
                          \hspace*{-0.075in}& &\hspace*{-0.075in}
                \left. \hspace*{0.175in}
              - {\cal O}^{\gamma\gamma_5}_{1324}
              + {\cal O}^{\gamma\gamma_5}_{2314}
              + {\cal O}^{\gamma\gamma_5}_{1423}
              - {\cal O}^{\gamma\gamma_5}_{2413}
              - {\cal O}^{\gamma\gamma_5}_{3142}
              + {\cal O}^{\gamma\gamma_5}_{4132}
              + {\cal O}^{\gamma\gamma_5}_{3241}
              - {\cal O}^{\gamma\gamma_5}_{4231}
                \right)
                                             \nonumber \\ 
                          \hspace*{-0.075in}& &\hspace*{-0.075in}
                       \hspace*{0.175in}
                 + {\cal O}^{\gamma\gamma_5}_{1243}
                 - {\cal O}^{\gamma\gamma_5}_{4213}
                 - {\cal O}^{\gamma\gamma_5}_{1342}
                 + {\cal O}^{\gamma\gamma_5}_{4312}
                 + {\cal O}^{\gamma\gamma_5}_{2134}
                 - {\cal O}^{\gamma\gamma_5}_{3124}
                 - {\cal O}^{\gamma\gamma_5}_{2431}
                 + {\cal O}^{\gamma\gamma_5}_{3421}
                                             \nonumber \\ 
            {\cal O}_{v_4}^{m_3} \hspace*{-0.075in}&=&\hspace*{-0.075in}
                \half i\left(
       + {\cal O}^{\sigma\gamma_5}_{2413} - {\cal O}^{\sigma\gamma_5}_{2431} 
       + {\cal O}^{\sigma\gamma_5}_{1324} - {\cal O}^{\sigma\gamma_5}_{1342}
       - {\cal O}^{\sigma\gamma_5}_{3412} + {\cal O}^{\sigma\gamma_5}_{3421} 
       - {\cal O}^{\sigma\gamma_5}_{1234} + {\cal O}^{\sigma\gamma_5}_{1243}
                       \right.
                                             \nonumber \\ 
                          \hspace*{-0.075in}& &\hspace*{-0.075in}
                \left. \hspace*{0.20in}
       + 2{\cal O}^{\sigma\gamma_5}_{2314} - 2{\cal O}^{\sigma\gamma_5}_{2341} 
       + 2{\cal O}^{\sigma\gamma_5}_{1423} - 2{\cal O}^{\sigma\gamma_5}_{1432}
                \right) \,.
         \end{eqnarray}
         ${\cal O}_{v_4}^{m_3}$ is an operator with one lower dimension 
         (and different chiral properties) than the other ${\cal O}_{v_4}$
         operators. It is also possible that four-fermion operators
         can mix: we shall not consider this here though.
\end{itemize}

In Appendix~\ref{group} we illustrate, by an example for ${\cal O}_{v_4}$,
how ${\cal O}_{v_4}^{m_1}$ and ${\cal O}_{v_4}^{m_2}$ can be derived.
The other mixing operators follow by similar considerations.

While this list contains all operators allowed by group theoretical
arguments, not all the operators contribute,
see section~\ref{renormalisation}.

% ----------------------------------------------------------------------

\subsubsection{Operator Improvement}

As we are using $O(a)$ improved fermions, to achieve the elimination
of $O(a)$ terms in matrix elements, the corresponding operators
must also include additional irrelevant terms, with coefficients
appropriately chosen. Presently only for the lowest moment (ie $v_2$)
are these extra operators known, \cite{capitani00a}.
In this case we should replace
the operators ${\cal O}^{\gamma}_{\mu\nu}$ in $v_{2a}$ or $v_{2b}$ by%
\footnote{Often, alternatively, the improved operator is re-written as
$(1 + am_q b_{\cal O})({\cal O} + \sum_{i=1}^3 a c^{\prime}_i
{\cal O}^{(i)})$. In the remainder of this article it should hopefully be
clear from context whether we are referring to the operator or its
improved version.}
\begin{eqnarray}
   {\cal O}^{\gamma}_{\mu\nu}
     &\to& \left( 1 + am_q c_0 \right) {\cal O}^{\gamma}_{\mu\nu}
           + \sum_{i=1}^3 a c_i {\cal O}^{(i)}_{\mu\nu}
                                             \nonumber \\
     &\equiv&
           \left( 1 + am_q c_0 \right) \overline{q} \gamma_\mu 
             \stackrel{\leftrightarrow}{D}_{\nu} q
                                             \nonumber \\
     & &   + a c_1 \left[ i\sum_{\lambda} \overline{q} \sigma_{\mu\lambda}
             \stackrel{\leftrightarrow}{D}_{\left[\nu\right.}
             \stackrel{\leftrightarrow}{D}_{\left.\lambda\right]} q
                   \right]
           + a c_2 \left[ - \overline{q}
             \stackrel{\leftrightarrow}{D}_{\left\{\mu\right.}
             \stackrel{\leftrightarrow}{D}_{\left.\nu\right\}} q
                   \right]
                                             \nonumber \\
    & &    + a c_3 \left[ i \sum_\lambda \partial_\lambda \left( 
             \overline{q} \sigma_{\mu\lambda}
             \stackrel{\leftrightarrow}{D}_{\nu} q \right)
                   \right] \,,
\label{op_improvement}
\end{eqnarray}
where $m_q$ is the bare lattice quark mass, related to the hopping
parameter $\kappa$ by
\begin{equation}
   am_q = {1 \over 2} \left( {1\over \kappa} - {1\over \kappa_c} \right) \,.
\label{bare_qm_def}
\end{equation}
So we see that there are potentially four additional improvement operators,
defined in eq.~(\ref{op_improvement})
together with four unknown improvement coefficients, $c_i(g_0)$,
for $i = 0, \ldots, 3$, which are functions of the gauge coupling
constant $g_0$. The $i=3$ operator only contributes for
non-forward matrix elements, which are not considered here
and so this term may be dropped.
Also ultimately $c_0$ will not concern us as we are only interested
in the result in the chiral limit.
For on-shell matrix elements the equation of motion may be used
to eliminate one of the improvement terms, for example
we can choose $c_0$ and $c_1$ as linear functions of $c_2$.
First order perturbation theory gives the relation between
the improvement coefficients for $v_{2b}$ of
\begin{eqnarray}
   c_0 &=& 1 - c_2 + {g_0^2C_F \over 16\pi^2} ( 17.20377 - 8.69045c_2) 
                   + O(g_0^4) \,,
                                         \nonumber \\
   c_1 &=& c_2 + O(g_0^2) \,,
\label{pert_improvement_coeff_v2b}
\end{eqnarray}
($C_F=4/3$) and a similar expression for $v_{2a}$, \cite{capitani00a}.
With these values of the improvement coefficients, $O(a)$
corrections to the $v_{2b}$ matrix element have been eliminated
(at least to lowest order perturbatively).
We see that we cannot determine the $O(g_0^2)$ term of $c_1$
because it is the coefficient of an operator that vanishes
at tree level. Non-perturbatively the improvement coefficients
have not yet been determined. However one might suspect,
that choosing $c_2=0$ also gives in this case a small $c_1$
coefficient.

For higher moments, the bases for the improved operators
become increasingly cumbersome, as not only would we expect
more possible irrelevant operators built from the original operator
together with an additional covariant derivative, but also
four-fermion operators may play a role. This does not necessarily
detract from the original operator though, because we can always attempt
to make a continuum extrapolation in $a$ rather than $a^2$
if we cannot motivate why the irrelevant matrix elements are small.

% ----------------------------------------------------------------------

\subsection{Determining the matrix element}
\label{determining_me}

Matrix elements are determined from the ratio $R$ of three point to
two point correlation functions,
\begin{equation}
   R(t,\tau; \vec{p}; {\cal O})
      = { C_{\half(1+\gamma_4)}(t,\tau; \vec{p}; {\cal O}) \over
              C_{\half(1+\gamma_4)}(t;\vec{p})} \,,
\label{R_def_general}
\end{equation}
where $C_{\half(1+\gamma_4)}(t;\vec{p})$ is the unpolarised nucleon two point
function with a source at time $0$ and sink at time $t$, while the also
unpolarised three point function
$C_{\half(1+\gamma_4)}(t,\tau; \vec{p}; {\cal O})$
has an operator insertion at time $\tau$.
If we consider a region $0 \ll \tau \ll t \lsim \half L_T$
(ie well inside the nucleon branch of the propagator) where $L_T$ is
the temporal extension of the lattice, then the transfer matrix
formalism upon projecting out the ground state nucleon leads to
\begin{eqnarray}
   R(t,\tau; \vec{p}; {\cal O}_{v_{2a}})
      &=& i p_1 v_{2a} \,,
                                              \nonumber  \\
   R(t,\tau; \vec{p}; {\cal O}_{v_{2b}})
      &=& - { E_{\vec{p}}^2 + \third \vec{p}^{\,2} \over E_{\vec{p}} }
            v_{2b} \,,
                                              \nonumber  \\
   R(t,\tau; \vec{p}_1; {\cal O}_{v_3})
      &=& - i p_1 E_{\vec{p}_1} v_3 \,,
                                              \nonumber  \\
   R(t,\tau; \vec{p}_1; {\cal O}_{v_4})
      &=& p_1^2 E_{\vec{p}_1} v_4 \,,
\label{R_def_practical}
\end{eqnarray}
for the bare matrix elements $v_n$.
For some more details see eg \cite{best97a,gockeler95a}.
In general we would expect
that the best signals with the smallest noise are seen for zero momentum.
We take (and have numerically checked) that the
standard dispersion relation $E^2_{\vec{p}} = m_N^2 + \vec{p}^{\,2}$
holds for $\vec{p}=\vec{p}_1$.

\begin{figure}[t]
   \hspace*{0.50in}
   \begin{tabular}{cc}
      \epsfxsize=5.00cm \epsfbox{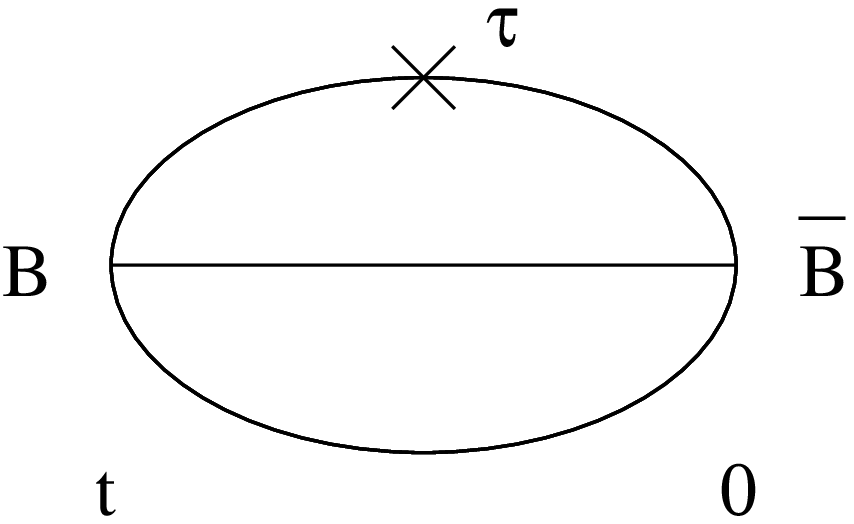}     &
      \hspace{1.0cm}
      \epsfxsize=5.00cm \epsfbox{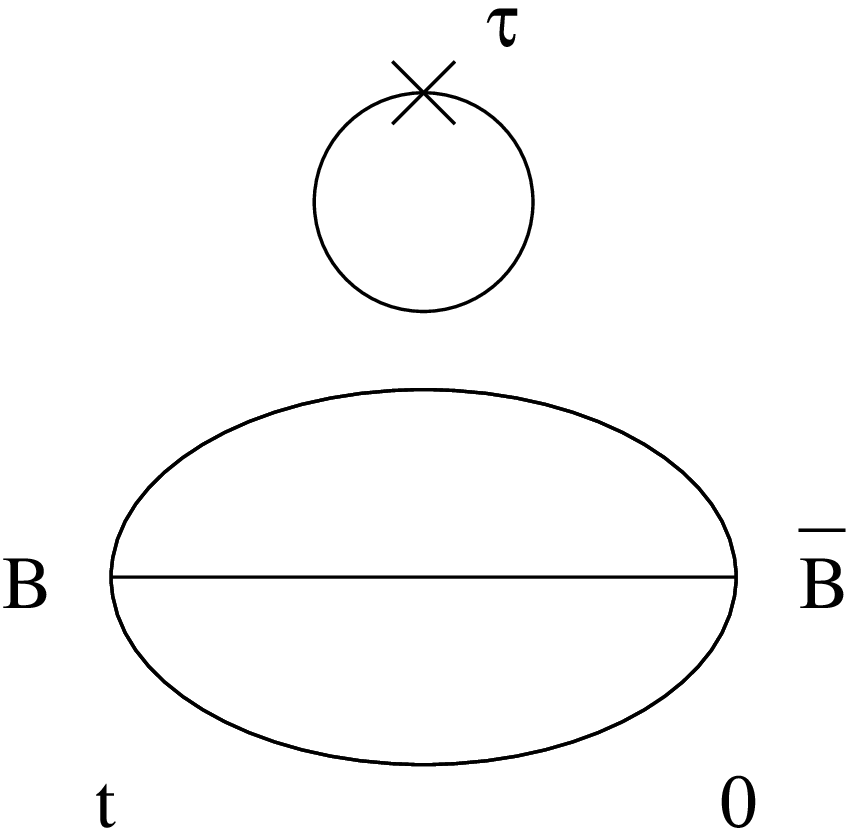}
   \end{tabular}
   \caption{The $3$-point quark correlation function for a baryon.}
   \label{fig_B_3pt}
\end{figure}

The nucleon three-point correlation function is depicted
in Fig.~\ref{fig_B_3pt}.
While the two-point correlation function consists
of one diagram, for the three-point correlation function
we have two diagrams - a `quark line connected' contribution,
and a `quark line disconnected' contribution, left and right diagrams 
in Fig.~\ref{fig_B_3pt} respectively.
This is not the usual field theoretic splitting of the
Green's function into connected and disconnected diagrams.
As quarks can travel backwards in time as well as forwards,
we would expect that the quark line connected term
would also give a contribution to the anti-quark
parton density defined in eq.~(\ref{parton_density_def}).
As quark line disconnected diagrams can, by definition, only interact
with the hadron via the exchange of gluons then the numerical
results suffer from large short distance (ie ultra-violet) fluctuations.
So a very large number of configurations is required, which is very
expensive in computer time. We have not computed this term here.
To cancel any effects of these disconnected terms, if flavour
$SU_F(2)$ symmetry is a good symmetry, it is
sufficient to consider NS matrix elements,
ie the $u$ quark matrix element minus the $d$ quark matrix element.
In Appendix~\ref{appendix_3pt}, for completeness, we give explicit
expressions for the relevant three-point correlation functions.

A further class of disconnected terms are given by gluon
matrix elements. Again on the lattice these are difficult to compute,
see \cite{gockeler96b}, but again they cancel upon considering
non-singlet matrix elements.

All these gluonic or sea-quark effects are concentrated at
small $x$, and thus for higher moments, ie $n = 3$ or $4$,
are naturally suppressed anyway. Thus disconnected contributions
may be less significant, so that the computation of singlet matrix
elements is then more reliable and we can consider just
a $u$ or $d$ operator matrix element. Although the quenched approximation
does not handle the sea-quarks correctly, we might also expect for
these higher moments that quenching has less effect.
But these statements are hard to quantify, and
as this is all less likely to be the case for the lowest moment anyway,
we shall consider mainly the non-singlet results here.

% ----------------------------------------------------------------------

\subsection{Raw results for lattice matrix elements}
\label{bare_results}

We now discuss our raw numerical results for the lattice operators
and the numerical significance of the additional improvement
operators and/or additional relevant operators to the nucleon
matrix element. Since our original publication \cite{gockeler95a} which
employed unimproved Wilson fermions at $\beta = 6.0$,
we have used quenched configurations with $O(a)$ improved
fermions: thus for the action we take the standard Wilson gluon
action, while for the fermion propagator the standard Wilson
fermion is used together with a `clover' improvement term.
The (non-perturbative) coefficient $c_{sw}$ of the improvement term
was taken from \cite{luscher96b}. This means that on-shell quantities,
such as masses, have only $O(a^2)$ discretisation effects.
Simulations were performed at three $\beta \equiv 6/g_0^2$ values,
$\beta = 6.0$, $6.2$, $6.4$. At each $\beta$ value, four or more
quark masses (degenerate in $u$ and $d$) were used,
at each mass a statistic of several hundred configurations was generated.
Antiperiodic fermion boundary conditions were taken 
in the time direction and periodic in the remaining
spatial directions. Further details of our runs are given in
Appendix~\ref{tables} in Table~\ref{table_run_params}.

To improve the overlap with the nucleon we employed Jacobi smearing,
and used a non-relativistic (NR) projection of the nucleon.
Jacobi smearing is described for example in \cite{best97a},
where a hopping parameter ($\kappa_s$) expansion (of order $n_s$)
of a Wilson fermion operator restricted to a time plane smears
out the original point quark source.
For $\beta = 6.0$, $6.2$ and $6.4$ we use $(\kappa_s, n_s)=$
$(0.21, 50)$, $(0.21, 100)$ and $(0.21, 150)$ respectively
giving a root-mean-square radius of about $0.4\,\mbox{fm}$,
a reasonable fraction of the nucleon radius $\sim 0.8\,\mbox{fm}$.
The NR projection, where in our Dirac gamma matrix representation
only the upper two components of the spinors are used,
is briefly described in \cite{gockeler94a} and Appendix~\ref{appendix_3pt}.

The nucleon sink positions were chosen as $t = 13a$, $17a$ and $23a$
for $\beta = 6.0$, $6.2$ and $6.4$ respectively and the fit ranges
in $\tau$ were taken as $[4a,9a]$, $[6a,11a]$ and $[7a,16a]$.
These fit range values were not so critical,
but allowed a splitting of the range $[0,t]$ into
roughly three equal parts each piece being roughly the same
physical size.

\begin{figure}[t]
   \hspace*{1.00cm}
   \epsfxsize=12.25cm \epsfbox{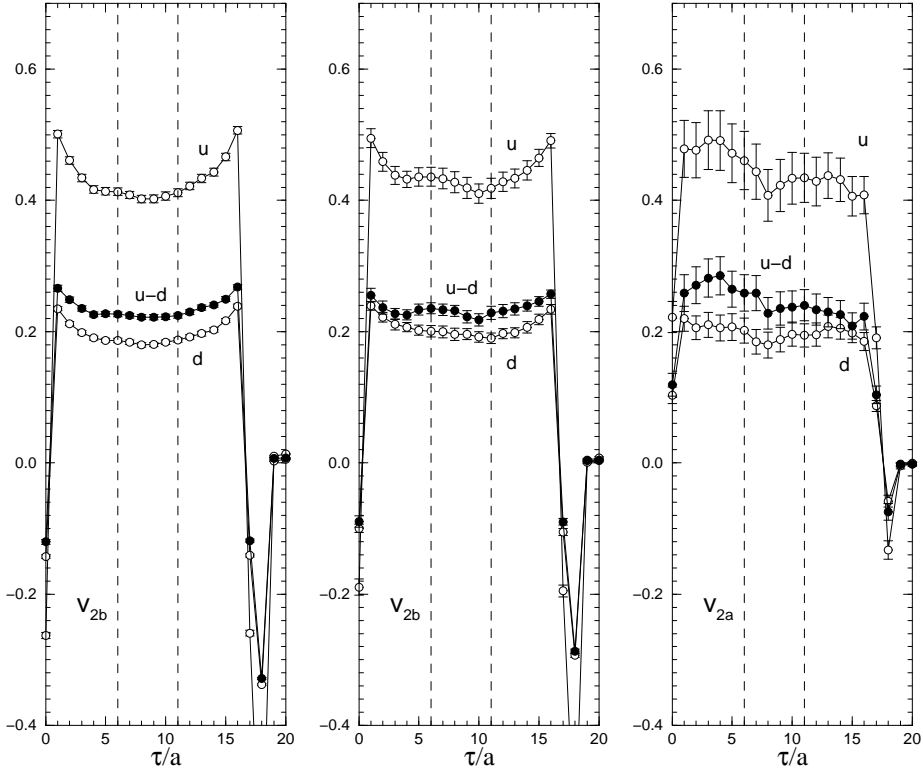}
   \caption{$v_{2b}$ versus $\tau/a$ from the ratios
            $R(17a,\tau;\vec{0};{\cal O}_{v_{2b}})$,
            $R(17a,\tau;\vec{p}_1;{\cal O}_{v_{2b}})$,
            eq.~(\ref{R_def_practical}) left and middle pictures
            respectively and $v_{2a}$ from
            $R(17a,\tau;\vec{p}_1;{\cal O}_{v_{2a}})$, right picture
            for ${\cal O}^{(u)}$, ${\cal O}^{(d)}$ (empty circles) and NS 
            (ie ${\cal O}^{(u)}-{\cal O}^{(d)}$)
            (filled circles) for $\beta = 6.2$ at $\kappa = 0.1344$.
            The chosen fit intervals are denoted by vertical dotted lines.}
   \label{Rratio_bare_b6p20kp1344_3pics}
\end{figure}

In Appendix~\ref{tables}, the data is presented in tables giving
separately the $u$ and $d$ contributions. As discussed previously though,
as the disconnected diagrams have not been computed,
the most physically significant result
is for the non-singlet matrix elements. 
Thus we have also repeated the data analysis directly for these
matrix elements. This allows, in particular, a better estimation
of the error. These numbers are also given in the tables in
Appendix~\ref{tables}. As an example of the typical ratios obtained,
in Fig.~\ref{Rratio_bare_b6p20kp1344_3pics} we show
$R(17a,\tau;\vec{0};{\cal O}_{v_{2b}})$,
$R(17a,\tau;\vec{p}_1;{\cal O}_{v_{2b}})$,
$R(17a,\tau;\vec{p}_1;{\cal O}_{v_{2a}})$, left to right pictures
respectively, for $\beta = 6.2$ and $\kappa = 0.1344$. We seek a plateau
in the region $0 \ll \tau/a \ll 17$. The region chosen is denoted by 
vertically dashed lines. Clearly the operator corresponding to $v_{2b}$
delivers a better signal for the ratio than the operator for $v_{2a}$,
although even for the $v_{2b}$ operator we see that it is better to
choose zero momentum rather than non-zero momentum.

In this section, as mentioned before, we wish to merely
estimate the numerical significance of extra operators,
as described in section~\ref{operator_mixing}.
To this end, as mixing coefficients and improvement coefficients
are not so well known, we make a series of plots either comparing
the ratio of the additional improvement operators
(ie the $v_n$ constructed using the ${\cal O}^{(i)}$ in
eqs.~(\ref{op_improvement}) and (\ref{R_def_practical}))
to the original operator,
\begin{equation}
   r^i_{v_n} \equiv { v_{n}^i \over v_{n} } \,,
\label{Rimprov}
\end{equation}
or compare directly the $v_n^{m_i}$ derived from
eqs.~(\ref{Ov3mi}), (\ref{Ov4mi}) to the original operator, $v_n$.
(It will be seen in section~\ref{renormalisation} that the ratios
$r^i_{v_n}$ have a physical significance.)

These are all plotted against the square of the pseudoscalar meson mass.
As $(am_{ps})^2$ is proportional to the quark mass for small 
quark mass, using this quantity avoided the necessity of
first determining the critical hopping parameter $\kappa_c$
in the quark mass definition, eq.~(\ref{bare_qm_def}).
We have thus used the (dimensionless) extrapolation parameter
$(r_0/a)^2 \times (am_{ps})^2 \equiv (r_0 m_{ps})^2$, with $r_0/a$
being given by the Pad{\'e} formula in \cite{guagnelli98a}.
Using $r_0 = 0.5\,\mbox{fm}$ this enables us to get an idea of
how close our simulation points are to the chiral limit in physical units.

% ----------------------------------------------------------------------

\subsubsection{$v_{2}$}

\begin{figure}[t]
   \hspace*{1.00cm}
   \epsfxsize=12.25cm \epsfbox{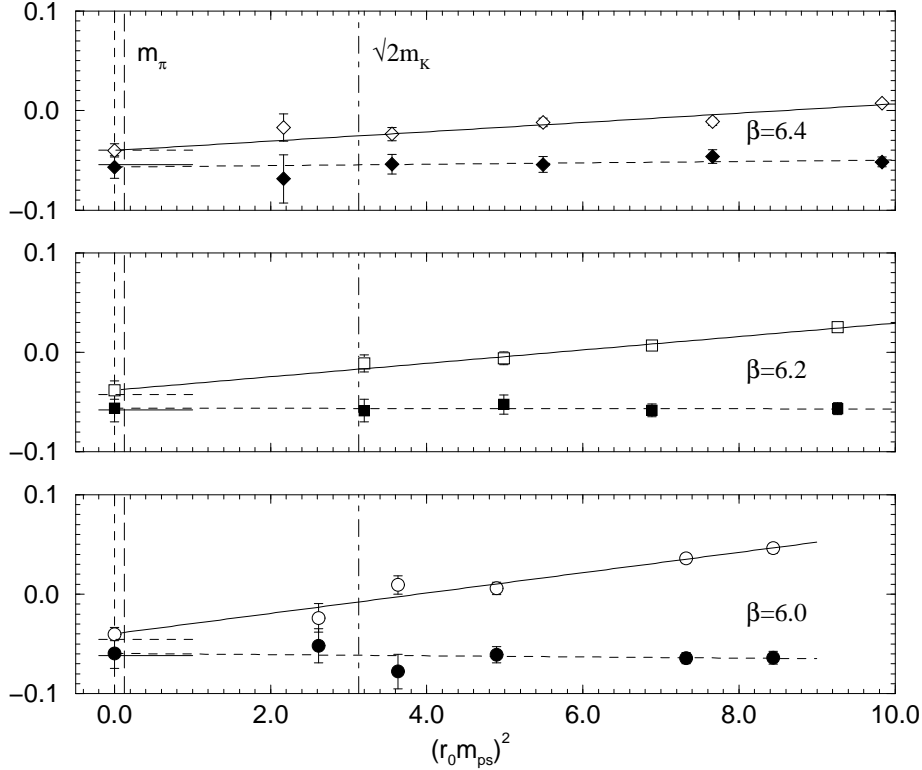}
   \caption{The ratios $ar^1_{v_{2a}}$, $ar^2_{v_{2a}}$
            plotted against $(r_0m_{ps})^2$. The filled symbols at
            non-zero $(r_0m_{ps})^2$ are for $ar^1_{v_{2a}}$,
            while the empty symbols denote $ar^2_{v_{2a}}$.
            Also shown is a linear chiral extrapolation
            (dashed line and full line for $ar^1_{v_{2a}}$, $ar^2_{v_{2a}}$
            respectively). The result in the chiral limit
            $(r_0m_{ps})^2 \equiv 0$ is also indicated, again using
            full, empty symbols for $ar^1_{v_{2a}}$, $ar^2_{v_{2a}}$
            respectively.
            The small horizontal lines around the chiral limit represent
            the perturbative estimate, see section~\ref{renormalisation},
            with again a dashed line corresponding to $ar^1_{v_{2a}}$
            and a full line to $ar^2_{v_{2a}}$ respectively.
            The rough values of the (hypothetical) pseudoscalar mass
            composed from $u/d$ or $s$ quark masses, evaluated
            from $m_\pi$ and $m_K$ respectively are shown as
            dashed vertical lines.}
   \label{fig_x1a_1u-1d.pm1_O1+O2_030814_1054}
\end{figure}

In Fig.~\ref{fig_x1a_1u-1d.pm1_O1+O2_030814_1054}
we show $ar_{2a}^1$, $ar_{2a}^2$ plotted against $(r_0m_{ps})^2$,
together with a linear chiral extrapolation. Also plotted is the
approximate value of the pseudoscalar mass corresponding to degenerate
$u/d$ or $s$ quark masses. So we see that our simulation runs over
the range from about two to three times the strange quark mass to a
little under the strange quark mass. It is also noticeable that
we are a long way from simulating with a light $u/d$ quark mass
-- in fact  within our errors, there is no difference between linearly
extrapolating to the chiral limit or to the pion mass.
Nevertheless we see that the effect of any extra improvement
operators is likely to be small, of the order of a few percent.

\begin{figure}[p]
   \hspace*{1.00cm}
   \epsfxsize=13.00cm \epsfbox{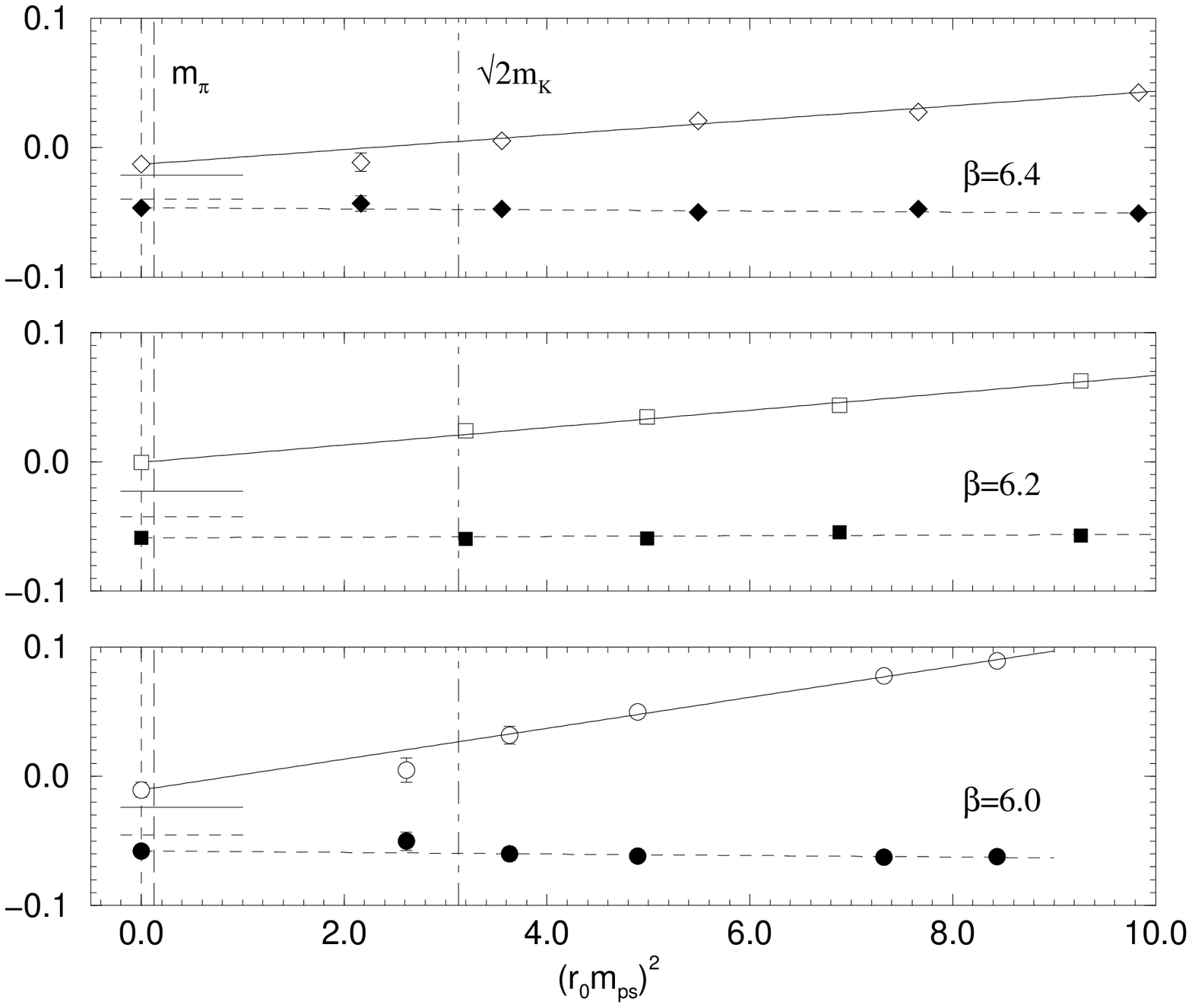}
   \caption{$ar^1_{v_{2b}}$, $ar^2_{v_{2b}}$ versus $(r_0m_{ps})^2$
            for $\beta =6.0$, $6.2$ and $6.4$ with $\vec{p}=\vec{0}$.
            The notation is the same as for
            Fig.~\ref{fig_x1a_1u-1d.pm1_O1+O2_030814_1054}.}
   \label{fig_x1b_1u-1d.p0_O1+O2_030813_2342}
\end{figure}

The above result is for the off-diagonal operator,
which needs a non-zero momentum in its evaluation.
This figure is to be compared with the result using
the diagonal operator, which advantageously may use zero momentum.
In Fig.~\ref{fig_x1b_1u-1d.p0_O1+O2_030813_2342}
we show this result. In comparison with the previous picture,
the errors are considerably reduced, as $\vec{p}=0$
is used (see also Fig.~\ref{Rratio_bare_b6p20kp1344_3pics}).
Indeed using $\vec{p}=\vec{p}_1$ (see Appendix~\ref{tables}
for the numbers) we see that the errors grow again, although they are
never as large as for the off-diagonal operator.
Again, the effect of any extra improvement operators is small.

Thus all the numerical results and linearly chirally
extrapolated results for $ar^i_{v_2}$ look small giving
$av_2^i \lsim 5\%$ of $v_2$ and some indeed are consistent with zero.
From eq.~(\ref{pert_improvement_coeff_v2b}), we see that if we choose
$c_2 = 0$ then $c_1 \equiv O(g_0^2)$, which is likely to be small.
So these results lead numerically to a small additional improvement
term and we conclude that the effect of these terms in
$v_2$ can be assumed to be negligible.

% ----------------------------------------------------------------------

\subsubsection{$v_{3}$ and $v_{4}$}
\label{v_3andv_4}

\begin{figure}[p]
   \hspace*{1.00cm}
   \epsfxsize=13.00cm 
           \epsfbox{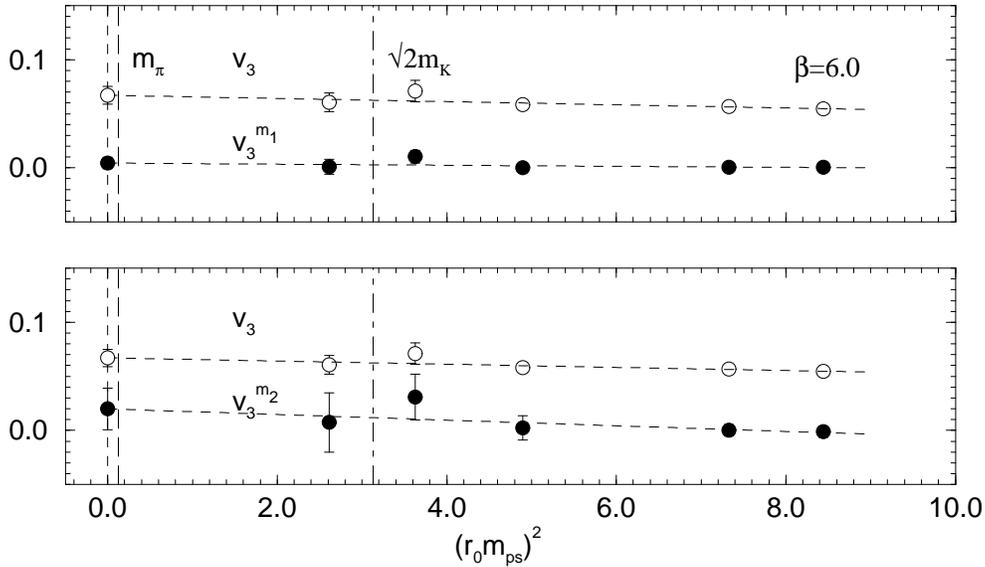}
   \caption{Mixing terms, $v_{3}^{m_1}$, $v_{3}^{m_2}$
            for $\beta = 6.0$ (filled circles).
            Also shown is $v_{3}$ (open circles)
            and linear chiral extrapolations. 
            The same notation as for
            Fig.~\ref{fig_x1a_1u-1d.pm1_O1+O2_030814_1054}.}
   \label{fig_x2bmi1+2_1u-1d.pm1_b6p0_040726_1325}
\end{figure}

We now present our results for the higher moments. 
In Fig.~\ref{fig_x2bmi1+2_1u-1d.pm1_b6p0_040726_1325}
we show $v_3^{m_1}$, $v_3^{m_2}$ for $\beta = 6.0$,
together with a linear chiral extrapolation.
Also shown, for comparison, is the operator $v_3$.
As expected while the magnitude of the noise has increased
in comparison with $v_2$ and scatters more, an acceptable signal
is still seen. We find a clear separation between $v_3$ and
the mixing operators (indeed they are consistent with $\approx 0$).
For higher $\beta$ values, the data fluctuates more
and it becomes more difficult to disentangle the results.

\begin{figure}[t]
   \hspace*{1.00cm}
   \epsfxsize=13.00cm \epsfbox{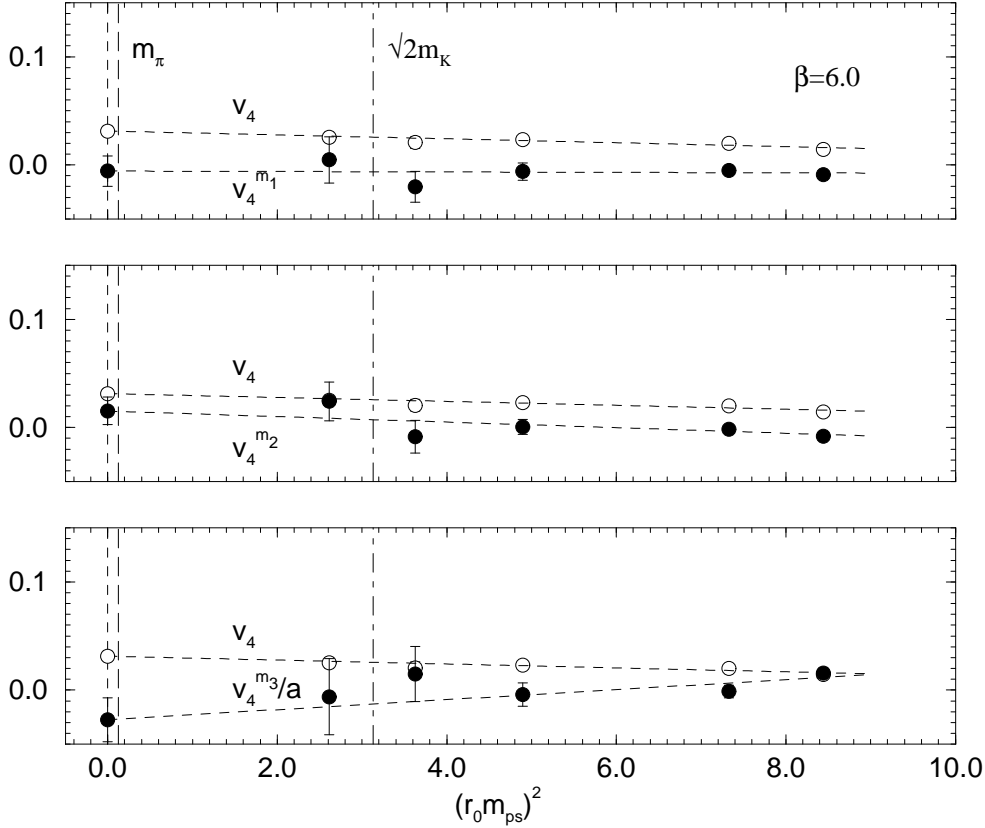}
   \caption{$v_4$ for $\beta =6.0$ together with $v_4^{m_1}$, 
            $v_4^{m_2}$, $v_4^{m_3}/a$. The notation is the same as for
            Fig.~\ref{fig_x1a_1u-1d.pm1_O1+O2_030814_1054}.}
    \label{fig_x3mi1+2+3_1u-1d.pm1_040927_1742}
\end{figure}

In Fig.~\ref{fig_x3mi1+2+3_1u-1d.pm1_040927_1742} we plot $v_4^{m_1}$,
$v_4^{m_2}$, $v_4^{m_3}/a$, together with $v_4$. While there is a
reasonable signal for $v_4$, the improvement terms fluctuate a lot
(again becoming worse the higher the value of $\beta$ is).
Indeed even finding a plateau for the ratio $R(t,\tau; \vec{p}; {\cal O})$
in eq.~(\ref{R_def_general}) becomes problematical. 
It would appear that while, numerically, $v_4^{m_1}$, $v_4^{m_2}$
are small in comparison with $v_4$ the situation is less clear cut for 
$v_4^{m_3}/a$ (although it is only including the heaviest quark mass
point that leads to a non-zero result in the chiral limit).
Nevertheless as most of the results for the mixing terms are much
smaller than $v_4$ we shall ignore any effects from them.

% ----------------------------------------------------------------------

\section{Operator renormalisation}
\label{renormalisation}

A lattice operator (or matrix element) must, in general, be renormalised.
Again, we shall discuss mixing and $O(a)$ operator improvement separately.

% ----------------------------------------------------------------------

\subsection{Operator mixing and renormalisation}
\label{operator_mixing+renormalisation}

For operator mixing we can generally write
\begin{equation}
   {\cal O}^{\cal S}_i(M) = \sum_j
   Z_{{\cal O}_i {\cal O}_j; \lat}^{\phantom{{\cal O}_i {\cal O}_j;}\cal S}
     (M, a) {\cal O}_j^{\lat}(a) \,,
\label{operator_mixing_def}
\end{equation}
working in a scheme ${\cal S}$ at scale $M$. So we must determine
a matrix of renormalisation constants. While for those mixing operators
of the same dimension as the original operator, low order
perturbation results can be calculated (at least in principle),
for lower dimensional operators this is not reliable: the
renormalisation constant is proportional to a positive power of $1/a$
and non-perturbative terms may contribute.
For the higher moments, both cases occur.
For $v_3$ both the $m_1$ and $m_2$ operators have the same dimension
as the original operator, and for $v_4$, $m_1$ and $m_2$ have the same
dimension, while $m_3$ is dimension one lower,
see eqs.~(\ref{Ov3mi}), (\ref{Ov4mi}).

% ----------------------------------------------------------------------

\subsubsection{Renormalisation and relevant operator mixing}
\label{renormalisation+rel_op}

In this section we want to make a few comments on the 
operator mixings seen for the operators we use in this paper, 
and contrast them with the mixing problems for the operators which 
we rejected. 

In the continuum, symmetry under  $O(4)$ in the Euclidean case 
(or the Lorentz group in the Minkowski case) 
imposes strong constraints on which operators can have 
unpolarised forward nucleon matrix elements. 
One way of stating the rule is to say that the only operators
with non-zero unpolarised forward  matrix elements are those
which have $J^P = 0^+$ in the nucleon's rest frame. The 
quantum numbers  $J^P = 0^+$ only occur in $O(4)$ representations
of the form $U_+^{(\mu, \mu)}$ in the notation of~\cite{baake83a}.
Considering the $O(4)$ classification of the operators is
relevant, because although bare lattice matrix elements only
respect the symmetries of the hypercubic group $H(4)$, 
renormalised quantities should respect the full continuum 
symmetry group. 

If we look at the mixing operators for $v_3$, as listed in
eq.~(\ref{Ov3mi}), we see that under the hypercubic group
they transform exactly the same way as $v_3$, but under $O(4)$ 
they both transform according to the representation
$U^{(\half,\threehalf)}$, which means that by the continuum
symmetries their renormalised forward matrix elements must be zero, 
\begin{equation}
   v_{3}^{m_i;{\cal S}} = 0 \,. 
\end{equation} 
Similarly the mixing operators for $v_4$, listed in eq.~(\ref{Ov4mi}),
all transform according to the $U^{(0,2)}$ representation of $O(4)$, so 
\begin{equation}
  v_{4}^{m_i;{\cal S}} = 0 \,. 
\end{equation} 

If the renormalised mixing operators all have the value zero,
we can express all the bare lattice matrix elements in terms of 
a single renormalised matrix element. Again taking the $v_3$ system  
as our exemplar, 
\begin{eqnarray}
   v_3 &=& \left(Z^{-1}\right)_{v_3;v_3} v_3^{\cal S} \,,
                                                 \nonumber \\
   v_3^{m_i}
       &=&  \left(Z^{-1}\right)_{v_3^{m_i};v_3} v_3^{\cal S} \,,
\label{v3_diag}
\end{eqnarray} 
(temporarily dropping, for clarity, other arguments
of the renormalisation constants).
Since all the lattice matrix elements are multiples of the
physically interesting matrix element $v_3^{\cal S}$ we have a 
choice in how we calculate the renormalised matrix element. 
We could add up all the terms in eq.~(\ref{operator_mixing_def}),
assuming that we know the mixing $Z$s well enough.
Or we could equally well just invert eq.~(\ref{v3_diag})
and calculate  $v_3^{\cal S}$ from $v_3$ alone,
\begin{equation}
   v_3^{\cal S} = { v_3 \over 
                    \left(Z^{-1}\right)_{v_3;v_3} } \,.
\label{renorm_practice} 
\end{equation} 
If we calculate the renormalised matrix element from $v_3$ alone, 
we should really renormalise by dividing by
$(Z^{-1})_{v_3;v_3}$ instead of multiplying by
$Z_{v_3;v_3}$. In practice the difference is minor,
\begin{equation}
   { 1 \over \left(Z^{-1}\right)_{v_3;v_3} }
    \approx Z_{v_3;v_3}
             - \sum_i { Z_{v_3;v_3^{m_i}} Z_{v_3^{m_i};v_3}
                        \over Z_{v_3^{m_i};v_3^{m_i}} } \,.
\label{Zapprox} 
\end{equation}
The difference involves the product of two off-diagonal $Z$s,
and so is ${\cal O}(g_0^4)$ in perturbation theory,
see section~\ref{perturbation_theory}, and is still likely
to be tiny non-perturbatively. Note that eq.~(\ref{Zapprox}) tells us
that mixing with lower dimension operators is no more dangerous than
mixing with operators of the same dimension, in a case like this where
the lower-dimensional operator has zero renormalised matrix element. 
This is because the product $Z_{v_3;v_3^{m_i}} Z_{v_3^{m_i};v_3}$
is dimensionless, even when $v_3^{m_i}$ is a lower dimensional 
operator. 

We conclude that for the operators we use in this paper, 
mixing is relatively benign, because continuum symmetry says that
the renormalised mixing matrix elements are zero.

Finally we want to contrast this with an example where this is not
so, to show the importance of choosing the operators carefully.  
We could have tried to measure $v_3$ with the  operator
\begin{equation}
   O^\gamma_{\{444\}} - 
      O^\gamma_{\{411\}}-  O^\gamma_{\{422\}} -  O^\gamma_{\{433\}} \;. 
\label{v3bad}
\end{equation} 
This however would have much worse mixing problems because it mixes with
operators which are allowed to have nucleon matrix elements in the
continuum, for example 
\begin{equation} 
   O^\gamma_{\{444\}} + 
     O^\gamma_{\{411\}}+  O^\gamma_{\{422\}} +  O^\gamma_{\{433\}} \,, \quad
     \overline{q} \gamma_4 q \quad \hbox{ or } \quad  
      \overline{q} \stackrel{\leftrightarrow}{D}_4 q \;. 
\end{equation}
Now we would not have the option of using eq.~(\ref{renorm_practice}),
we would  have to use eq.~(\ref{operator_mixing_def})
and to get a reliable answer we would need to
know the off-diagonal elements of $Z$ accurately (especially those
for the operators of lower dimension). This is why we rejected the 
operator eq.~(\ref{v3bad}), and why we think that the mixing operators
of subsection~\ref{op_mising_rel} are not a problem.

% ----------------------------------------------------------------------

\subsubsection{Renormalisation and operator improvement}
\label{renormalisation+op_imp}

For $O(a)$ improved operators, we know that the operator depends
linearly on the improvement coefficients $c_i$ as shown in
eq.~(\ref{op_improvement}). Our experience from extensive
perturbative calculations in \cite{capitani00a} leads us to expect
a result of the form 
 \begin{equation}
    \Lambda_{{\cal O}^i}^{\lat} = {\zeta_i \over a} \Lambda_{\cal O}^{\lat}
                                     + O(a^0) \,,
\label{zeta_def}
\end{equation}  
where $\Lambda^{\lat}$ is the lattice amputated
three-point Green's function, ${\cal O}$ is the relevant operator
and ${\cal O}^i$ are the improvement or irrelevant operators 
in eq.~(\ref{op_improvement}). (Note that there is nothing forbidding
the improvement terms mixing with ${\cal O}/a$.) The only case that
does not mix is the mass improvement term $a m_q {\cal O}$, 
\begin{equation}
   a m_q \Lambda_{\cal O}^{\lat} = O(a) \,.
\end{equation}
We can calculate the $c_i$ dependence of the renormalisation constant 
by requiring that the renormalised $\Lambda^{\cal S}_{\cal O}$ 
should be independent of the $c_i$ at leading order in $a$
and thus we have (temporarily suppressing additional $M$, $a$ arguments)
\begin{eqnarray}
   \Lambda^{\cal S}_{\cal O}
         &=& 
       { Z_{\cal O;\lat}^{\phantom{{\cal O};}\cal S}(\{c_i\}) \over
         Z_{q; \lat}^{\phantom{q;}{\cal S}} }
           \Big[ \Lambda^{\lat}_{\cal O}
                 + \sum_{i\not=0} c_i \zeta_i  \Lambda^{\lat}_{\cal O} +O(a)
           \Big] \,,
                                                  \nonumber \\
        &=&
      { Z_{\cal O;\lat}^{\phantom{{\cal O};}\cal S}(\{c_i\}) \over
         Z_{q; \lat}^{\phantom{q;}{\cal S}} }
            \Big( 1 + \sum_{i\not=0} c_i \zeta_i \Big) \Lambda^{\lat}_{\cal O}
                 + O(a) \,,
                                                  \nonumber \\
        &\equiv&
      { Z_{\cal O;\lat}^{\phantom{{\cal O};}\cal S}(\{0\}) \over
         Z_{q; \lat}^{\phantom{q;}{\cal S}} }
               \Lambda^{\lat}_{\cal O} + O(a) \,,
\end{eqnarray} 
and so we can write
\begin{equation}
   {\cal O}^{\cal S}(M)
      = Z_{\cal O;\lat}^{\phantom{{\cal O};}\cal S}(M, a, \{c_i\})
          {\cal O}(a, \{c_i\}, c_0)
      \equiv
        { Z_{\cal O;\lat}^{\phantom{{\cal O};}\cal S}(M,a, \{0\})
          \over 1 + \sum_{i \ne 0} c_i \zeta_i}
               {\cal O}(a; \{c_i\}, c_0 ) \,,
\label{renorm_op}
\end{equation}
where the coefficients $\zeta_i \equiv \zeta_i(g_0) = O(g_0^2)$
have to be determined.
Again in \cite{capitani00a} we have given these coefficients
for $v_{2a}$ and $v_{2b}$ in lowest order perturbation theory
using eq.~(\ref{zeta_def}). (Requiring $O(a)$ improvement also
determines the $c_i$ as given for example in
eq.~(\ref{pert_improvement_coeff_v2b}).)
Numerically these coefficients turn out to be quite small.
In addition to the perturbative $\zeta_i$'s we can also find
non-perturbative $\zeta_i$'s by looking at our measured nucleon
matrix elements and requiring that the improved result and the
unimproved agree to leading order in $a$,
ie from eq.~(\ref{renorm_op}) we have 
\begin{equation}
   { v_n + a\sum_{i \ne 0} c_i v_n^i
          \over 1 + \sum_{i \ne 0} c_i \zeta_i }  =  v_n + O(a) \,,
\end{equation}
giving, cf eq.~(\ref{zeta_def}),
\begin{equation}
   v_n^i = {1\over a} \zeta_i v_n + O(a^0)  \,,
\end{equation}
or  
\begin{equation}
   \zeta_i = \left. ar^i_{v_n} \right|_{am_q=0}  + O(a) \,,
\end{equation}
(see eq.~(\ref{Rimprov})). We would expect $\zeta_i$ calculated in
this way to agree up to ambiguities of $O(a)$.
In Fig.~\ref{fig_x1a_1u-1d.pm1_O1+O2_030814_1054} and 
Fig.~\ref{fig_x1b_1u-1d.p0_O1+O2_030813_2342} we compare
the values for $\zeta_i$ obtained this way with the $1$-loop
perturbative results from \cite{capitani00a} (shown in the
figures as short horizontal lines about the chiral limit).
Although, not unexpectedly, there are differences to the perturbative
result, the ratios remain small. From
eq.~(\ref{pert_improvement_coeff_v2b}) we may choose $c_2=0$
and so $c_1=O(g_0^2)$ is also small. Hence the change in the
denominator of eq.~(\ref{renorm_op}) from $1$ is small
and so does not change the renormalisation constant perceptibly.
Thus we shall ignore any small effects here
(just as in section~\ref{bare_results}, where our conclusion
was that we could numerically drop the $O(a)$ improvement terms).

% ----------------------------------------------------------------------

\subsection{Determining the renormalisation constants}
\label{determine_ren_const}

To define the renormalisation constants a renormalisation procedure
must be prescribed. Often the renormalisation constants
are defined first in a MOM scheme by computing the (Landau) gauge fixed
two-quark Green's function with one operator insertion
and setting
\begin{equation}
   \left[ \Lambda^{\mom}_{{\cal O}} \right]_{p^2 = \mu_p^2}
      = \left. \Lambda^{\mom}_{{\cal O}} \right|_{\born} \,,
\label{Z_definition}
\end{equation}
where $\Lambda^{\mom}_{{\cal O}}$ is the (renormalised) amputated
three-point Green's function for the operator $\cal O$.
The RHS of this equation is the tree level value
(or Born approximation) of the amputated function.
This definition may be used for perturbative computations,
see eg \cite{gockeler96c}, where $[ \ldots ]$ in eq.~(\ref{Z_definition})
means that, in a given basis, we drop terms not proportional to the
Born term. This may be modified, using a trace condition
for the definition of the renormalisation constants,
to give the alternative $\rm{RI}^\prime - \rm{MOM}$ scheme
\cite{martinelli94a} which is also suitable for non-perturbative
calculations. A discussion and some results for this non-perturbative
method will be given in section~\ref{np_Z_determination}.

The resulting $Z_{{\cal O}_i; \lat}^{\phantom{{\cal O};} \mom}$ may,
if wished, be converted to another scheme,
ie to the $\overline{MS}$ scheme using eq.~(\ref{Z_O_Sp_S_def}) where
$Z_{{\cal O}_i;{\cal O}_j ; \mom}^{\phantom{{\cal O}_i{\cal O}_j ;} \msbar}
(\mu, \mu_p)$ is (perturbatively) calculable.
The ${\overline{MS}}$ scheme is particularly convenient,
as the renormalisation constants are independent of the gauge fixing
condition chosen.

Unfortunately, the definition given in eq.~(\ref{Z_definition})
has its limitations: ${\cal O}^{m_2}_{v_3}$, ${\cal O}^{m_1}_{v_4}$
${\cal O}^{m_2}_{v_4}$ and ${\cal O}^{m_3}_{v_4}$
all have vanishing Born matrix elements between quark states
(as they involve commutators of covariant derivatives).
So in order to be able to compute the renormalisation constants
we would have to consider more general Green's functions
(eg quark-gluon). At present we must simply ignore this problem though.

After these more general remarks, we shall now give
various results for the renormalisation constants for (one-loop)
perturbation theory, TRB perturbation theory and finally a
non-perturbative determination of the relevant constants.

% ----------------------------------------------------------------------

\subsubsection{Perturbation Theory}
\label{perturbation_theory}

One loop perturbation theory%
\footnote{For a general review of lattice perturbation theory, 
see \cite{capitani02a}.}
yields%
\footnote{For diagonal elements we write
$B^{\cal S}_{{\cal O}_i;{\cal O}_i} \equiv B^{\cal S}_{{\cal O}_i}$ and
$d_{{\cal O}_i;{\cal O}_i;0} \equiv d_{{\cal O}_i;0}$.}
\begin{equation}
  Z_{{\cal O}_i;{\cal O}_j;\lat}^{\phantom{{\cal O}_i;{\cal O}_j;}\cal S}(M,a)
     = \delta_{{\cal O}_i;{\cal O}_j} + g_0^2 
       \left[ d_{{\cal O}_i;{\cal O}_j;0} \ln(aM)
                           - B^{\cal S}_{{\cal O}_i;{\cal O}_j}(c_{sw})
                         \right] + O(g_0^4) \,.
\label{one_loop_Z}
\end{equation}
In the $\overline{MS}$ scheme ($M \equiv \mu$) we have, \cite{capitani00a},
\begin{eqnarray}
   B^{\msbar}_{v_{2a}}(c_{sw})
                       &=& { C_F \over (4\pi)^2 }
                           \left( 1.27959 - 3.87297c_{sw} - 0.67826c_{sw}^2
                           \right) \,,
                                              \nonumber  \\
   B^{\msbar}_{v_{2b}}(c_{sw})
                       &=& { C_F \over (4\pi)^2 }
                           \left( 2.56184 - 3.96980c_{sw} - 1.03973c_{sw}^2
                           \right) \,,
\label{pert_B_results}
\end{eqnarray}
with $C_F=4/3$. The calculations have been extended by S. Capitani
\cite{capitani03a} to now include $B^{\msbar}_{v_3}$, $B^{\msbar}_{v_4}$.
For ${\cal O}_{v_3}$, off-diagonal elements in eq.~(\ref{one_loop_Z})
have also been computed
\begin{eqnarray}
   B^{\msbar}_{v_3}(c_{sw})
                       &=& { C_F \over (4\pi)^2 }
                           \left( -12.12740 - 2.92169c_{sw} - 0.98166c_{sw}^2
                           \right) \,,
                                              \nonumber  \\
   B^{\msbar}_{v_3;v_3^{m_1}}(c_{sw})
                       &=& { C_F \over (4\pi)^2 }
                           \left( -0.36848 - 0.032760c_{sw} + 0.029137c_{sw}^2
                           \right) \,,
                                              \nonumber  \\
   B^{\msbar}_{v_3^{m_1}}(c_{sw})
                       &=& { C_F \over (4\pi)^2 }
                          \left( - 14.85157 - 2.15228c_{sw} - 1.70741c_{sw}^2
                          \right) \,,
                                              \nonumber  \\
   B^{\msbar}_{v_3^{m_1};v_3}(c_{sw})
                       &=& { C_F \over (4\pi)^2 }
                           \left( -3.30605 + 0.33335c_{sw} - 0.37050c_{sw}^2
                           \right) \,.
\label{pert_Bv3_results}
\end{eqnarray}
For $B^{\msbar}_{v_4}$ we have%
\footnote{The number for $c_{sw}=0$ was incorrectly found in
\cite{gockeler95a,gockeler96c}. We shall use the result of \cite{capitani00b}
in the following.}
\begin{equation}
   B^{\msbar}_{v_4}(c_{sw})
                        = { C_F \over (4\pi)^2 }
                           \left( -25.50303 - 2.41788c_{sw} - 1.12826c_{sw}^2
                           \right) \,.
\end{equation}
The lowest order anomalous dimension coefficient, being universal,
is given for the three moments by eq.~(\ref{gammaN=2_res})
and in our basis the coefficients $d_{v_3; v_3^{m_1};0}$ 
and $d_{v_3^{m_1};v_3;0} = 0$ vanish, while
$d_{v_3^{m_1};0} = - 28/(9(4\pi)^2)$.

For consistency at this order in perturbation theory, we take $c_{sw}=1$
(the tree level value). Setting $\mu = 1/a$ to avoid large logarithmic
factors gives, for example at $\beta = 6.0$, the results
\begin{eqnarray}
   Z_{v_{2a}; \lat}^{\phantom{v_{2a};} \msbar} &=& 1.0276 \,,
                                              \nonumber  \\
   Z_{v_{2b}; \lat}^{\phantom{v_{2b};} \msbar} &=& 1.0207 \,,
                                              \nonumber  \\
   Z_{v_3; \lat}^{\phantom{v_3;} \msbar}       &=& 1.1354 , \qquad
          Z_{v_3;v_3^{m_1}; \lat}^{\phantom{v_3;v_3^{m_1};} \msbar}
                                                = 0.003142 \,,
                                              \nonumber  \\
   Z_{v_4; \lat}^{\phantom{v_4 ;} \msbar}      &=& 1.2453 \,.
\label{pert_B_b=6.0_results}
\end{eqnarray}
While we see that for $v_2$ first order perturbation theory
changes the tree level result ($\equiv 1$) very little, there
are perceptible differences for the higher moments.
Note also that the mixing renormalisation constant for $v_3$
is very small in comparison to the diagonal renormalisation constant,
$Z_{v_3; \lat}^{\phantom{v_3;} \msbar}$. In addition, although the mixing
operator signal is rather noisy, $v_3^{m_1} \ll v_3$
as we have seen in section~\ref{v_3andv_4}.
Thus assuming that for a non-perturbative evaluation
$Z_{v_3; \lat}^{\phantom{v_3;} \msbar} \gg
Z_{v_3;v_3^{m_1}; \lat}^{\phantom{v_3;v_3^{m_1};} \msbar}$
(as is the case for the perturbative result),
we can ignore the effects of the mixing term in the future.
   
We also use the three loop result from Table~\ref{table_v2sbar_values}
for $\Delta Z_{v_n}^{\msbar}(\mu)$ to find the RGI factor
$Z^{\rgi}_{v_n}$, eq.~(\ref{Zrgi_def}).

% ----------------------------------------------------------------------

\subsubsection{TRB perturbation theory}
\label{TRB-PT}

To improve the perturbative renormalisation results
of the last section, we shall apply tadpole-improved
renormalisation-group-improved boosted perturbation theory or
TRB-PT, \cite{capitani01a}, which we shall now describe.
The renormalised operator is given by
\begin{equation}
   {\cal O}^{\msbar}(\mu) 
       = Z^{\phantom{{\cal O};}\msbar}_{{\cal O};\lat} ( \mu, a)
                              {\cal O}(a) \,.
\label{ren_lat2msbar}
\end{equation}
As in eq.~(\ref{gamma_fun_def}), we may define a $\gamma$ function
either in the $\overline{MS}$ scheme or what we shall formally call
here the $LAT$ scheme. Additionally as expansions in the bare coupling
constant seem to be badly convergent, we choose to expand in the boosted
coupling constant and thus we have
\begin{equation}
   \left. 
     { \partial \over \partial \log a} 
         \log Z^{\phantom{{\cal O};}\msbar}_{{\cal O};\lat} (\mu, a)
                                                       \right|_{\mu}
      = \gamma^{\lat}_{\cal O}(g_{\plaq})
      =  d_{{\cal O};0} g_{\plaq}^2 + d^{\lat}_{{\cal O};1} g_{\plaq}^4
              + \ldots \,,
\label{gam_lat_def}
\end{equation}
where
\begin{equation}
   g^2_{\plaq} = { g_0^2 \over u_0^4 } \,,
       \qquad u_0^4 =\langle \third \mbox{Tr} U^{\plaq}\rangle \,,
\end{equation}
and $U^{\plaq}$ is the product of links around an elementary plaquette.
Expanding%
\footnote{Appropriate numerical values of
$u_0^4 =\langle \third \mbox{Tr} U_{\plaq}\rangle$
are given in \cite{booth01a}. For other $g_0^2$ numbers a simple
interpolation between these values can be performed, or alternatively
a Pad{\'e} fit, including the known first three coefficients
of the plaquette expansion, can be made.}
$u_0$ we have $u_0 = 1 - \quart g_0^2 p_1 + O(g_0^4)$ where $p_1 = 1/3$.

From eq.~(\ref{Z_O_Sp_S_res}), ie integrating eq.~(\ref{gamma_fun_def}) for
$({\cal S},M) \equiv (\overline{MS},\mu)$ and $(LAT, a^{-1})$,
eq.~(\ref{gam_lat_def}), gives
\begin{equation}
   Z^{\phantom{{\cal O};}\msbar}_{{\cal O};\lat} (\mu, a)
      = {\Delta Z^{\lat}_{\cal O}(a) \over \Delta Z^{\msbar}_{\cal O}(\mu)}\,,
\label{Zmsbarplaq}
\end{equation}
and thus from eq.~(\ref{ren_lat2msbar}),
the RGI quantity may be written as
\begin{equation}
   {\cal O}^{\rgi} = \Delta Z^{\msbar}_{\cal O}(\mu) {\cal O}^{\msbar}(\mu)
                   = \Delta Z^{\lat}_{\cal O}(a) {\cal O}(a) \,.
\end{equation}
Expanding eq.~(\ref{Zmsbarplaq}) and comparing with
eq.~(\ref{one_loop_Z}) enables an expression to be found
for $d_{{\cal O};1}^{\lat}$ of
\begin{equation}
   d_{{\cal O};1}^{\lat} = d_{{\cal O};1}^{\msbar} + d_{{\cal O};0}
                                                        (t_1 - p_1) 
                   - 2b_0 B^{\msbar}_{\cal O}(1) \,,
\label{d1plaq_def}
\end{equation}
for $O(a)$ improved Wilson fermions and where
\begin{equation}
   g^{\msbar} = g_0 \left( 1 + \half t_1 g_0^2 + \ldots \right) \,,
\end{equation}
at the scale $\mu = 1/a$. $t_1$ is known and is given by $t_1 = 0.468201$,
\cite{luscher95a}. Hence $d_{{\cal O};1}^{\lat}$ may be computed. Values
are given in Table~\ref{table_d1_values}.
\begin{table}[t]
   \begin{center}
      \begin{tabular}{||c|c||}
         \hline
         \hline
         ${\cal O}$ & $d_{{\cal O};1}^{\lat}$         \\
         \hline
         \hline
         $v_{2a}$& $ - 152.14 / (4\pi)^4 $  \\
         $v_{2b}$& $ - 176.31 / (4\pi)^4 $  \\
         $v_{3}$ & $   91.828 / (4\pi)^4 $  \\
         $v_{4}$ & $   382.32 / (4\pi)^4 $  \\
         \hline
         \hline
      \end{tabular}
   \end{center}  
\caption{Values of $d_{{\cal O};1}^{\lat}$ for $O(a)$ improved fermions
         from eq.~(\ref{d1plaq_def}).}
\label{table_d1_values}
\end{table}

For two loops a simple exact analytic expression is possible for
$\Delta Z^{\cal S}_{\cal O}(M)$ of
\begin{equation}
   \Delta Z^{\lat}_{\cal O}(a) =
      \left[ 2 b_0 g^2_{\plaq} \right]^{d_{{\cal O};0}\over 2b_0}
      \left[ 1 + {b_1 \over b_0} g^2_{\plaq}
      \right]^{b_0 d_{{\cal O};1}^{\lat} - b_1 d_{{\cal O};0}
                        \over 2 b_0 b_1 } \,.
\label{DelZ_lat}
\end{equation}

The expression in eq.~(\ref{DelZ_lat}) is the result of
renormalisation-group-improved boosted perturbation theory.
We can finally tadpole improve it to obtain to this order
\begin{equation}
   \Delta Z^{\lat}_{\cal O}(a) = u_0^{1-n_D}
      \left[ 2 b_0 g^2_{\plaq} \right]^{d_{{\cal O};0}\over 2b_0}
      \left[ 1 + {b_1 \over b_0} g^2_{\plaq}
      \right]^{{b_0 d_{{\cal O};1}^{\lat} - b_1 d_{{\cal O};0}
                                               \over 2 b_0 b_1} +
               { p_1 \over 4 } { b_0 \over b_1} \left( 1 - n_D \right)} \,,
\label{DelZ_lat_TI}
\end{equation}
where $n_D$ is the number of derivatives in the operator.
Note that for one derivative operators, TI has no effect,
ie eq.~(\ref{DelZ_lat}) is the same as eq.~(\ref{DelZ_lat_TI}).
Again we have the factor, eq.~(\ref{Zrgi_def}),
\begin{equation}
   Z^{\rgi}_{\cal O}(a) \equiv \Delta Z^{\lat}_{\cal O}(a) .
\end{equation}
Thus $\Delta Z^{\lat}_{\cal O}(a)$ is the function
that takes you directly from the lattice result to the RGI result.

Finally if we additionally wish to TI the improvement coefficients,
\cite{capitani00a}, then we replace $g_0^2$ by $g_{\plaq}^2$ in
eq.~(\ref{pert_improvement_coeff_v2b}). Numerical results from this
procedure for $c_0$ in $v_2$ with $c_2=0$ are given in
Table~\ref{table_Zrgi_vn}. The associated $\kappa_c$, necessary for the
computation of $am_q$, eq.~(\ref{bare_qm_def}) are given in
\cite{bakeyev03a}.

% ----------------------------------------------------------------------

\subsubsection{Non-perturbative $Z$ determinations}
\label{np_Z_determination}

We now look at the $\rm{RI}^\prime - \rm{MOM}$ non-perturbative
determination of renormalisation constants, using the method proposed by
Martinelli et al \cite{martinelli94a} which mimics (up to a point)
the approach of the perturbative lattice procedure,
by defining
\begin{equation}
   Z_{\cal O; \lat}^{\phantom{{\cal O};}\ripmom}(\mu_p,a)
           = \left. { Z_{q; \lat}^{\phantom{q;}\ripmom}(p) \over
                        \twelfth \mbox{Tr}
                           \left[\Lambda_{\cal O}(p)
                                 \Lambda_{{\cal O},\born}^{-1}(p)
                           \right]} \right|_{p^2=\mu_p^2} \,,
\label{non_pert_ren_cond}
\end{equation}
where the wave function renormalisation constant
$Z_{q; \lat}^{\phantom{q;}\ripmom}(p)$ can be fixed from the
conserved vector current or from the (Fourier transformed)
quark propagator $S_q(p)$ by
\begin{eqnarray}
   Z_{q; \lat}^{\phantom{q;}\ripmom}(p)
          = { \mbox{\rm Tr} ( - i\sum_\lambda \gamma_\lambda 
                                            \sin (ap_\lambda) aS_q^{-1}(p))
            \over
            12 \sum_\lambda \sin^2 (ap_\lambda) } \,.
\end{eqnarray}
(There are still various possibilities for $Z_{q; \lat}$,
see eg \cite{chetyrkin99a} for different definitions.
Again $\Lambda_{\cal O}$ is the amputated three-point
Green's function for the operator ${\cal O}$.)
For our implementation using a `momentum source' see \cite{gockeler98a}.
For the higher derivative operators considered here, this is a
non-covariant renormalisation condition, depending on the momentum
direction, \cite{gockeler98a,gockeler98b} (numerically this is a
small effect for the momenta considered here).

The non-perturbative results for
$Z_{\cal O; \lat}^{\phantom{{\cal O};}\ripmom}$ should now be brought
to an RGI form, which can only be done perturbatively.
In order to avoid problems caused by the non-covariance of the
renormalisation condition eq.~(\ref{non_pert_ren_cond}) we first transform 
$Z_{\cal O; \lat}^{\phantom{{\cal O};}\ripmom}$
(perturbatively) to a covariant scheme $\cal S$ like
$\overline{MS}$ or $MOM$ employing a conversion factor of the form
\begin{equation} 
    1 + c^{\cal S}_1  (g^{\msbar})^2 + 
        c^{\cal S}_2 \left( g^{\msbar} \right)^4 + \ldots \,.
\end{equation} 
For ${\cal S} = \overline{MS}$ the general expression to one-loop
order can be found in \cite{gockeler98a}, while an explicit formula
for $v_{2b}$ is given in \cite{gockeler98b}.The one-loop expressions
for ${\cal S} = MOM$ are also known. In the case of $v_{2a}$ and $v_{2b}$,
three-loop expressions can be derived from \cite{gracey03a} and will
be used in the following.

In a second step, multiplying the resulting numbers by 
$\Delta Z_{\cal O}^{\cal S}$, we obtain $Z_{\cal O}^{\rgi}$ 
(see eq.~(\ref{Zrgi_def})). Thus $\Delta Z_{\cal O}^{\cal S}$ has
to be found, which as in section~\ref{TRB-PT} is again the computation
of the perturbative coefficients of the anomalous dimension
$\gamma^{\cal S}$. For ${\cal S} = \overline{MS}$ the anomalous
dimension is known to three loops for $v_2$, $v_3$ and $v_4$.
For ${\cal S} = MOM$ only the first two loops (or coefficients)
are available in the case of $v_3$ and $v_4$, while in the case of 
$v_2$ we can make use of the three-loop calculation in
\cite{gracey03a}. For example, expanding in $g^{\msbar}$, 
$\gamma_{\cal O}^{\cal S} = d_{{\cal O};0} (g^{\msbar})^2
+ d_{{\cal O};1}^{\cal S} (g^{\msbar})^4  + \ldots$,
we have to two loops, similarly to eq.~(\ref{DelZ_lat_TI}),
\begin{equation}
   \Delta Z_{\cal O}^{\cal S}  = 
        \left[ 2 b_0 (g^{\msbar})^2 \right]^{ d_{{\cal O};0} \over 2 b_0 }
        \left[ 1 + {b_1 \over b_0} (g^{\msbar})^2 \right]
           ^{ b_0 d_{{\cal O};1}^{\cal S} - b_1 d_{{\cal O};0}
              \over 2 b_0 b_1 } \,.
\end{equation}
Depending on the choice of $\cal S$ and the coupling in which 
$\gamma^{\cal S}$ is expanded, there are several possibilities
(of course equivalent if one knew the whole power series). We shall 
briefly describe two methods here (I and II), whose difference 
we shall use to estimate the potential error due to (unknown)
higher terms in the perturbative expansion. In method I we choose 
${\cal S} = \overline{MS}$ and expand in $g^{\msbar}$.
In method II we work in the $MOM$ scheme (see \cite{gockeler98a}),
and therefore it may be more natural to expand in other coupling
constants defined using momentum renormalisation conditions.
In \cite{chetyrkin00a} several possibilities are given.
We shall use here the coupling defined by using the three
gluon vertex, the $\widetilde{MOM}gg$ scheme, in the notation of
\cite{chetyrkin00a}.

If we plot $Z^{\rgi}_{v_n}$ against $(r_0\mu_p)^2$ and the perturbative
expressions are sufficently well known, we would expect to see a
plateau where $Z^{\rgi}_{v_n}$ is independent of $(r_0\mu_p)^2$.
This region occurs when $\mu_p$ is not too small $\sim \Lambda$
otherwise non-perturbative effects play a role, nor when it is too
large as lattice artifacts then occur. Unfortunately these
$O(r_0\mu_p)^2$ effects may become large (but depend very much on the
operator considered). In perturbation theory in the chiral limit
$O(a)$ terms of the Green's function have opposite chirality to
the leading term, they disappear when the trace in
eq.~(\ref{non_pert_ren_cond}) is taken. (For explicit lowest order
results see \cite{capitani00a}.) Condensates may spoil this at
low $\mu_p^2$, but here we are looking at higher
momentum scales. Thus we shall take the plateaus in the
chiral limit as our renormalisation constants.

 We have made determinations of
$Z_{{\cal O}; \lat}^{\phantom{{\cal O};}\ripmom}$
using eq.~(\ref{non_pert_ren_cond}) for $\beta = 6.0$, $6.2$ and $6.4$
on $24^3\times 48$, $24^3\times 48$, $32^3\times 40$ lattices respectively.
For each $\beta$ three or more quark masses were used and a linear
extrapolation in $am_q$ was performed to the chiral limit. More details
will be given in \cite{gockeler04a}.

% ----------------------------------------------------------------------

\subsection{Comparison of $Z^{\rgi}$ results}
\label{Z_reults}

\begin{figure}[p]
   \hspace*{1.0in}
   \epsfxsize=10.00cm \epsfbox{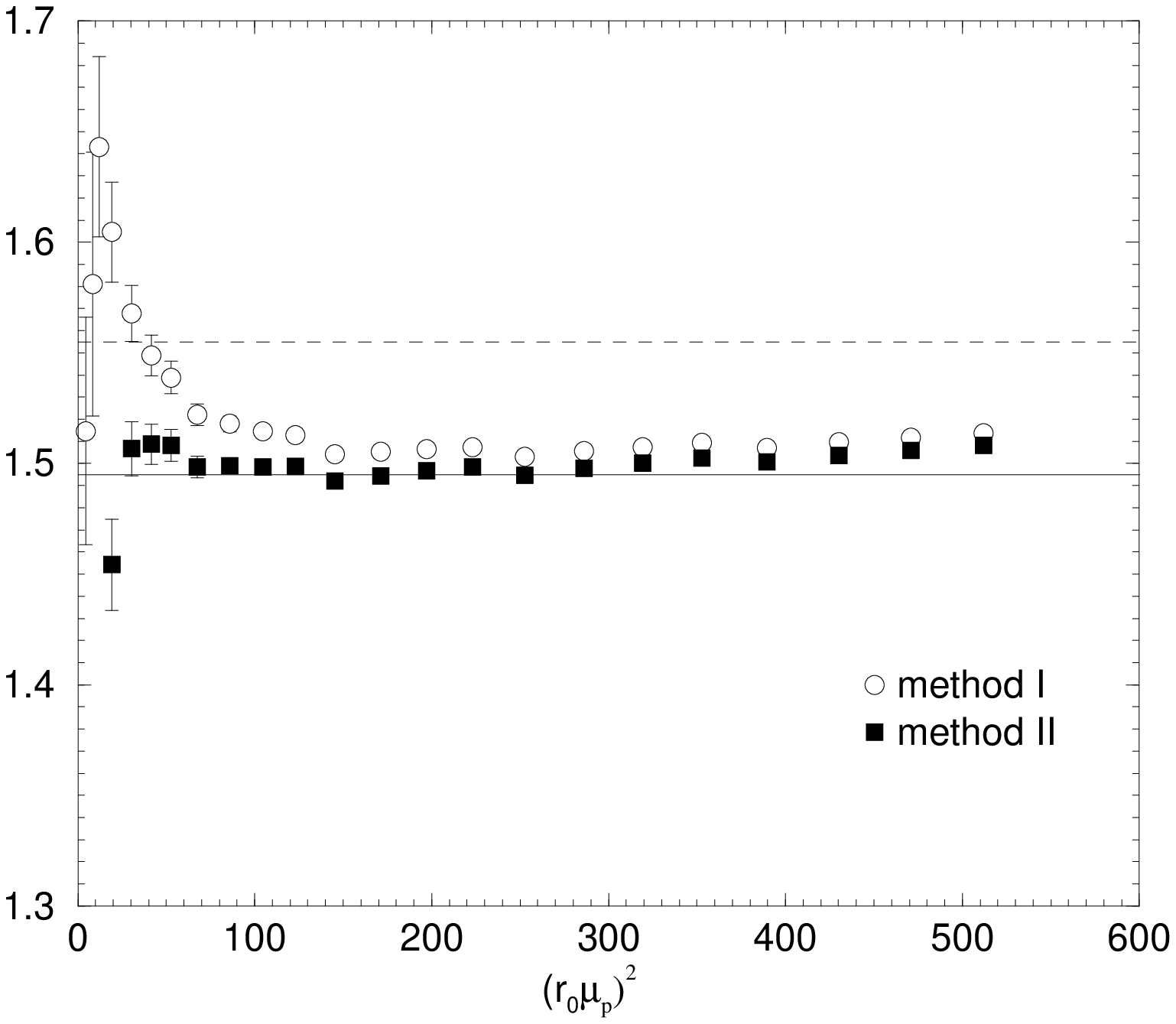}
   \caption{$Z^{\rgi}_{v_{2b}}$ at $\beta = 6.2$
            versus $(r_0\mu_p)^2$. NP method I is denoted by
            circles, while method II is given by filled squares.
            The dashed line is the TRB-PT result, while the full
            line is the NP estimate from method II.}
\label{fig_Z_v2b_rgi_r0p2_pa_mein_b6p2I+II}
\end{figure}

\begin{figure}[p]
   \hspace*{1.0in}
   \epsfxsize=10.00cm \epsfbox{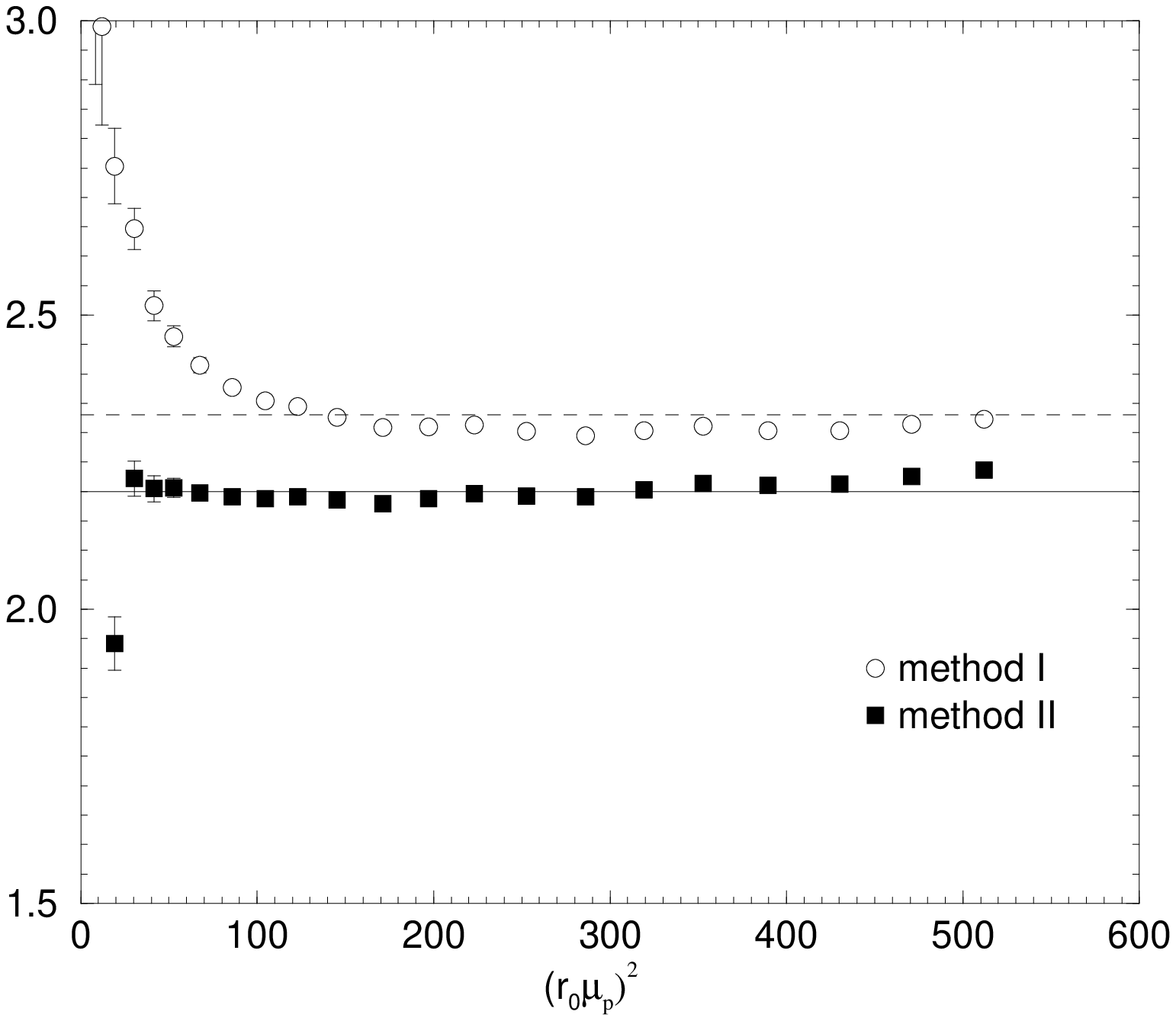}
   \caption{$Z^{\rgi}_{v_{3}}$ at $\beta = 6.2$
            versus $(r_0\mu_p)^2$. The same notation as 
            for Fig.~\ref{fig_Z_v2b_rgi_r0p2_pa_mein_b6p2I+II}.}
\label{fig_Z_v3_rgi_r0p2_pa_mein_b6p2I+II}
\end{figure}

\begin{figure}[p]
   \hspace*{1.0in}
   \epsfxsize=10.00cm \epsfbox{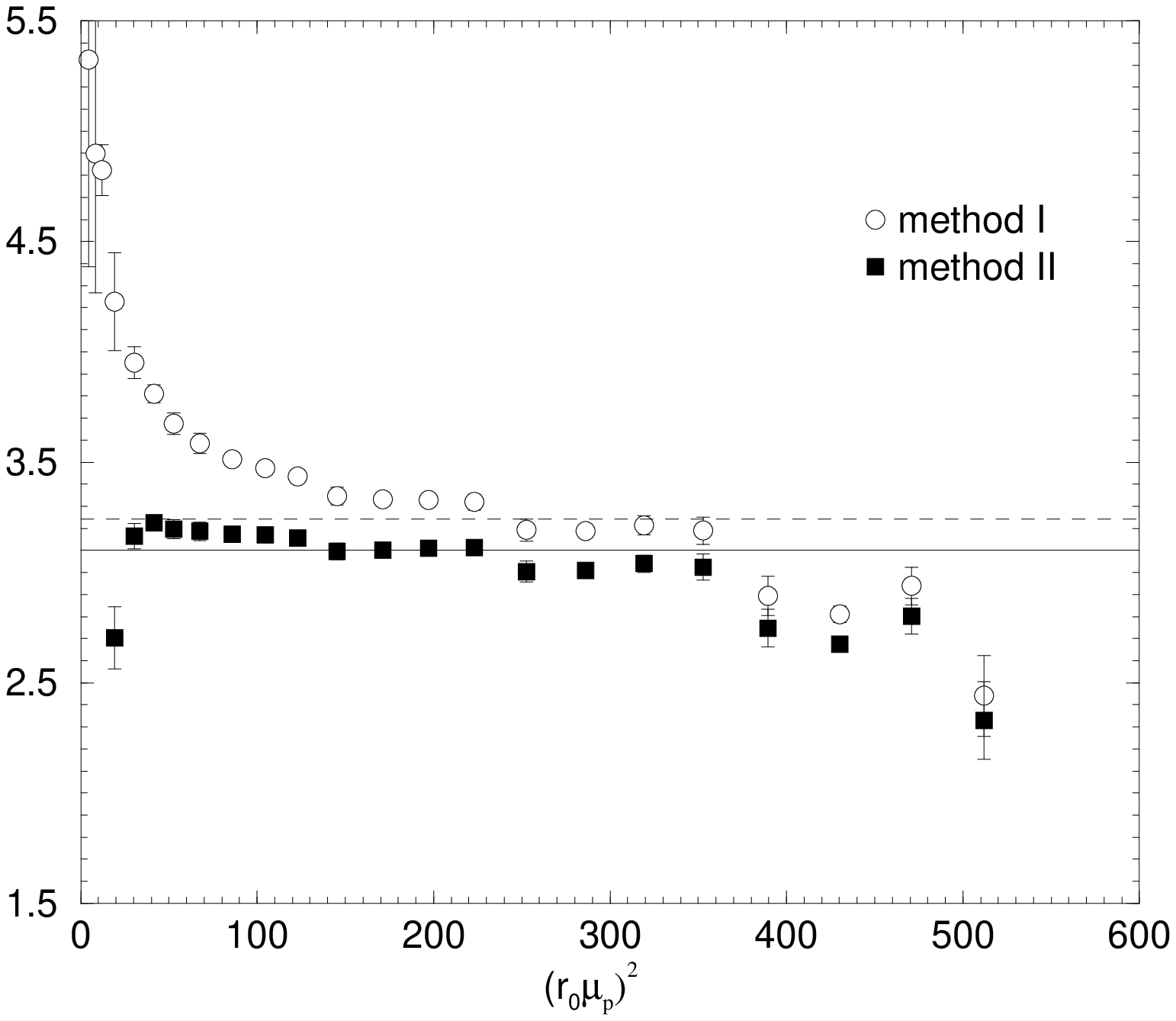}
   \caption{$Z^{\rgi}_{v_{4}}$ at $\beta = 6.2$
            versus $(r_0\mu_p)^2$. The same notation as 
            for Fig.~\ref{fig_Z_v2b_rgi_r0p2_pa_mein_b6p2I+II}.}
\label{fig_Z_v4_rgi_r0p2_pa_mein_b6p2I+II}
\end{figure}

\begin{figure}[p]
   \hspace*{1.0in}
   \epsfxsize=10.0cm \epsfbox{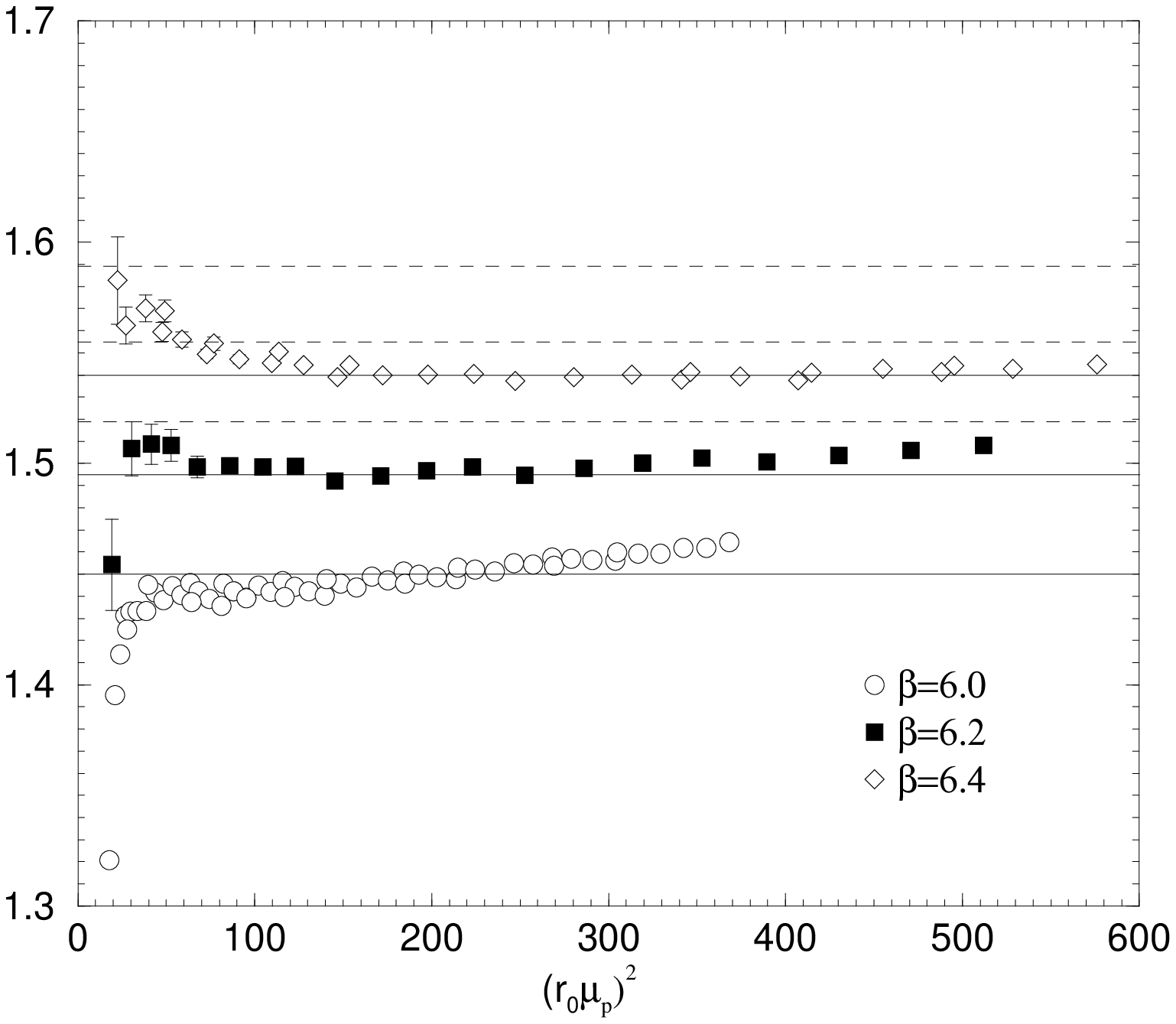}
   \caption{$Z^{\rgi}_{v_{2b}}$ versus $(r_0\mu_p)^2$ for
            $\beta = 6.0$ (circles), $6.2$ (filled squares),
            $6.4$ (diamonds) using method II.
            The corresponding dashed lines are the TRB-PT results,
            while the full lines are the NP estimates.}
\label{fig_Z_v2b_rgi_r0p2_pa_mein_b6p0-6.4II}
\end{figure}

We shall now use these results to find $Z^{\rgi}_{v_2}$, $Z^{\rgi}_{v_3}$
and $Z^{\rgi}_{v_4}$. In Figs.~\ref{fig_Z_v2b_rgi_r0p2_pa_mein_b6p2I+II},
\ref{fig_Z_v3_rgi_r0p2_pa_mein_b6p2I+II},
\ref{fig_Z_v4_rgi_r0p2_pa_mein_b6p2I+II} we plot
$Z^{\rgi}_{v_2}$, $Z^{\rgi}_{v_3}$ and $Z^{\rgi}_{v_4}$
for $\beta = 6.2$ as computed from these techniques.
In the case of the non-perturbative $Z$s a reasonable plateau is seen
for large $(r_0\mu_p)^2$ enabling a value for the renormalisation
constant to be found. Method II seems to reach a plateau faster
than method I, so we shall use these results
(taken around $\mu_p = 5 \, \mbox{GeV}$), with the difference
to method I giving a rough estimate of the error.
Also shown in the plots are the TRB-PT results.
The NP and TRB-PT results lie close to each other,
with a maximum discrepancy of about $8\%$. In the NP
determination of the $Z$s, the plateaus become better as $\beta$
increases. This is shown in
Fig.~\ref{fig_Z_v2b_rgi_r0p2_pa_mein_b6p0-6.4II}.

In Table~\ref{table_Zrgi_vn} we give the results from PT,
\begin{table}[t]
   \begin{center}
      \begin{tabular}{||c|l|l|l|l||}
         \hline
         \hline
\multicolumn{2}{||c|}{$\beta$}                                & 
\multicolumn{1}{c|}{$6.0$}                                    &
\multicolumn{1}{c|}{$6.2$}                                    &
\multicolumn{1}{c||}{$6.4$}                                   \\
         \hline
                    & PT         & 1.416    & 1.475    & 1.522       \\
$Z^{\rgi}_{v_{2a}}$ & TRB-PT     & 1.536    & 1.571    & 1.604       \\
                    & NP -- I    & 1.46     & 1.51     & 1.56        \\
                    & NP -- II   & 1.45     & 1.50     & 1.55        \\
         \hline
$c^{\ti}_{v_{2a};0}$& TI         & 1.232    & 1.218    & 1.205       \\
         \hline
                    & PT         & 1.407    & 1.465    & 1.513       \\
$Z^{\rgi}_{v_{2b}}$ & TRB-PT     & 1.519    & 1.555    & 1.589       \\
                    & NP -- I    & 1.46     & 1.50     & 1.55        \\
                    & NP -- II   & 1.45     & 1.49     & 1.54        \\
         \hline
$c^{\ti}_{v_{2b};0}$& TI         & 1.245    & 1.229    & 1.216       \\
         \hline
                    & PT         & 1.928    & 2.038    & 2.129       \\
$Z^{\rgi}_{v_3}$    & TRB-PT     & 2.268    & 2.330    & 2.389       \\
                    & NP -- I    & 2.2      & 2.3      & 2.4         \\
                    & NP -- II   & 2.1      & 2.2      & 2.3         \\
         \hline
                    & PT         & 2.367    & 2.548    & 2.700       \\
$Z^{\rgi}_{v_4}$    & TRB-PT     & 3.156    & 3.242    & 3.325       \\
                    & NP -- I    & 3.1      & 3.3      & 3.5         \\
                    & NP -- II   & 2.9      & 3.1      & 3.2         \\
         \hline
         \hline
      \end{tabular}
   \end{center}
\caption{$Z^{\rgi}$ results (and some $c_0$ results) at
         $\beta = 6.0$, $6.2$ and $6.4$,
         for $v_n$, $n = 2$, $3$ and $4$.
         PT denotes the perturbative results from
         section~\ref{perturbation_theory} and TRB-PT
         from section~\ref{TRB-PT}.
         Note that to obtain $Z^{\rgi}$ for
         PT the three-loop results from
         Table~\ref{table_v2sbar_values} have been used.
         Results from both NP variations, I and II are shown.}
\label{table_Zrgi_vn}
\end{table}
TRB-PT and NP method (both variants) for various renormalisation constants%
\footnote{For an independent nonperturbative calculation of
$Z_{v_{2b}}^{RGI}$ see \cite{guagnelli04a}.}
at $\beta = 6.0$, $6.2$ and $6.4$.

% ----------------------------------------------------------------------

\section{Results: chiral and continuum extrapolations}
\label{chiral_extraps}

% ----------------------------------------------------------------------

\subsection{The phenomenological approach}
\label{phenomenological_approach}

\begin{figure}[t]
   \hspace*{1.00cm}
   \epsfxsize=12.25cm \epsfbox{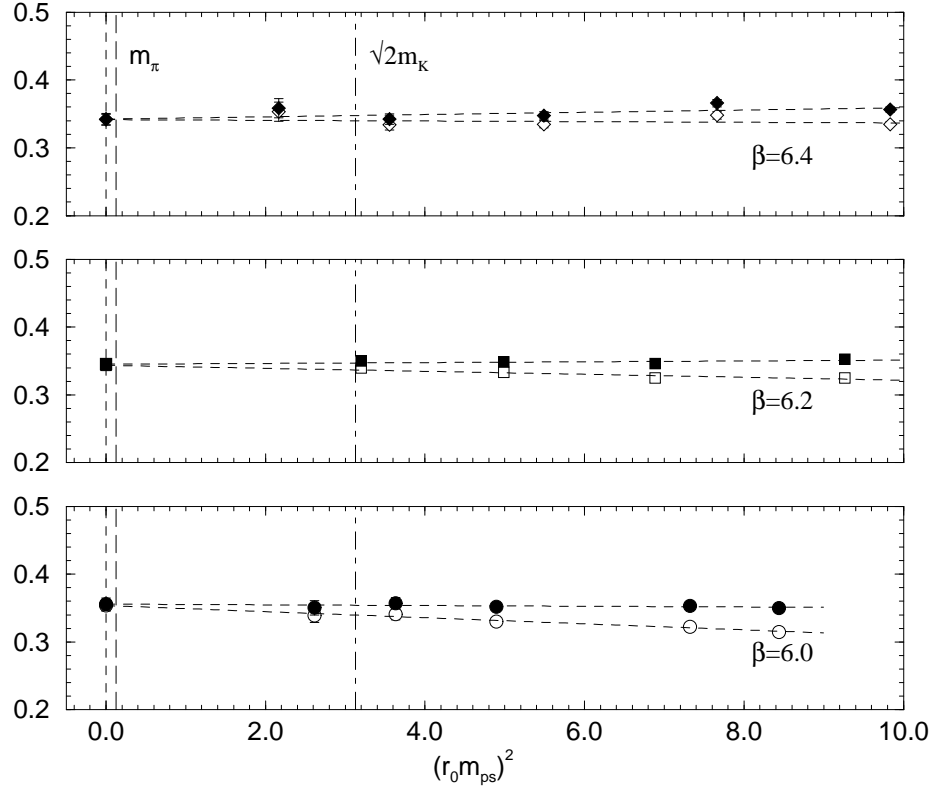}
   \caption{$v^{\rgi}_{2b}$ versus $(r_0m_{ps})^2$ for $\beta = 6.0$
            (circles), $6.2$ (squares) and $6.4$ (diamonds).
            The filled symbols are obtained when using the TI value for
            $c_0$ while the empty symbols have $c_0 = 0$.
            Also shown are linear extrapolations to the chiral limit
            (dashed lines). Other notation as in
            Fig.~\ref{fig_x1a_1u-1d.pm1_O1+O2_030814_1054}.}
   \label{fig_x1b_1u-1d.p0_030904_1718_pap}
\end{figure}

From the bare results given in Appendix~\ref{tables} and the Zs in
Table~\ref{table_Zrgi_vn} we can now construct our renormalised 
matrix elements. As well as the continuum extrapolation $a \to 0$,
as the quark masses presently used are rather heavy we also need
to extrapolate these renormalised results to the chiral limit.
In this section we shall consider both these limits.
In Fig.~\ref{fig_x1b_1u-1d.p0_030904_1718_pap}
we show the results for $v_{2b}$ for $\beta = 6.0$, $6.2$ and $6.4$
with both $c_0 = c_0^{\ti}$, as given in Table~\ref{table_Zrgi_vn},
and $c_0 = 0$. (We would expect that for $c_0$ set to the TI value the
results are practically $O(a)$ improved for non-zero quark mass.)

Also shown in the figure is a linear extrapolation 
$v_{n} \equiv F^{(n)}_\chi(r_0 m_{ps})$ where
\begin{equation}
   F_\chi^{(n)}(x) = a_n x^2 + b_n \,,
\label{linear_ch_fit}
\end{equation}
(with $n=2$). This fit describes the data well and (not surprisingly)
using either $c_0$ value gives the same result in the chiral limit.
Indeed the $v_{2b}$ $O(a)$ improved results seem to be independent of
the quark mass.

\begin{figure}[t]
   \hspace*{1.00cm}
   \epsfxsize=12.25cm \epsfbox{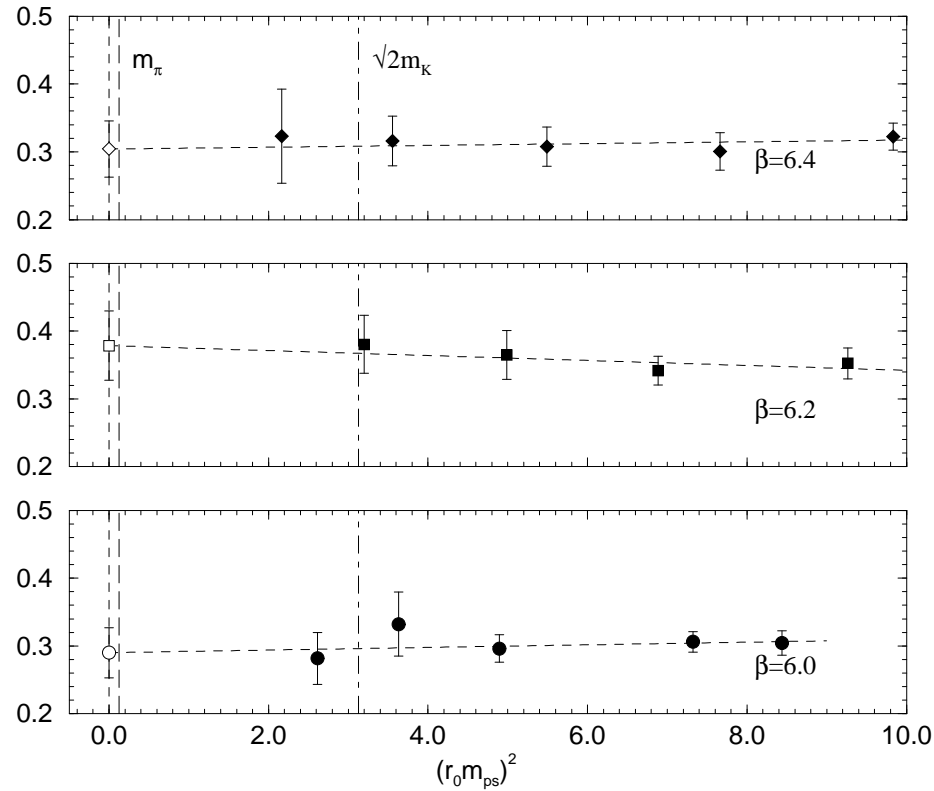}
   \caption{$v^{\rgi}_{2a}$ versus $(r_0m_{ps})^2$ for $\beta = 6.0$
            (circles), $6.2$ (squares) and $6.4$ (diamonds).
            Also shown are linear extrapolations to the chiral limit
            (dashed lines). Other notation as in
            Fig.~\ref{fig_x1a_1u-1d.pm1_O1+O2_030814_1054}.}
   \label{fig_x1a_1u-1d.pm1_030905_1514_pap}
\end{figure}

A similar situation holds for $v_{2b}$, $v_{2a}$, $v_3$ and $v_4$
all evaluated with non-zero momentum but with larger (and increasing)
error bars. In Fig.~\ref{fig_x1a_1u-1d.pm1_030905_1514_pap}
we show the results for $v_{2a}$ for $\beta = 6.0$, $6.2$ and $6.4$
together with a linear chiral extrapolation (using the TI value for $c_0$).
Immediately noticeable when comparing with
Fig.~\ref{fig_x1b_1u-1d.p0_030904_1718_pap} is the large increase in the
error bars and less consistent ordering of gradients with increasing $\beta$. 

\begin{figure}[t]
   \hspace*{1.00cm}
   \epsfxsize=12.25cm \epsfbox{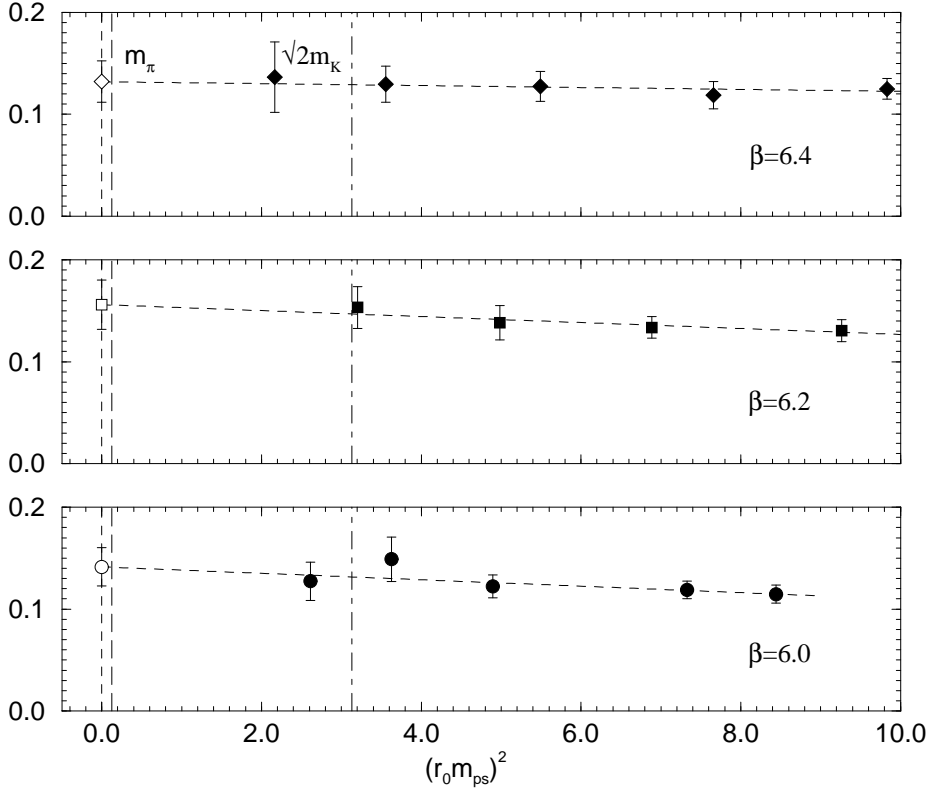}
   \caption{$v^{\rgi}_3$ versus $(r_0m_{ps})^2$ for $\beta = 6.0$ (circles),
            $6.2$ (squares) and $6.4$ (diamonds). Also shown are linear
            extrapolations to the chiral limit (dashed lines). Other
            notation as in Fig.~\ref{fig_x1a_1u-1d.pm1_O1+O2_030814_1054}.}
   \label{fig_x2b_1u-1d.pm1_040429_1554}
\end{figure}

\begin{figure}[t]
   \hspace*{1.00cm}
   \epsfxsize=12.25cm \epsfbox{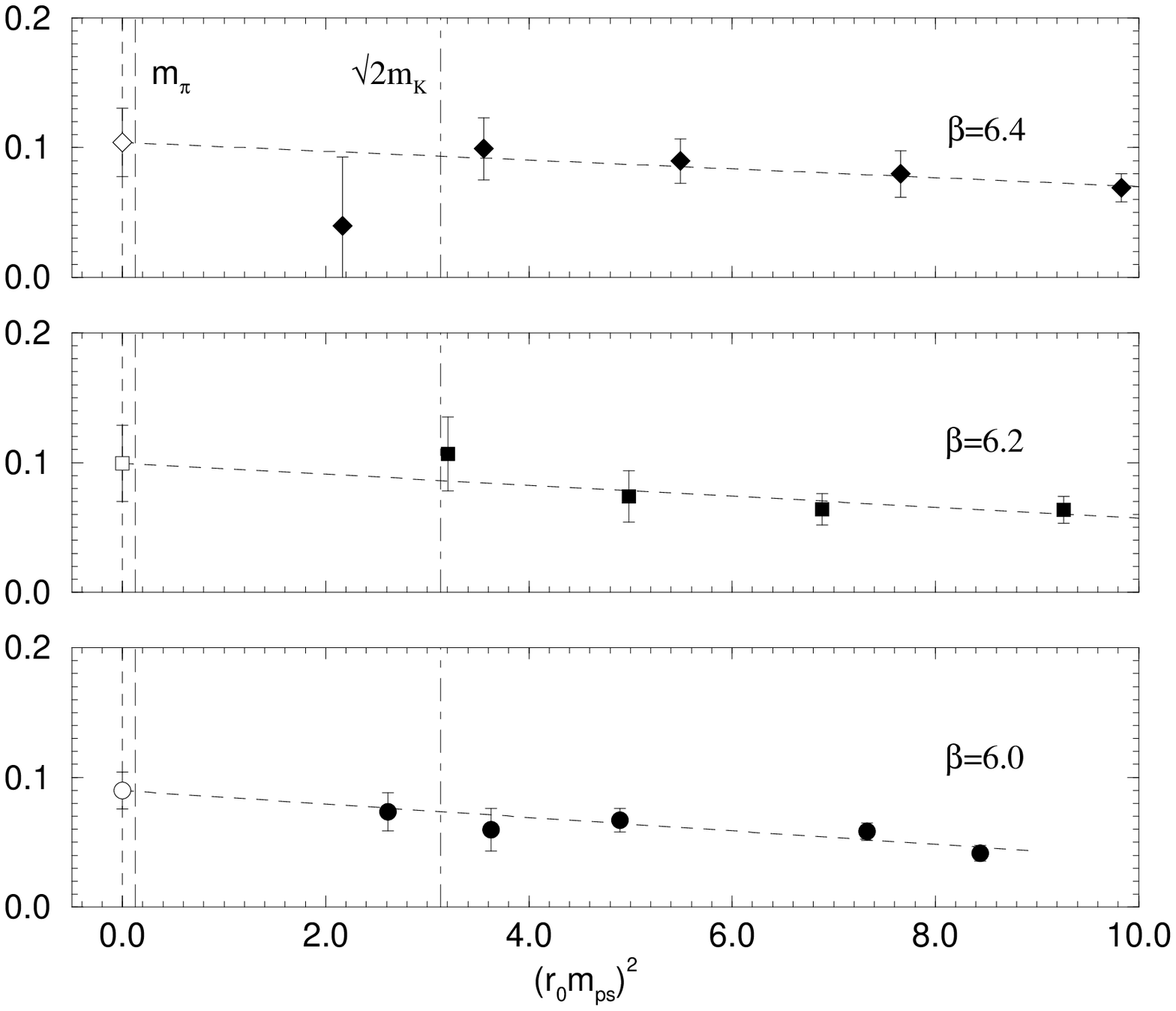}
   \caption{$v^{\rgi}_4$ versus $(r_0m_{ps})^2$ for $\beta = 6.0$ (circles),
            $6.2$ (squares) and $6.4$ (diamonds). Also shown are linear
            extrapolations to the chiral limit (dashed lines). Other
            notation as in Fig.~\ref{fig_x1a_1u-1d.pm1_O1+O2_030814_1054}.}
   \label{fig_x3b_1u-1d.pm1_040517_1247}
\end{figure}

In Fig.~\ref{fig_x2b_1u-1d.pm1_040429_1554} we show the results for
$v_3$ and in Fig.~\ref{fig_x3b_1u-1d.pm1_040517_1247}
the analogous results for $v_4$. As expected the results become noisier for
increasing $\beta$. Perhaps surprisingly the results for $v_4$ seem
to be as consistent over our $\beta$-range as those of $v_3$ (we have
no real explanation for this).

The last limit to be taken is the continuum limit, $a \to 0$.
As discussed in section~\ref{lattice} we believe that for $v_2$
the improvement terms are numerically small and so can be neglected.
Thus we can make an extrapolation in $a^2$ (rather than $a$).
While we cannot be so confident in this for the higher moments,
based on the experience with the lowest moment, we shall also
assume this for these higher moments.
In Table~\ref{table_rgi_vn} we give first the RGI values
\begin{table}[t]
   \begin{center}
      \begin{tabular}{||c|c||l|l|l||l||}
         \hline
         \hline
\multicolumn{1}{||c|}{$\beta$}                                & 
\multicolumn{1}{c||}{$\vec{p}$}                               & 
\multicolumn{1}{c|}{$6.0$}                                    &
\multicolumn{1}{c|}{$6.2$}                                    &
\multicolumn{1}{c||}{$6.4$}                                   &
\multicolumn{1}{c||}{$\infty$}                                \\
         \hline
  $v^{\rgi}_{2a}$  & $\vec{p}_1$ 
               & 0.290(37) & 0.379(51) & 0.304(41) & 0.343(56)  \\
  $v^{\rgi}_{2b}$  & $\vec{p}_1$
               & 0.338(16) & 0.328(16) & 0.340(16) & 0.335(22)  \\
  $v^{\rgi}_{2b}$  & $\vec{0}$  
               &0.354(8)   & 0.344(8)  & 0.342(8)  & 0.335(11)  \\
  $v^{\rgi}_{3}$   & $\vec{p}_1$
               &0.141(19)  & 0.156(24) & 0.132(20) & 0.137(28)  \\
  $v^{\rgi}_{4}$   & $\vec{p}_1$
               & 0.090(14) & 0.099(29) & 0.104(26) & 0.110(33)  \\
         \hline
         \hline
      \end{tabular}
   \end{center}
\caption{$v^{\rgi}_{n}$ results for
         $O(a)$ improved fermions at $\beta = 6.0$, $6.2$ and $6.4$,
         for $v^{\rgi}_n$, $n = 2$, $3$ and $4$.
         `$\infty$' denotes the continuum extrapolation ($a = 0$)
         limit.}
\label{table_rgi_vn}
\end{table}
at $\beta = 6.0$, $6.2$, $6.4$, for $v^{\rgi}_{2a}$, $v^{\rgi}_{2b}$
both for non-zero and zero momentum, $v^{\rgi}_{3}$ and $v^{\rgi}_{4}$.
(These have all been obtained using the NP method II
results, as given in Table~\ref{table_Zrgi_vn}.)

\begin{figure}[p]
   \hspace*{1.00cm}
   \epsfxsize=12.25cm \epsfbox{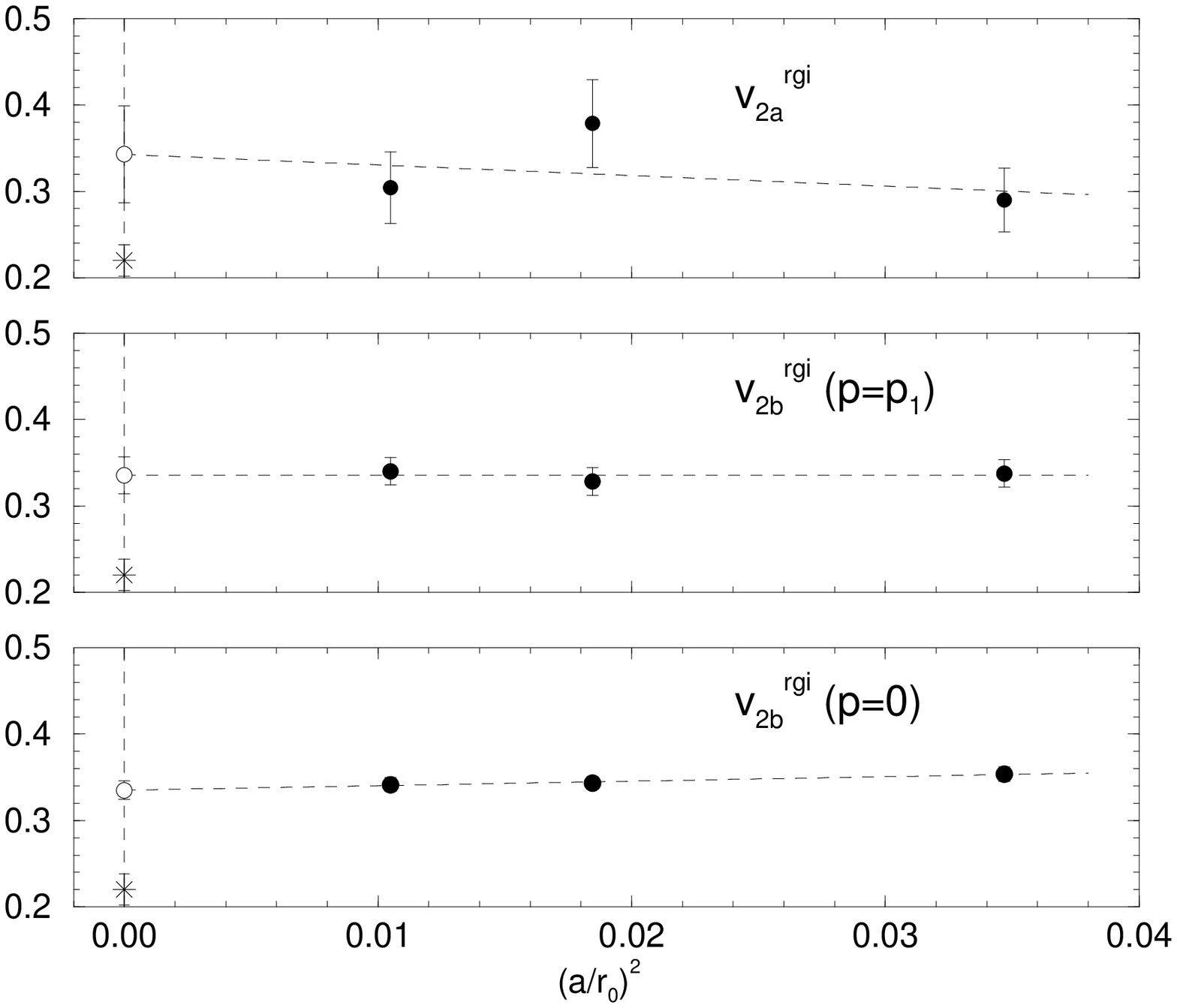}
   \caption{$v^{\rgi}_{2a}$ and $v^{\rgi}_{2b}$ (both for
            $\vec{p}=\vec{p}_1$ and $\vec{0}$) versus $(a/r_0)^2$,
            using the results at $\beta = 6.0$, $6.2$, $6.4$ (filled circles).
            A linear continuum extrapolation in $a^2$ is
            also given (dashed line and empty circle). The star
            is the MRST value given in Table~\ref{table_expt_moments}.}
   \label{fig_x1a+b_aor02_pap}
\end{figure}

\begin{figure}[p]
   \hspace*{1.00cm}
   \epsfxsize=12.25cm \epsfbox{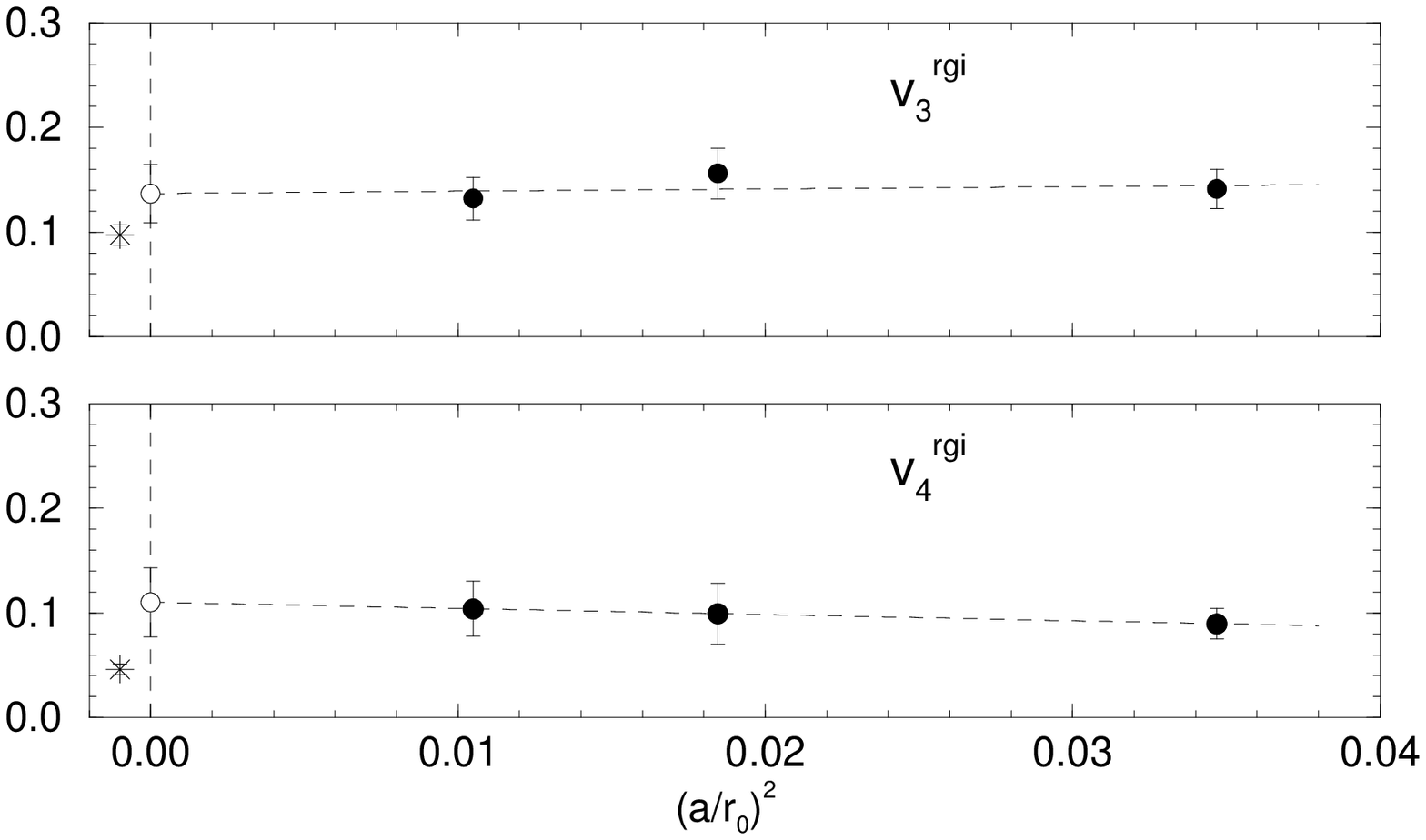}
   \caption{$v^{\rgi}_{3}$ and $v^{\rgi}_{4}$ versus $(a/r_0)^2$,
            using the results at $\beta = 6.0$, $6.2$, $6.4$.
            A continuum extrapolation is also given.
            The same notation as for Fig.~\ref{fig_x1a+b_aor02_pap}.}
   \label{fig_x2+x3_aor02_pm1_040517}
\end{figure}

We now use these results to perform a continuum extrapolation.
In Fig.~\ref{fig_x1a+b_aor02_pap}
we plot the continuum extrapolations for the various $v^{\rgi}_{2}$.
A very consistent picture is obtained firstly
between the different representations (`$a$' and `$b$')
and secondly between the different momenta used in the `$b$' representation.
As expected though using a non-zero momentum gives a much noisier signal:
in the extreme case between the off-diagonal and diagonal
representations the error is about $\sim O(2)$ larger.
We shall present our final result using $v^{\rgi}_{2b}$
for $\vec{p}=\vec{0}$ only. In Fig.~\ref{fig_x2+x3_aor02_pm1_040517}
we show the results for $v^{\rgi}_{3}$ and $v^{\rgi}_{4}$.
Using the modified operators of eq.~(\ref{lattice_ops}) enables
a relatively smooth extrapolation to the continuum limit, giving
results with about a $20\%$ -- $30\%$ error.

Finally, we convert our results to the $\overline{MS}$ scheme at
a scale of $\mu = 2\,\mbox{GeV}$, using the three loop result
for $[\Delta Z_{v_n}^{\msbar}(\mu)]^{-1}$ given in
Table~\ref{table_v2sbar_values}. We find
\begin{eqnarray}
   v_{2}^{\msbar}(2\,\mbox{GeV}) &=& 0.245(9)  \,,
                                                \nonumber  \\
   v_{3}^{\msbar}(2\,\mbox{GeV}) &=& 0.083(17) \,,
                                                \nonumber  \\
   v_{4}^{\msbar}(2\,\mbox{GeV}) &=& 0.059(18) \,.
\label{final_results}
\end{eqnarray}
The total error are the combined errors from the three
point functions and chiral/continuum fits together 
with the error for the renormalisation constant.
For $v_{2b}$ this indicates that the dominant error is now possibly
coming from the renormalisation constant; the opposite is so for the
higher moments.

How reasonable are the results in comparison with experiment
or the MRST phenomenological fits?
The MRST numbers from section~\ref{experiment_theory} in
Table~\ref{table_expt_moments} are also plotted in
Figs.~\ref{fig_x1a+b_aor02_pap} and \ref{fig_x2+x3_aor02_pm1_040517}.
We see that for $v_{2b}^{\rgi}$, the discrepancy between the
experimental result and the lattice result stubbornly remains - and has
persisted ever since the first pioneering works in the field
\cite{martinelli89a}. It is also notable that $v_4^{\rgi}$
in particular is too large in comparison with the phenomenological
result. For both $v_3^{\rgi}$ and $v_4^{\rgi}$ the chiral 
and continuum extrapolations are more problematic than for $v_{2b}$.
This can probably only be resolved by increasing the statistic
of the ensembles and by additional simulations at other $\beta$ values.

Finally for completeness we give the results for $v_n^{(q)\msbar}$
for $q = u$, $d$ separately. We find
\begin{eqnarray}
   v_{2}^{(u)\msbar}(2\,\mbox{GeV}) &=& 0.436(18)  \,,
                                                \nonumber  \\
   v_{3}^{(u)\msbar}(2\,\mbox{GeV}) &=& 0.136(25)  \,,
                                                \nonumber  \\
   v_{4}^{(u)\msbar}(2\,\mbox{GeV}) &=& 0.096(25)  \,,
\label{final_results_u}
\end{eqnarray}
for the $u$ quark and
\begin{eqnarray}
   v_{2}^{(d)\msbar}(2\,\mbox{GeV}) &=& 0.191(7)   \,,
                                                \nonumber  \\
   v_{3}^{(d)\msbar}(2\,\mbox{GeV}) &=& 0.052(11)  \,,
                                                \nonumber  \\
   v_{4}^{(d)\msbar}(2\,\mbox{GeV}) &=& 0.027(15)  \,,
\label{final_results_d}
\end{eqnarray}
for the $d$ quark. As discussed previously in
section~\ref{determining_me}, the quark line disconnected ($qldis$)
contribution to the matrix element is not computed. Thus on the RHS of
eqs.~(\ref{final_results_u}), (\ref{final_results_d}) there should
be an extra term $v_n^{\msbar}|_{qldis}$, which is the same for
$u$ and $d$ quarks, and of course, cancels for the NS results
of eq.~(\ref{final_results}).

% ----------------------------------------------------------------------

\subsection{Chiral perturbation theory}
\label{chiral_perturbation_theory}

\begin{figure}[t]
   \epsfxsize=14.00cm \epsfbox{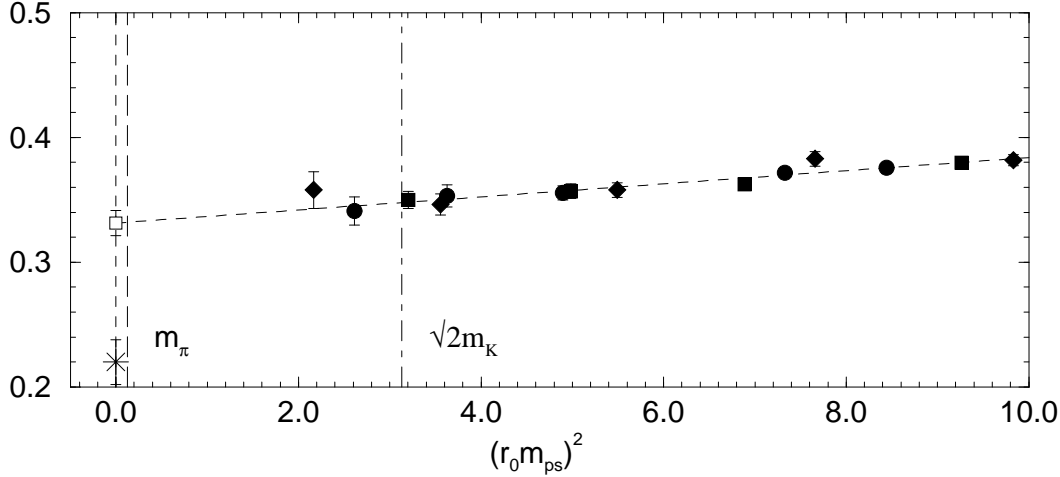}
   \caption{$v^{\rgi}_{2b}- c_2 (a/r_0)^2 - d_2 ar_0m_{ps}^2
            \equiv F^{(2)}_\chi = a_2 (r_0m_{ps})^2 + b_2$
            (ie using the chiral function of eq.~(\ref{linear_ch_fit}))
            versus $(r_0m_{ps})^2$, dashed line. 
            Filled circles, squares and diamonds represent
            $\beta = 6.0$, $6.2$ and $6.4$ respectively.
            The empty square represents the chirally extrapolated
            value. Again the MRST phenomenological value
            of $v^{\rgi}_2$ is represented by a star.}
\label{fig_x1b_1u-1d.p0_040119_1938}
\end{figure}

Although linear extrapolations in $m_{ps}^2$
seem to describe the results presented earlier quite well,
it is not clear that for the quark masses used here
(and in other simulations) higher order effects and/or chiral
logarithms can be neglected. There has recently been a flurry of
interest in this direction. In \cite{detmold01a},
based on chiral perturbation theory proposed in \cite{thomas00a},
a fit function model is used which tries to take into
account the `pion cloud' around the nucleon, giving with 
$v^{\rgi}_n \equiv F_\chi^{(n)}(r_0m_{ps})$,
\begin{eqnarray}
   F_\chi^{(n)}(x)  &=&  a_n x^2 +
               b_n \left( 1 - c x^2 \ln { x^2 \over
                                         (x^2 + (r_0 \Lambda_\chi)^2) }
                           \right) ,
\label{ch_log_fit}
\end{eqnarray}
where $\Lambda_\chi$ is a parameter, the chiral scale, usually taken
to be of $O(4\pi f_\pi) \sim O(1\,\mbox{GeV})$
(where $f_\pi \approx 93\,\mbox{MeV}$). For $\Lambda_\chi = 0$
or large pseudoscalar mass the equation reduces to the linear case,
eq.~(\ref{linear_ch_fit}). (Unfortunately as most of the masses used
in the present simulation are larger than the strange quark mass,
eq.~(\ref{ch_log_fit}) may need higher order terms in chiral
perturbation theory to be valid at these large masses.)
These first results have been confirmed by further chiral
perturbation computations, \cite{chen01a,chen01b,arndt01a,chen01c}%
\footnote{These works find the leading chiral logarithm behaviour,
\begin{eqnarray}
   F_\chi^{(n)}(x)  &\sim&  b_n ( 1 - c x^2 \ln (
                                    x^2 / (r_0 \Lambda_\chi)^2 ) ) \,,
                                                        \nonumber
\end{eqnarray}
which is built into the model of eq.~(\ref{ch_log_fit}).}.
In particular in \cite{chen01c}, quenched QCD was considered,
with the result that at least for the nucleon there are
no additional quenched chiral logarithms present,
so-called `hairpin diagrams', so the structure of the result in
eq.~(\ref{ch_log_fit}) remains unchanged. Furthermore in
(unquenched) QCD we have
\begin{equation}
   c \equiv {3g_A^2 + 1 \over (4\pi r_0 f_\pi)^2} \sim 0.66 \,,
\end{equation}
while for quenched QCD, assuming that $\alpha^{(2)} \sim \alpha^{(1)}$,
$\beta^{(2)} \sim \beta^{(1)}$ in \cite{chen01c}, then
$c \sim 0.28$. Of course, in principle these formulae,
eq.~(\ref{ch_log_fit}), are valid in the continuum
so we should first take the continuum limit and then apply
chiral perturbation theory.
Thus we should interpolate the values for $v_n$ in
Figs.~\ref{fig_x1b_1u-1d.p0_030904_1718_pap},
\ref{fig_x1a_1u-1d.pm1_030905_1514_pap},
\ref{fig_x2b_1u-1d.pm1_040429_1554} and
\ref{fig_x3b_1u-1d.pm1_040517_1247}
to a set of constant $(r_0 m_{ps})^2$ for each value of $\beta$.
For each of these values of $(r_0m_{ps})^2$ a continuum extrapolation
should be performed. A chiral extrapolation of the data can then be
attempted. Unfortunately our `grid' of data points is not fine enough and
also for the higher moments is too noisy for this precedure.
Thus we shall try a `half way house' approach and attempt a
simultaneous continuum and chiral extrapolation of the data,
\begin{equation}
   v_n^{\rgi} = F^{(n)}_\chi(r_0m_{ps}) 
               + c_n \left( {a\over r_0} \right)^2 + d_n ar_0 m_{ps}^2 \,,
\label{joint_fit}
\end{equation}
where the first term represents chiral physics, given by
eq.~(\ref{linear_ch_fit}) or eq.~(\ref{ch_log_fit}),
the second term discretisation effects and the last term accounts
for residual $O(a)$ effects $\propto am_q$, see eq.~(\ref{op_improvement}).
With this type of fit the number of free parameters
is slightly reduced in comparison with the previous fit procedure
given in section~\ref{phenomenological_approach} and so tends
to produce smaller error bars. We shall restrict our results
here to our best data set -- $v^{\rgi}_{2b}$ and first check
that using eq.~(\ref{linear_ch_fit}) for $F^{(2)}_\chi$
reproduces our previous results.
In Fig.~\ref{fig_x1b_1u-1d.p0_040119_1938}
we first fit $v^{\rgi}_{2b}$ to eq.~(\ref{joint_fit}) with
$F^{(2)}_\chi$ given by the linear function of eq.~(\ref{linear_ch_fit}),
$F^{(2)}_\chi = a_2 (r_0m_{ps})^2 + b_2$, and then plot
$v^{\rgi}_{2b} - c_2(a/r_0)^2 - d_2 a r_0 m_{ps}^2$ for our three
$\beta$ values. (There is no perceptible difference between using
$c_0 = c_0^{\ti}$ or $c_0 = 0$, for definiteness we show the result
for $c_0 = 0$.) The points lie reasonably well on a straight line,
with extrapolated result $0.331(10)$ consistent
with our previously obtained value in Table~\ref{table_rgi_vn}.
The alternative possibility of
$v_2^{\rgi}-(F_\chi^{(2)}(r_0m_{ps}) - d_2ar_0m_{ps}^2)$ 
versus $(a/r_0)^2$ would display the $O(a^2)$ lattice discretisation
errors. However from Fig.~\ref{fig_x1a+b_aor02_pap} we see that the
$O(a^2)$ effects are small and this alternative plot just reproduces
them again.

\begin{figure}[t]
   \epsfxsize=14.00cm \epsfbox{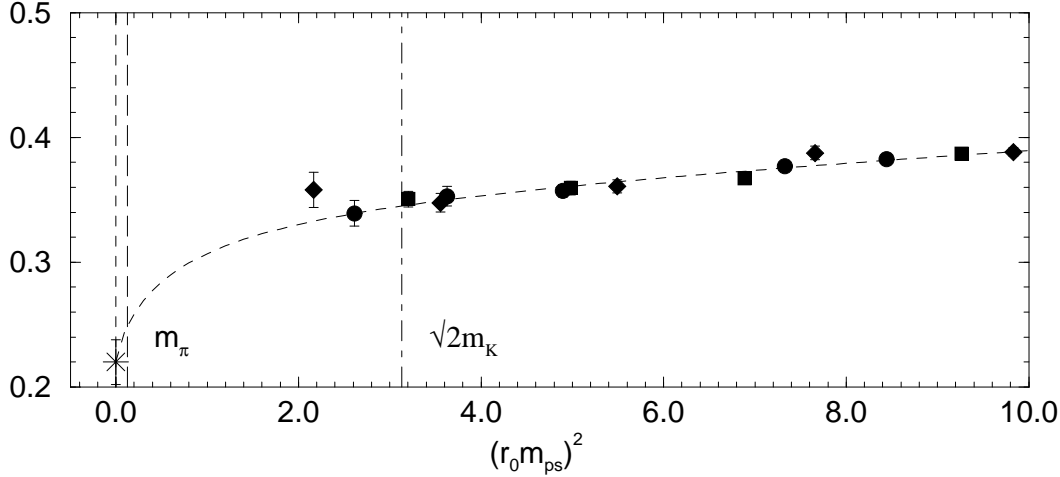}
   \caption{$v^{\rgi}_{2b}- c_2 (a/r_0)^2 - d_2 ar_0m_{ps}^2
            \equiv F^{(2)}_\chi$ (using the chiral function
            of eq.~(\ref{ch_log_fit})) versus $(r_0m_{ps})^2$,
            dashed line. The chiral limit was fixed to the MRST
            phenomenological value. Other notation is the same as in
            Fig.~\ref{fig_x1b_1u-1d.p0_040119_1938}.}
\label{fig_x1b_1u-1d.p0_040119_1930}
\end{figure}

Bolstered by this result, we now try to use $F_\chi^{(2)}$ from
eq.~(\ref{ch_log_fit}) as shown in Fig.~\ref{fig_x1b_1u-1d.p0_040119_1930}.
However it is difficult to detect any non-linearities in the data
and a $6$-parameter fit ($a_2$, $b_2$, $c$, $\Lambda_\chi$, $c_2$, $d_2$)
fails. We were forced to see if such a 6-parameter fit could be
plausible, by fixing the chiral limit, $b_2$ to be the MRST
phenomenological value given in Table~\ref{table_expt_moments}.
(Note that there is no reason that
the quenched QCD value must be the same as the QCD value, however for
many hadronic quantities there appears to be little difference between
the quenched and unquenched QCD values.) As expected, while the fit 
(dashed) line and the numerical results are in reasonable agreement,
all of the curvature of the fit takes place at small quark mass values
where there is no data. Also the fitted parameter result
for the chiral scale, $\Lambda_\chi \sim 505(48)\,\mbox{MeV}$, is
small in comparison with the expected value discussed earlier.
For $c$ we found $0.39(6)$ which lies between the unquenched
and quenched values.

So it would seem that any possible non-linearities can only show
up at rather small quark mass outside the present range of data.
Teraflop simulations will be necessary to reach more
physical pion masses. At present we shall stick to
the simplest extrapolation possible.

% ----------------------------------------------------------------------

\section{Conclusions}
\label{conclusions}

In this article we have computed $v_{n}$, $n = 2, 3, 4$.
There is still a difference between the lattice results and
the phenomenological results, particularly apparent for $v_2$
where there is a $\sim 40\%$ discrepancy.

We have tried here to narrow down the range of possibilities for the
disagreement, on the experimental side by comparing the
global MRST/CTEQ fits with the experimental data: there is good agreement.

On the lattice side, we have $O(a)$ improved the lowest moment,
investigated possible mixing operators for the higher moments
and discussed renormalisation, both perturbatively and non-perturbatively.
At least for $v_2$ there do not seem to be large $O(a^2)$
corrections and so a continuum extrapolation can be reliably performed.
For the higher moments, we have introduced modified operators
for $v_3$, $v_4$, which reduces the numerical noise and improves the signal.
It is difficult to see if there are any $O(a)$ corrections.
However at present the simplest assumption that these corrections
are small also fits with the data.

Although this is only a partial study here of mixing operators
for the higher operators (due in part to the present difficulty
of even defining renormalisation constants for several
of these operators) presently we find little sign of problems.
Indeed even for $v_{4}^{m_3}$ the mixing with a lower dimensional
operator appears harmless. In other situations this is not the case
and there may be significant changes, see eg \cite{gockeler00a}.

We have discussed and compared various renormalisation procedures
ranging from simple perturbation theory to TRB perturbation theory
to a non-perturbative method. Using NP results eliminates
uncertainty concerning the renormalisation constant. It is seen though
that between TRB perturbation theory and non-perturbative results there
can be up to an $8\%$ difference -- far less than the difference
between the lattice and phenomenological result for $v_2^{\rgi}$.

Finally although at present we see little numerical evidence of
chiral logarithms, this is perhaps telling us that we must go to
significantly smaller quark masses before the chiral extrapolation
`bends' over. However preliminary results at lighter quark mass for
unimproved Wilson fermions also show a linear behaviour and
so do not seem to improve the situation, \cite{gockeler02a,bakeyev03b}.

Whether quenching effects are significant remains unclear,
but recent unquenched results, \cite{bakeyev03b,dolgov02a}, do
not seem to reveal any significant differences between quenching
and unquenching, at least in the quark mass range considered.
Finally there are hints of a different situation for overlap fermions,
\cite{galletly03a}, which might suggest again that one has to
simulate at light quark masses close to the chiral limit
 -- a challenge for the lattice.

% ----------------------------------------------------------------------

\section*{ACKNOWLEDGEMENTS}
\label{acknowledgement}

The numerical calculations were performed on the
APE1000 and Quadrics at DESY (Zeuthen) and the Cray T3E at ZIB (Berlin).
We thank the operating staff for their support.
This work is supported in part by the
EU Integrated Infrastructure Initiative Hadron Physics (I3HP) 
and by the DFG (Forschergruppe Gitter-Hadronen-Ph\"anomenologie).
We also wish to thank L. Mankiewicz for much help with obtaining
the experimental data used in section~\ref{experiment_theory}
and S. Capitani for one-loop perturbative results for $v_3$ and $v_4$.

% ----------------------------------------------------------------------

\clearpage

\appendix

\section*{Appendices}

% ----------------------------------------------------------------------

\section{Some values of $[\Delta Z_{v_n}^{\msbar}(\mu)]^{-1}$}
\label{table_v2sbar}

\begin{table}[h]
   \begin{center}
      \begin{tabular}{||l||l|l|l||}
         \hline
         \hline
         \multicolumn{1}{||c||}{$\mu$}
                        & one-loop & two-loop & three-loop \\
         \hline
         \hline
         \multicolumn{1}{||c||}{} &
           \multicolumn{3}{c||}{} \\[-0.9em]
         \multicolumn{1}{||c||}{} &
           \multicolumn{3}{c||}{$[\Delta Z_{v_2}^{\msbar}(\mu)]^{-1}$} \\
         \multicolumn{1}{||c||}{} &
           \multicolumn{3}{c||}{} \\[-0.9em]
         \hline
         \hline
 $2.00\,\mbox{GeV}$            & $0.783(10)$ & $0.721(8)$
                               & $0.732(9)$ \\
 $2.12\,\mbox{GeV}$ ($1/a$ at $\beta = 6.0$)
                               & $0.776(10)$ & $0.715(8)$
                               & $0.726(8)$  \\
 $2.90\,\mbox{GeV}$ ($1/a$ at $\beta = 6.2$)
                               & $0.743(8)$  & $0.688(7)$
                               & $0.696(7)$  \\
 $3.85\,\mbox{GeV}$ ($1/a$ at $\beta = 6.4$)
                               & $0.718(7)$  & $0.668(6)$
                               & $0.674(6)$  \\
         \hline
         \hline
         \multicolumn{1}{||c||}{} &
           \multicolumn{3}{c||}{} \\[-0.9em]
         \multicolumn{1}{||c||}{} &
           \multicolumn{3}{c||}{$[\Delta Z_{v_3}^{\msbar}(\mu)]^{-1}$} \\
         \multicolumn{1}{||c||}{} &
           \multicolumn{3}{c||}{} \\[-0.9em]
         \hline
         \hline
 $2.00\,\mbox{GeV}$            & $0.682(14)$ & $0.596(10)$
                               & $0.609(11)$ \\
 $2.12\,\mbox{GeV}$ ($1/a$ at $\beta = 6.0$)
                               & $0.673(13)$ & $0.589(10)$
                               & $0.602(10)$ \\
 $2.90\,\mbox{GeV}$ ($1/a$ at $\beta = 6.2$)
                               & $0.629(11)$ & $0.555(8)$
                               & $0.564(8)$  \\
 $3.85\,\mbox{GeV}$ ($1/a$ at $\beta = 6.4$)
                               & $0.596(9)$  & $0.529(7)$
                               & $0.537(7)$  \\
         \hline
         \hline
         \multicolumn{1}{||c||}{} &
           \multicolumn{3}{c||}{} \\[-0.9em]
         \multicolumn{1}{||c||}{} &
           \multicolumn{3}{c||}{$[\Delta Z_{v_4}^{\msbar}(\mu)]^{-1}$} \\
         \multicolumn{1}{||c||}{} &
           \multicolumn{3}{c||}{} \\[-0.9em]
         \hline
         \hline
 $2.00\,\mbox{GeV}$            & $0.619(15)$ & $0.520(11)$
                               & $0.534(13)$ \\
 $2.12\,\mbox{GeV}$ ($1/a$ at $\beta = 6.0$)
                               & $0.609(15)$ & $0.512(10)$
                               & $0.526(12)$  \\
 $2.90\,\mbox{GeV}$ ($1/a$ at $\beta = 6.2$)
                               & $0.559(12)$  & $0.476(9)$
                               & $0.486(10)$  \\
 $3.85\,\mbox{GeV}$ ($1/a$ at $\beta = 6.4$)
                               & $0.522(10)$  & $0.448(8)$
                               & $0.456(8)$  \\
         \hline
         \hline
         \multicolumn{1}{||c||}{} &
           \multicolumn{3}{c||}{} \\[-0.9em]
         \multicolumn{1}{||c||}{} &
           \multicolumn{3}{c||}{$\alpha_s^{\msbar}(\mu)$} \\
         \multicolumn{1}{||c||}{} &
           \multicolumn{3}{c||}{} \\[-0.9em]
         \hline
 $2.00\,\mbox{GeV}$            & $0.268(10)$ & $0.195(6)$
                               & $0.201(6)$  \\
 $2.12\,\mbox{GeV}$ ($1/a$ at $\beta = 6.0$)
                               & $0.261(9)$  & $0.191(5)$
                               & $0.196(6)$  \\
 $2.90\,\mbox{GeV}$ ($1/a$ at $\beta = 6.2$)
                               & $0.228(7)$  & $0.170(5)$
                               & $0.174(5)$  \\
 $3.85\,\mbox{GeV}$ ($1/a$ at $\beta = 6.4$)
                               & $0.205(6)$  & $0.156(3)$
                               & $0.159(4)$  \\
         \hline
         \hline
      \end{tabular}
   \end{center}
\caption{Useful values of $[\Delta Z_{v_n}^{\msbar}(\mu)]^{-1}$
         ($n = 2$, $3$, $4$) and $\alpha_s^{\msbar}(\mu)
         \equiv (g^{\msbar}(\mu))^2/4\pi$. The errors are 
         a reflection of the error in $\Lambda^{\msbar} r_0$.
         The lattice simulations performed here, as described in
         section~\ref{lattice}, give the above $1/a$ values
         (found from using $r_0/a$ in \cite{guagnelli98a},
         namely $r_0/a = 5.37$, $7.36$, $9.76$ at $\beta = 6.0$,
         $6.2$, $6.4$ respectively,
         together with the scale $r_0 = 0.5\,\mbox{fm}$).}
\label{table_v2sbar_values}
\end{table}

% ----------------------------------------------------------------------

\section{Group properties of mixing operators - an example}
\label{group}
 
When QCD is put on the lattice, the necessary analytic continuation
from Mink\-ow\-ski to Euclidean space replaces the Lorentz group by the
orthogonal group $O(4)$, which by the discretisation of space-time
is further reduced to the hypercubic group $H(4) \subset O(4)$.
Since $H(4)$ is only a finite group, the restrictions imposed
by symmetry are less stringent than in the continuum and the
possibilities for mixing increase, sometimes in a way which is not easily
anticipated by our (continuum) intuition. In this appendix we illustrate
the problem by an example. For a more complete and systematic treatment
we refer to \cite{gockeler96a}.

As a further symmetry we have charge conjugation. It 
operates on the fermion fields $q (x)$,
$\overline{q} (x)$ and on the lattice gauge field $U_\mu(x)$
according to
\begin{eqnarray}
   q (x)       & \stackrel{C}{\to} & C \overline{q} (x) ^T  \,, \nonumber \\
   \overline{q} (x)
               & \stackrel{C}{\to} & - q(x) ^T C^{-1} \,,       \nonumber \\
   U_\mu(x)    & \stackrel{C}{\to} & U_\mu(x)^*
\end{eqnarray}
with the charge conjugation matrix $C$ satisfying
\begin{equation}
   C \gamma_\mu^T C^{-1} = - \gamma_\mu \,.
\label{charge_conj_def}
\end{equation}
So we get, eg,
\begin{eqnarray}
   {\mathcal O}^\gamma_{\mu_1 \mu_2 \ldots \mu_n} \stackrel{C}{\to}
   (-1)^n {\mathcal O}^\gamma_{\mu_1 \mu_n \mu_{n-1} \ldots \mu_2 }
                                        \,,  \nonumber \\
   {\mathcal O}^{\gamma \gamma_5}_{\mu_1 \mu_2 \ldots \mu_n} \stackrel{C}{\to}
   (-1)^{n-1} {\mathcal O}^{\gamma \gamma_5}
   _{\mu_1 \mu_n \mu_{n-1} \ldots \mu_2 }  \,.
\end{eqnarray}

Identifying elements $R$ of $H(4)$ with the corresponding $4 \times 4$ 
matrices in the defining representation we find that $R$ acts on 
$\overline{q} \gamma_\mu q$ as follows:
\begin{equation}
   \overline{q} \gamma_\mu q \stackrel{R}{\to} 
   \sum_\nu R_{\nu \mu} \bar{q} \gamma_\nu q \,,
\end{equation}
ie, the four operators $\overline{q} \gamma_\mu q$ form a basis for the 
defining representation of $H(4)$. More generally, we get for the action 
of $R$:
\begin{equation} \label{trans}
   {\mathcal O}^\gamma_{\mu_1 \mu_2 \ldots \mu_n} \stackrel{R}{\to} 
   \sum_{\nu_1, \nu_2, \ldots, \nu_n} R_{\nu_1 \mu_1} R_{\nu_2 \mu_2}
   \cdots R_{\nu_n \mu_n} {\mathcal O}^\gamma_{\nu_1 \nu_2 \ldots \nu_n} 
\end{equation}
and
\begin{equation}  \label{trans5}
   {\mathcal O}^{\gamma \gamma_5}_{\mu_1 \mu_2 \ldots \mu_n} 
   \stackrel{R}{\to} \; \det R \sum_{\nu_1, \nu_2, \ldots, \nu_n} 
   R_{\nu_1 \mu_1} R_{\nu_2 \mu_2} \cdots R_{\nu_n \mu_n} 
   {\mathcal O}^{\gamma \gamma_5}_{\nu_1 \nu_2 \ldots \nu_n} \,.
\end{equation}
Thus the $4^n$ operators ${\mathcal O}^\gamma_{\mu_1 \mu_2 \ldots \mu_n}$
as well as the operators 
${\mathcal O}^{\gamma \gamma_5}_{\mu_1 \mu_2 \ldots \mu_n}$
form a basis for a representation of $H(4)$, which for $n > 1$ is reducible.
It is helpful to consider these operators as forming an orthonormal
basis of the representation space. (Orthonormal) bases of irreducible 
subspaces have been given in \cite{gockeler96a}.

Operators transforming according to the same irreducible representation
of $H(4)$ (and having the same ${\cal C}$-parity) will in general
mix with each other so that one has to consider appropriate linear
combinations. Writing down these linear combinations one has to choose
the bases in the different (equivalent) representation spaces
such that they transform {\em identically} under $H(4)$
and not just equivalently.

Consider two bases which are known to transform according 
to the same irreducible representation. In order to check whether they
even transform identically it is sufficient to compare their transformation
behaviour under a set of generators. For $H(4)$ there is a set of three
generators $\{ \alpha, \beta, \gamma \}$ given by 
\begin{eqnarray}
\alpha & = & \left (
   \begin{array}{cccc}
       0 & 1 & 0 & 0  \\
       1 & 0 & 0 & 0  \\
       0 & 0 & 1 & 0  \\
       0 & 0 & 0 & 1    \end{array}   \right ) \,,  \nonumber \\
\beta  & = & \left (
   \begin{array}{cccc}
       0 & 0 & 0 & 1  \\
       1 & 0 & 0 & 0  \\
       0 & 1 & 0 & 0  \\
       0 & 0 & 1 & 0    \end{array}   \right ) \,,  \nonumber \\
\gamma & = & \left (
   \begin{array}{cccc}
      -1 & 0 & 0 & 0  \\
       0 & 1 & 0 & 0  \\
       0 & 0 & 1 & 0  \\
       0 & 0 & 0 & 1    \end{array}   \right ) \,.
\end{eqnarray}
According to (\ref{trans}), $\alpha$ interchanges 1 and 2, $\beta$ 
performs a cyclic permutation of the values of the indices, and $\gamma$
produces a factor of $-1$ for every index 1, eg, for the operator
${\mathcal O}^\gamma_{3341}$ one finds
\begin{eqnarray}
{\mathcal O}^\gamma_{3341} & \stackrel{\alpha}{\to} & 
{\mathcal O}^\gamma_{3342}   \,, \nonumber \\
{\mathcal O}^\gamma_{3341} & \stackrel{\beta}{\to} & 
{\mathcal O}^\gamma_{4412}   \,, \nonumber \\
{\mathcal O}^\gamma_{3341} & \stackrel{\gamma}{\to} & 
- {\mathcal O}^\gamma_{3341}   \,.
\end{eqnarray}

As an example for mixing operators, let us consider $\mathcal O_{v_4}$. 
It belongs to a doublet of operators transforming according to the 
$H(4)$ representation $\tau^{(2)}_1$ (in the notation of
\cite{baake82a} and Table~\ref{table_H4_transformation})
and has positive ${\cal C}$-parity. Indeed, the two operators
\begin{equation}
u_1 :=  \frac{\sqrt{6}}{2} \left(
{\mathcal O}^\gamma_{\{1122 \}} + {\mathcal O}^\gamma_{\{3344 \}} 
- {\mathcal O}^\gamma_{\{1133 \}} - {\mathcal O}^\gamma_{\{2244 \}} \right)
                                                                  \,,
\end{equation}
and
\begin{equation}
u_2 := \frac{1}{\sqrt{2}} \left(
2 {\mathcal O}^\gamma_{\{1144 \}} + 2 {\mathcal O}^\gamma_{\{2233 \}} 
- {\mathcal O}^\gamma_{\{1122 \}} - {\mathcal O}^\gamma_{\{3344 \}} 
- {\mathcal O}^\gamma_{\{1133 \}} - {\mathcal O}^\gamma_{\{2244 \}} \right)
= \sqrt{2} \mathcal O_{v_4}
\end{equation}
form an orthonormal basis for this representation. It is obvious that 
both operators remain unchanged under the action of $\gamma$. The 
generator $\alpha$ acts on them as follows:
\begin{eqnarray}
u_1 & \stackrel{\alpha}{\to} & \half u_1 - \half \sqrt{3} u_2 \,, \nonumber \\
u_2 & \stackrel{\alpha}{\to} & - \half \sqrt{3} u_1 - \half u_2  \,, 
\end{eqnarray}
while $\beta$ gives
\begin{eqnarray}
u_1 & \stackrel{\beta}{\to} &  \half u_1 + \half \sqrt{3} u_2 \,, \nonumber \\
u_2 & \stackrel{\beta}{\to} &  \half \sqrt{3} u_1 - \half u_2 \,. 
\end{eqnarray}
It is straightforward to check that the operators
\begin{eqnarray}
w_1 & := & \frac{1}{4 \sqrt{3}} \Big( {}
 - {\mathcal O}^\gamma_{1122} - {\mathcal O}^\gamma_{2112}
 - {\mathcal O}^\gamma_{1221} - {\mathcal O}^\gamma_{2211} 
 + 2 {\mathcal O}^\gamma_{1212} + 2 {\mathcal O}^\gamma_{2121}
\nonumber \\ {} & {} & \hphantom{\frac{1}{4 \sqrt{3}} \Big(} {} 
 - {\mathcal O}^\gamma_{3344} - {\mathcal O}^\gamma_{4334}
 - {\mathcal O}^\gamma_{3443} - {\mathcal O}^\gamma_{4433} 
 + 2 {\mathcal O}^\gamma_{3434} + 2 {\mathcal O}^\gamma_{4343} 
\nonumber \\ {} & {} & \hphantom{\frac{1}{4 \sqrt{3}} \Big(} {}
 + {\mathcal O}^\gamma_{1133} + {\mathcal O}^\gamma_{3113}
 + {\mathcal O}^\gamma_{1331} + {\mathcal O}^\gamma_{3311} 
 - 2 {\mathcal O}^\gamma_{1313} - 2 {\mathcal O}^\gamma_{3131} 
\nonumber \\ {} & {} & \hphantom{\frac{1}{4 \sqrt{3}} \Big(} {} 
 + {\mathcal O}^\gamma_{2244} + {\mathcal O}^\gamma_{4224}
 + {\mathcal O}^\gamma_{2442} + {\mathcal O}^\gamma_{4422} 
 - 2 {\mathcal O}^\gamma_{2424} - 2 {\mathcal O}^\gamma_{4242} \Big) \,,
\end{eqnarray}
and
\begin{eqnarray}
w_2 & := & \frac{1}{12} \Big( 
 + {\mathcal O}^\gamma_{1122} + {\mathcal O}^\gamma_{2112}
 + {\mathcal O}^\gamma_{1221} + {\mathcal O}^\gamma_{2211} 
 - 2 {\mathcal O}^\gamma_{1212} - 2 {\mathcal O}^\gamma_{2121} 
\nonumber \\ {} & {} & \hphantom{\frac{1}{12} \Big(} {} 
 + {\mathcal O}^\gamma_{3344} + {\mathcal O}^\gamma_{4334}
 + {\mathcal O}^\gamma_{3443} + {\mathcal O}^\gamma_{4433} 
 - 2 {\mathcal O}^\gamma_{3434} - 2 {\mathcal O}^\gamma_{4343} 
\nonumber \\ {} & {} & \hphantom{\frac{1}{12} \Big(} {}
 + {\mathcal O}^\gamma_{1133} + {\mathcal O}^\gamma_{3113}
 + {\mathcal O}^\gamma_{1331} + {\mathcal O}^\gamma_{3311} 
 - 2 {\mathcal O}^\gamma_{1313} - 2 {\mathcal O}^\gamma_{3131} 
\nonumber \\ {} & {} & \hphantom{\frac{1}{12} \Big(} {}
 + {\mathcal O}^\gamma_{2244} + {\mathcal O}^\gamma_{4224}
 + {\mathcal O}^\gamma_{2442} + {\mathcal O}^\gamma_{4422} 
 - 2 {\mathcal O}^\gamma_{2424} - 2 {\mathcal O}^\gamma_{4242}
\nonumber \\ {} & {} & \hphantom{\frac{1}{12} \Big(} {} 
 - 2 {\mathcal O}^\gamma_{1144} - 2 {\mathcal O}^\gamma_{4114}
 - 2 {\mathcal O}^\gamma_{1441} - 2 {\mathcal O}^\gamma_{4411} 
 + 4 {\mathcal O}^\gamma_{1414} + 4 {\mathcal O}^\gamma_{4141} 
\nonumber \\ {} & {} & \hphantom{\frac{1}{12} \Big(} {} 
 - 2 {\mathcal O}^\gamma_{2233} - 2 {\mathcal O}^\gamma_{3223}
 - 2 {\mathcal O}^\gamma_{2332} - 2 {\mathcal O}^\gamma_{3322} 
 + 4 {\mathcal O}^\gamma_{2323} + 4 {\mathcal O}^\gamma_{3232}  \Big) \,,
\nonumber \\ {} & \hphantom{:}= & 
 \frac{1}{6} {\mathcal O}^{m_1}_{v_4} \,,
\end{eqnarray}
transform identically to $u_1$, $u_2$: 
\begin{eqnarray}
   w_1 & \stackrel{\alpha}{\to} & \half w_1 - \half \sqrt{3} w_2 
                                                           \,, \nonumber \\
   w_2 & \stackrel{\alpha}{\to} & - \half \sqrt{3} w_1 - \half w_2
                                                           \,, 
\end{eqnarray}
\begin{eqnarray}
   w_1 & \stackrel{\beta}{\to} &  \half w_1 + \half \sqrt{3} w_2
                                                          \,, \nonumber \\
   w_2 & \stackrel{\beta}{\to} &  \half \sqrt{3} w_1 - \half w_2
                                                          \,,
\end{eqnarray}
and they are invariant under the action of $\gamma$.
Hence any linear combination $r \cdot u_i + s \cdot w_i$ ($i=1,2$) has the 
same transformation properties under $H(4)$ as $u_i$. In particular,
${\mathcal O}_{v_4}$ may mix with ${\mathcal O}^{m_1}_{v_4}$, and the
renormalisation of ${\mathcal O}_{v_4}$ will in general involve also 
${\mathcal O}^{m_1}_{v_4}$.

Also the operators
\begin{eqnarray}
w'_1 & := & \frac{1}{4} \Big( {}
 + {\mathcal O}^{\gamma \gamma_5}_{1324} 
 - {\mathcal O}^{\gamma \gamma_5}_{2314} 
 - {\mathcal O}^{\gamma \gamma_5}_{1423} 
 + {\mathcal O}^{\gamma \gamma_5}_{2413} 
 + {\mathcal O}^{\gamma \gamma_5}_{3142} 
 - {\mathcal O}^{\gamma \gamma_5}_{4132} 
 - {\mathcal O}^{\gamma \gamma_5}_{3241} 
 +{\mathcal O}^{\gamma \gamma_5}_{4231}  \hspace*{0.25in}
\nonumber \\ {} & {} &  {} \hspace*{-0.10in} 
 + {\mathcal O}^{\gamma \gamma_5}_{1234} 
 - {\mathcal O}^{\gamma \gamma_5}_{3214} 
 - {\mathcal O}^{\gamma \gamma_5}_{1432} 
 + {\mathcal O}^{\gamma \gamma_5}_{3412} 
 + {\mathcal O}^{\gamma \gamma_5}_{2143} 
 - {\mathcal O}^{\gamma \gamma_5}_{4123} 
 - {\mathcal O}^{\gamma \gamma_5}_{2341} 
 + {\mathcal O}^{\gamma \gamma_5}_{4321} \Big) \,,
\end{eqnarray}
and
\begin{eqnarray}
w'_2 & := & \frac{1}{4 \sqrt{3}} \Big( {}
 + {\mathcal O}^{\gamma \gamma_5}_{1234} 
 - {\mathcal O}^{\gamma \gamma_5}_{3214} 
 - {\mathcal O}^{\gamma \gamma_5}_{1432} 
 + {\mathcal O}^{\gamma \gamma_5}_{3412} 
 + {\mathcal O}^{\gamma \gamma_5}_{2143} 
 - {\mathcal O}^{\gamma \gamma_5}_{4123} 
 - {\mathcal O}^{\gamma \gamma_5}_{2341} 
 + {\mathcal O}^{\gamma \gamma_5}_{4321} 
\nonumber \\ {} & {} & \hphantom{\frac{1}{4 \sqrt{3}} \Big(} {} 
 - {\mathcal O}^{\gamma \gamma_5}_{1324} 
 + {\mathcal O}^{\gamma \gamma_5}_{2314} 
 + {\mathcal O}^{\gamma \gamma_5}_{1423} 
 - {\mathcal O}^{\gamma \gamma_5}_{2413} 
 - {\mathcal O}^{\gamma \gamma_5}_{3142} 
 + {\mathcal O}^{\gamma \gamma_5}_{4132} 
 + {\mathcal O}^{\gamma \gamma_5}_{3241} 
 - {\mathcal O}^{\gamma \gamma_5}_{4231} 
\nonumber \\ {} & {} & \hphantom{\frac{1}{4 \sqrt{3}} \Big(} {} 
 + 2{\mathcal O}^{\gamma \gamma_5}_{1243} 
 - 2{\mathcal O}^{\gamma \gamma_5}_{4213} 
 - 2{\mathcal O}^{\gamma \gamma_5}_{1342} 
 + 2{\mathcal O}^{\gamma \gamma_5}_{4312} 
 + 2{\mathcal O}^{\gamma \gamma_5}_{2134} 
 - 2{\mathcal O}^{\gamma \gamma_5}_{3124} 
\nonumber \\ {} & {} &  \hphantom{\frac{1}{4 \sqrt{3}} \Big( 
 + 2{\mathcal O}^{\gamma \gamma_5}_{1243} 
 - 2{\mathcal O}^{\gamma \gamma_5}_{4213} 
 - 2{\mathcal O}^{\gamma \gamma_5}_{1342} 
 + 2{\mathcal O}^{\gamma \gamma_5}_{4312} } {}
 - 2{\mathcal O}^{\gamma \gamma_5}_{2431} 
 + 2{\mathcal O}^{\gamma \gamma_5}_{3421}  \Big)  \,,
\nonumber \\ {} &  \hphantom{:}=  &  
             \frac{1}{2 \sqrt{3}} {\mathcal O}^{m_2}_{v_4} \,,
\end{eqnarray}
transform in exactly the same way under $\alpha$, $\beta$, and $\gamma$,
ie under $H(4)$, as follows from (\ref{trans5}). Therefore 
${\mathcal O}^{m_2}_{v_4}$ may mix with ${\mathcal O}_{v_4}$, too.

% ----------------------------------------------------------------------

\section{Two- and three-point correlation functions}
\label{appendix_3pt}

% ----------------------------------------------------------------------

\subsection{General formulae}

In this appendix we shall give explicit expressions for the
correlation functions in terms of quark propagators.
We start with the two-point correlation function.
A suitable proton operator is
\begin{equation}
   B_\alpha(t; \vec{p})
             =  \sum_{\vec{x}} e^{-i\vec{p} \cdot \vec{x}} \epsilon^{abc}
                             u^a_\alpha(\vec{x},t) 
                      \left[ u^b(\vec{x},t)^{T_D}
                    C\gamma_5 d^c(\vec{x},t) \right] \,,
\label{baryon_op}
\end{equation}
($a \ldots$ denote colour indices, $\alpha\ldots$ Dirac indices,
$T_D$ is the transpose in Dirac space and
$\gamma_5 = \gamma_1\gamma_2\gamma_3\gamma_4$). $C$, the charge
conjugation matrix, has the defining property given in
eq.~(\ref{charge_conj_def}) and is antisymmetric. As we take our
gamma matrices to be hermitian then $C$ may be taken as as unitary.
Thus $-C = C^{T_D}$ and $C^{-1} = C^{\dagger}$. One possible choice
(used in our computer programme) is $C = \gamma_4\gamma_2$
(so that $-C = C^{-1}$) and the Dirac basis
\begin{equation}
   \gamma_i = \left( \begin{array}[c]{cc}
                        0           & -i \sigma_i \\
                        i \sigma_i & 0
                     \end{array}
              \right) \,, \qquad
   \gamma_4 = \left( \begin{array}[c]{cc}
                        I & 0 \\
                        0 & -I
                     \end{array}
              \right) \,,
\label{gamma_def}
\end{equation}
(but the results given here should not depend on this particular
choice). The last two quark fields in eq.~(\ref{baryon_op}) form a
di-quark structure, while the first quark carries the spin index.
With the help of
\begin{equation}
   \overline{B}_\alpha(t; \vec{p})
        =  \sum_{\vec{x}} e^{i\vec{p}.\vec{x}} \epsilon^{abc}
             \overline{u}^a_\alpha(\vec{x},t) 
                \left[ \overline{d}^b(\vec{x},t)
                             \gamma_5 C\overline{u}^c(\vec{x},t)^{T_D}
                \right] \,,
\end{equation}
the two-point correlation function is formed in the usual way
\begin{equation}
   C_{\Gamma}(t;\vec{p}) = \mbox{Tr}_D \Gamma
                                \langle B(t,\vec{p})
                                        \overline{B}(0,\vec{p})
                                \rangle \,,
\label{2pt_correlation}
\end{equation}
where we have introduced an arbitrary Dirac matrix, $\Gamma$, which
for unpolarised nucleons, the case considered here, should be taken as
$\half (1 + \gamma_4)$. This equation may be re-written using quark
propagators propagating from a (point) source to a sink.
As we are averaging over the gauge fields we may shift all
sources to $(\vec{0},0)$. Some algebra then yields
\begin{eqnarray}
   \lefteqn{C_\Gamma(t;\vec{p}) =
            - V_s \sum_{\vec{x}} e^{-i\vec{p}\cdot\vec{x}} \epsilon^{abc} 
                                      \epsilon^{a^\prime b^\prime c^\prime} }
         &                                   \nonumber \\
         &                \left\langle
    \mbox{Tr}_D \left[ \Gamma G^{(u)aa^{\prime}}(\vec{x},t;\vec{0},0) \right]
    \mbox{Tr}_D \left[ \tilde{G}^{(d)bb^{\prime}}(\vec{x},t;\vec{0},0)
                              G^{(u)cc^{\prime}}(\vec{x},t;\vec{0},0)
                                                           \right] \right.
                                              \nonumber \\
         & \qquad \quad + \left.
    \mbox{Tr}_D \left[ \Gamma G^{(u)aa^{\prime}}(\vec{x},t;\vec{0},0)
                       \tilde{G}^{(d)bb^{\prime}}(\vec{x},t;\vec{0},0)
                              G^{(u)cc^{\prime}}(\vec{x},t;\vec{0},0) \right]
                         \right\rangle_{\{U\}} \,,
\label{baryon_gory_detail}
\end{eqnarray}
where we have defined a tilde in Dirac space by
\begin{equation}
   \tilde{X} = (C\gamma_5 X \gamma_5 C)^{T_D} \,.
\label{tilde_def}
\end{equation}
The problem is thus reduced to finding the propagator or
Green's function for quark $q$ from a source $(\vec{0},0)$ to
$(\vec{x},t)$. In general, the quark propagator from $y$ to $x$
is defined by
\begin{equation}
   G^{(q) ab}_{\alpha\beta}(x; y) =
        \langle {q}_\alpha^{a}(x) \overline{{q}}_\beta^{b} (y)
        \rangle_q \,,
\label{green_integrate_out_psi}
\end{equation}
and can be computed from
\begin{equation}
   \sum_w M^{(q)}(x; w) G^{(q)}(w; y) = \delta_{xy} \,,\
\label{invert_M}
\end{equation}
where $M_{\alpha\beta}^{ab}$ is the Wilson (clover) matrix, given
in Appendix~\ref{tables}.

For the three-point correlation function,
\begin{equation}
   C_{\Gamma}(t,\tau; \vec{p}; {\cal O}_q) = 
      \mbox{Tr}_D \Gamma \langle B(t,\vec{p})  {\cal O}_q(\tau)
                                        \overline{B}(0,\vec{p})
                        \rangle \,,
\label{3pt_correlation}
\end{equation}
we shall only consider the quark line connected term
(ie the left diagram of Fig.~\ref{fig_B_3pt}).
First we re-write the operator insertion generally as
\begin{eqnarray}
   {\cal O}_q(\tau) &=& \sum_{\vec{y}} {\cal O}_q(\vec{y},\tau)
                                                \nonumber \\
                    &=& \sum_{\vec{y},v,w}
                         \overline{q}^a_\alpha(v)
                          O_{\alpha\beta}^{ab}(v,w; \vec{y},\tau)
                         q_\beta^b(w)
\label{latt_op_matrix}
\end{eqnarray}
(ie sum over spatial planes, where $(\vec{y},\tau)$ is the `center of
mass of the operator'). For operators without derivatives and with
exactly one derivative we have $\overline{q} \gamma q$ and
$\half \overline{q}
\gamma ( \stackrel{\rightarrow}{D} - \stackrel{\leftarrow}{D})q$
for ${\cal O}_q$ respectively,
($\stackrel{\rightarrow}{D}$ and $\stackrel{\leftarrow}{D}$ are defined
in eq.~(\ref{Ddefs})) while for two and three derivative  operators,
to minimise the extension on the lattice we `integrate by parts' and choose
$-\overline{q} \gamma \stackrel{\leftarrow}{D} \stackrel{\rightarrow}{D} q$
and $\half \overline{q} \gamma ( \stackrel{\leftarrow}{D}
\stackrel{\leftarrow}{D} \stackrel{\rightarrow}{D} -
\stackrel{\leftarrow}{D} \stackrel{\rightarrow}{D}
\stackrel{\rightarrow}{D} ) q$
respectively. This also allows the higher derivative operators
to be built from the previously constructed lower derivative
operators.
   
Some algebra yields the result for eq.~(\ref{3pt_correlation}) of
\begin{eqnarray}
   \lefteqn{C_\Gamma(t,\tau; \vec{p};{\cal O}_q)}
      & &                                        \nonumber  \\         
      &=& - V_s \sum_{\vec{y},v,w}
             \left\langle \mbox{Tr}_{CD} 
                 \left[ \Sigma_\Gamma^{(q)}(\vec{0},0; v;\vec{p},t)
                        O(v,w;\vec{y},\tau)
                        G^{(q)}(w;0,0)
                 \right]
             \right\rangle_{\{U\}} \,,
\end{eqnarray}
where $\Sigma_\Gamma^{(q)}(\vec{0},0;v;\vec{p},t)$ is given by
\begin{equation}
   \Sigma_\Gamma^{(q)}(\vec{0},0;v;\vec{p},t)
             = \sum_{\vec{x}} S_\Gamma^{(q)}(\vec{x},t;\vec{0},0;\vec{p})
               G^{(q)}(\vec{x},t; v) \,,
\label{source+gf_standard}
\end{equation}
in terms of
\begin{eqnarray}
   \lefteqn{
   S^{(u)a^\prime a}_\Gamma(\vec{x},t;\vec{0},0;\vec{p})
    = e^{-i\vec{p}\cdot\vec{x}} \epsilon^{abc}
                           \epsilon^{a^\prime b^\prime c^\prime} \times }
       & &                                \label{source_u}  \\
       & & \left[ \widetilde{G}^{(d)bb^\prime}(\vec{x},t;\vec{0},0)
                              G^{(u)cc^\prime}(\vec{x},t;\vec{0},0)\Gamma +
                  \mbox{Tr}_D [
                  \widetilde{G}^{(d)bb^\prime}(\vec{x},t;\vec{0},0)
                              G^{(u)cc^\prime}(\vec{x},t;\vec{0},0) ] \Gamma +
          \right.                                 \nonumber \\
       & & \left. \Gamma G^{(u)bb^\prime}(\vec{x},t;\vec{0},0)
                        \widetilde{G}^{(d)cc^\prime}(\vec{x},t;\vec{0},0) +
                  \mbox{Tr}_D [
                  \Gamma G^{(u)bb^\prime}(\vec{x},t;\vec{0},0) ]
                  \widetilde{G}^{(d)cc^\prime}(\vec{x},t;\vec{0},0) ]
           \right] \,,
                                                  \nonumber
\end{eqnarray}
when $q = u$ and a slightly simpler expression for $S_\Gamma^{(d)}$
namely
\begin{eqnarray}
   \lefteqn{
    S^{(d)a^\prime a}_\Gamma(\vec{x},t;\vec{0},0;\vec{p})
     = e^{-i\vec{p}\cdot\vec{x}} \epsilon^{abc}
                                \epsilon^{a^\prime b^\prime c^\prime} \times}
       & &                                \label{source_d}  \\
       & & \left[ \widetilde{G}^{(u)bb^\prime}(\vec{x},t;\vec{0},0)
                  \widetilde{\Gamma}
                  \widetilde{G}^{(u)cc^\prime}(\vec{x},t;\vec{0},0) +
                  \mbox{Tr}_D [
                  \Gamma G^{(u)bb^\prime}(\vec{x},t;\vec{0},0)
                  \widetilde{G}^{(u)cc^\prime}(\vec{x},t;\vec{0},0) ]
          \right] \,.
                                                  \nonumber
\end{eqnarray}
Practically we must find $\Sigma_\Gamma^{(q)}$ from a second Green's
function using these rather ugly looking $S_\Gamma^{(q)}$ expressions
as sources. By considering
\begin{equation}
   \sum_v \Sigma_\Gamma^{(q)}(\vec{0},0;v;\vec{p},t)
              M^{(q)}(v; v^\prime) = S_\Gamma^{(q)}(\vec{v}^{\,\prime} t;
                                                   \vec{0},0;\vec{p})
                                       \delta_{v_4^\prime t} \,,
\end{equation}
we see that this is the wrong way around for the inversion in
eq.~(\ref{invert_M}) but taking colour/spin components and using
$M^{ab}_{\alpha\beta}(x;y) = (\gamma_5 M^*(y;x)^{ba}\gamma_5)_{\beta\alpha}$
the equation for $\Sigma$ can be re-written as
\begin{equation}
   \sum_v M^{(q)}(v^\prime ; v) \gamma_5 
      \Sigma_\Gamma^{(q)\dagger_{CD}}(\vec{0},0;v;\vec{p},t)
             = \gamma_5 S_\Gamma^{(q)\dagger_{CD}}(\vec{v}^{\,\prime}, t;
                              \vec{0},0 ; \vec{p}) \delta_{v_4^\prime,t} \,,
\label{sig_invert}
\end{equation}
where $\dagger_{CD}$ is the Hermitian conjugate in colour and
spin space. We see that in this form $\Sigma$ is rather like a
Green's function from the source given on the RHS of the equation. 

Thus finding the three point correlation functions is a two
step process: first the usual Green's function from $(\vec{0},0)$ to any
point $x$ is found, stage I.
Then a second inversion (stage II) is made with the source given
from eq.~(\ref{source_u}) if the inserted operator consists of $u$ quarks
or using eq.~(\ref{source_d}) for $d$ quarks.

The advantage of this procedure is that by tying together the
two Green's functions appropriately any operator can be inserted
with no additional computational cost.
(There is no restriction on the derivative structure
and Dirac matrix $\gamma$). The disadvantage is that for each
nucleon state (ie $\Gamma$ determining whether the nucleon is
unpolarised or polarised), momentum $\vec{p}$ and
nucleon sink position ($t$) a separate inversion
is required. Thus results for a range of $t$ values are
expensive and practically we must restrict ourselves to one value.

% ----------------------------------------------------------------------

\subsection{The Non-Relativistic Projection}

To improve the overlap with the nucleon we have used
Jacobi smearing, \cite{best97a}, and non-relativistic, NR, 
projection of the nucleon wavefunction. For completeness we now briefly
describe our implementation of this projection. Rather than using
the nucleon operator of eq.~(\ref{baryon_op}) we shall replace it by 
\begin{equation}
   B^{\NR}_\alpha(t; \vec{p})
             =  \sum_{\vec{x}} e^{-i\vec{p} \cdot \vec{x}} \epsilon^{abc}
                             u^a_\alpha(\vec{x},t) 
                      \left[ u^b_\beta(\vec{x},t)
                      \left( C\gamma_5 \half ( 1 + \gamma_4)
                      \right)_{\beta\gamma} d_\gamma^c(\vec{x},t) \right] \,,
\label{baryon_op_NR}
\end{equation}
ie we replace the matrix  $C\gamma_5 \to C\gamma_5 \half (1+\gamma_4)$.
Both operators (eqs.~(\ref{baryon_op_NR}) and (\ref{baryon_op}))
behave the same way under rotations and reflections in the spatial
directions, and both have the same quantum numbers (colour neutral,
spin $\half$, isospin $\half$ and parity $+$), and both will therefore
overlap with the proton.

As we shall see, not only is $B^{\NR}$ computationally
cheaper by a factor two, but it has a better overlap
with the proton. This can be easiest shown if we use
the Dirac basis eq.~(\ref{gamma_def}) first giving
\begin{equation}
   C \gamma_5 = \gamma_4 \gamma_2 \gamma_5 
              = \left( \begin{array}{cccc}
                          0  & 1 & 0 & 0 \\
                          -1 & 0 & 0 & 0 \\
                          0  & 0 & 0 & 1 \\
                          0  & 0 & -1& 0 \\
                       \end{array} \right) \,,  \quad
   C \gamma_5 \half(1 + \gamma_4)
              = \left( \begin{array}{cccc}
                          0  & 1 & 0 & 0 \\
                          -1 & 0 & 0 & 0 \\
                          0  & 0 & 0 & 0 \\
                          0  & 0 & 0 & 0 \\
                       \end{array} \right) \,.
 \end{equation} 
If we now write out eq.~(\ref{baryon_op_NR}) for a spin-up (ie $\alpha=1$)
proton we have 
\begin{equation}
   B^{\NR}_1 = \epsilon^{abc} 
               \left( u^a_1  u^b_1  d^c_2 -  u^a_1  u^b_2  d^c_1 \right)
                                       \,,
\label{baryon_op_NR_nonsym} 
\end{equation}
setting $\vec{p}=\vec{0}$ and suppressing the co-ordinate index
for simplicity. This is not quite the final form. When  we sum over
all fermion line diagrams, only the part of the operator which is 
antisymmetric under the interchange of the two $u$ quarks
makes any contribution to the measured Green's function. Since the
colour wavefunction $\epsilon^{abc}$  is completely antisymmetric,
this means that  the part of eq.~(\ref{baryon_op_NR_nonsym})
which survives is
\begin{equation} 
    B^{\NR}_1 = \epsilon^{abc}
                \left( u^a_1  u^b_1  d^c_2 - \half  u^a_1  u^b_2  d^c_1
                - \half  u^a_2  u^b_1  d^c_1 \right) \,.
\label{baryon_op_NR_explicit}
\end{equation}
Since the mid-sixties it has been known that the lowest-lying octet
and decuplet baryons  are well described as a \underline{\bf 56} of
$SU(6)$. According to $SU(6)$ the flavour/spin wavefunction of the
spin up proton is (eg \cite{close79a})
\begin{equation}
   p_\uparrow =  \sqrt{\sixth} \left(
                 2 u_\uparrow  u_\uparrow d_\downarrow -  
                   u_\uparrow u_\downarrow d_\uparrow  -
                   u_\downarrow  u_\uparrow d_\uparrow \right) \,. 
\label{baryon_op_NR_SU6}
\end{equation} 
Comparing this with eq.~(\ref{baryon_op_NR_explicit}) we see that this
is exactly the wavefunction we have been using. The wavefunction
of eq.~(\ref{baryon_op_NR_SU6}) makes some very successful predictions, 
for example that the ratio of proton to neutron magnetic moments 
should be $-\threehalf$. The experimental value is $-1.460$,
in good agreement with the $SU(6)$ prediction. 
This success suggests that eq.~(\ref{baryon_op_NR_explicit})
is close to the true proton wavefunction. We would expect
this wavefunction to work even better with heavier quarks,
so it is an appropriate wavefunction to use on the lattice.  
If we carry out the same exercise with eq.~(\ref{baryon_op})
we obtain
\begin{eqnarray}
   B_1 &=& \epsilon^{abc}
          \big( u^a_1  u^b_1  d^c_2
          - \half  u^a_1  u^b_2  d^c_1 - \half  u^a_2  u^b_1  d^c_1
                                                       \nonumber \\
       & & \hspace*{0.80in}
          + \half  u^a_1  u^b_3  d^c_4 + \half  u^a_3  u^b_1  d^c_4 
          - \half  u^a_1  u^b_4  d^c_3 - \half  u^a_4  u^b_1  d^c_3 \Big)  
                                                            \,,
\end{eqnarray}
as the explicit component expression.
This has many terms involving the 3rd and 4th Dirac components with
amplitudes just as large as the terms with upper components only. 
In any sort of constituent quark model we would expect these terms
to be small in the ground state. Adding them into the wavefunction
not only increases the cost of the computation, it also degrades 
the signal by adding terms which are likely to have more 
overlap with excited baryon states.

\begin{figure}[t]
   \hspace*{0.75in}
   \epsfxsize=11.00cm \epsfbox{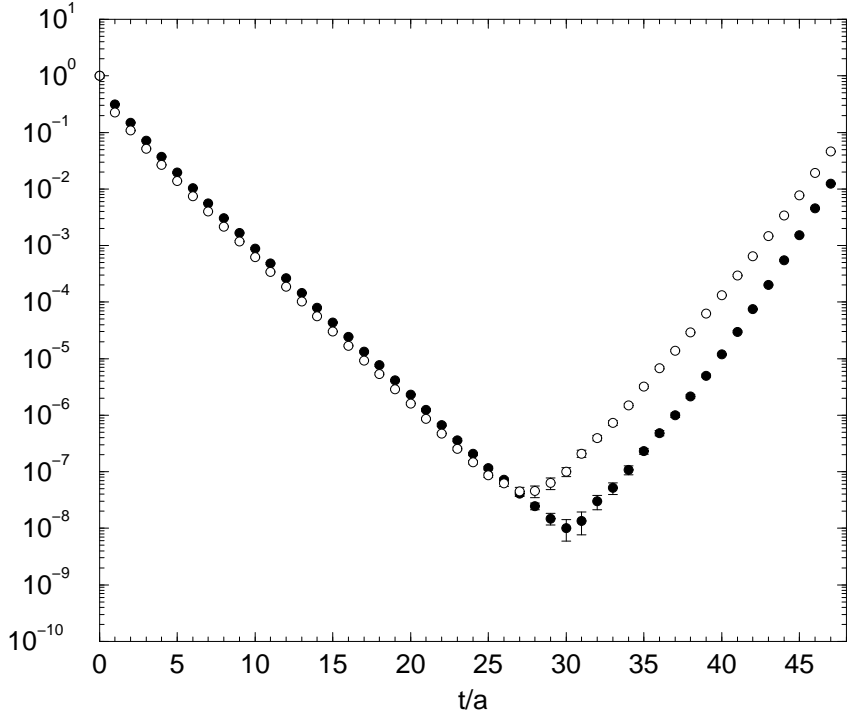}
   \caption{Relativistic (open circles) and non-relativistic
           (filled circles) two-point operators,
           eq.~(\ref{2pt_correlation}) with $\vec{p}=\vec{0}$,
           for $\beta = 6.2$ at $\kappa = 0.1344$.}
\label{fig_nrel_v_rel}
\end{figure}

The correlation functions are constructed also using
\begin{equation}
   \overline{B}^{\NR}_\alpha(t; \vec{p})
        =  \sum_{\vec{x}} e^{i\vec{p}.\vec{x}} \epsilon^{abc}
             \overline{u}^a_\alpha(\vec{x},t) 
                \left[ \overline{d}^b(\vec{x},t) \half (1+\gamma_4)
                             \gamma_5 C\overline{u}^c(\vec{x},t)^{T_D}
                \right] \,,
\end{equation}
ie replacing $\gamma_5 C \to \half (1+\gamma_4) \gamma_5 C$.
Thus the NR projection can be obtained by projecting out the positive
eigenvalues of $\gamma_4$, ie by replacing each quark field, $q$, by
\begin{equation}
   q \to \half (1 + \gamma_4 ) q \,, \qquad
   \overline{q} \to \overline{q} \half (1 + \gamma_4 ) \,,
\end{equation}
everywhere and considering polarisation matrices $\Gamma$ which satisfy
\begin{equation}
   \half ( 1 + \gamma_4) \Gamma = \Gamma = \Gamma \half ( 1 + \gamma_4) \,.
\label{GammaNR}
\end{equation}
This gives the NR nucleon two-point function, eq.~(\ref{2pt_correlation}).
(So for the Dirac basis, eq.~(\ref{gamma_def}), only the components
$\alpha = 1$, $2$ are non-zero, as discussed above.) 
In Fig.~\ref{fig_nrel_v_rel}
we show a comparison for the nucleon two-point correlation function
using the relativistic and non-relativistic operators
when applied to eq.~(\ref{2pt_correlation}).
The gradient of the left branch measures the nucleon mass.
It can be seen that this branch has been extended
by about $5$ units of $t/a$ when using the NR operator as opposed
to the relativistic operator.

For the three-point functions the new tilde,
replacing eq.~(\ref{tilde_def}), is given by
\begin{equation}
   \tilde{X} = (C\gamma_5 \half(1+\gamma_4) X \half(1+\gamma_4)
                                \gamma_5 C)^{T_D} \,,
\end{equation}
and obeys
\begin{equation}
   \half (1+\gamma_4) \tilde{X} = \tilde{X} 
                               = \tilde{X} \half (1+\gamma_4) \,.
\label{XtildeNR}
\end{equation}
Considering eqs.~(\ref{source_u}), (\ref{source_d})
then eqs.~(\ref{GammaNR}), (\ref{XtildeNR}) imply that 
\begin{equation}
   \half ( 1 + \gamma_4) S^{(q)} = S^{(q)} = S^{(q)} \half ( 1 + \gamma_4)\,.
\end{equation}
This identity reduces the number of independent components from $4$
to $2$ for the source for the stage II inversion, eq.~(\ref{sig_invert}).
Again, for the Dirac representation eq.~(\ref{gamma_def})
only the first two components are needed.

% ----------------------------------------------------------------------

\section{Tables}
\label{tables}

The action used here (in the quenched limit) is
\begin{equation}
   S = \third \beta \sum_{\plaq} \mbox{Tr}_C \mbox{Re} 
                 \left[ 1 - U^{\plaq}_{\mu\nu} \right] + 
       a^4 \sum_{xy;q=u,d} \overline{q}(x) M^{(q)}(x;y) q(y) \,,
\end{equation}
where $U^{\plaq}_{\mu\nu}$ is the product of links around an
elementary plaquette in the $\mu - \nu$ plane
and the Wilson (clover) fermion matrix is given by
\begin{eqnarray} 
    \lefteqn{\sum_{xy} \overline{q}(x) M^{(q)}(x;y) q(y) = } 
       & &                        
                                                  \nonumber \\
       & & \sum_x  \Big\{ 
            {1\over a}\overline{q}(x) q(x) 
             - {\kappa \over a} \sum_\mu \overline{q}(x) \,
             U^\dagger_\mu(x-a\hat{\mu})
                   \left[1 + \gamma_\mu\right] q(x-a\hat{\mu})
                                                  \nonumber \\
       & &  \phantom{ \sum_x  \Big\{ 
              {1\over a}\overline{q}(x) q(x) }
            - {\kappa \over a} \sum_\mu \overline{q}(x) \,
            U_\mu(x) \left[1 - \gamma_\mu\right] q(x+a\hat{\mu})
                                                  \nonumber \\
      & &   \hspace*{0.40in}
            - 2 \kappa a \,c_{sw} \,g_0\,  \sum_{\mu\nu}{1 \over 4}
            \overline{q}(x)\,\sigma_{\mu\nu} F_{\mu\nu}^{\rm clover}(x)
                      q(x)
                    \Big\} \,,
\label{action}
\end{eqnarray}  
where the hopping parameter, $\kappa$, is related to the (bare)
quark mass via eq.~(\ref{bare_qm_def}), and we are taking mass
degenerate $u$ and $d$ quarks. In eq.~(\ref{action}) the quark
fields are normalised according to the lattice conventions
ie they correspond to the continuum fields by rescaling
$q \to 1/\sqrt{2\kappa} q$. (This introduces a further
factor $1/(2\kappa)$ on the RHS of eq.~(\ref{R_def_practical})
when using the raw output for the two- and three- point functions.)
The last term in eq.~(\ref{action}), sufficient for on-shell $O(a)$
improvement (with a to be determined function $c_{sw}(g_0)$)
has a `clover' field strength tensor given by
\begin{equation}
   F^{\rm clover}_{\mu\nu}(x) = {1 \over 8ig_0 a^2} \sum_{\pm\mu,\pm\nu}
       \left[ U^{\plaq}_{\mu\nu}(x) - U^{\plaq}_{\mu\nu}(x)^\dagger \right]
   \,,
\end{equation}
where we have extended the definition of the plaquette, so that the
$\mu$, $\nu$ directions can be negative.

Thus we can re-write eq.~(\ref{action}) as
\begin{eqnarray} 
    \lefteqn{\sum_{xy} \overline{q}(x) M^{(q)}(x;y) q(y) \to } 
       & &                        
                                                  \nonumber \\
       & & \sum_x  \Big\{ 
            \overline{q}(x) ( \stackrel{\rightarrow}{\slashed{D}} + m_q )q(x)
             + m_{0c} \overline{q}(x) q(x)
             - \half a \sum_\mu \overline{q}(x) 
               \stackrel{\rightarrow}{\Delta_\mu^-}
               \stackrel{\rightarrow}{\Delta_\mu^+} q(x)
                                                  \nonumber \\
      & &   \hspace*{0.40in}
            - a \,c_{sw} \,g_0\,  \sum_{\mu\nu}{1 \over 4}
            \overline{q}(x)\,\sigma_{\mu\nu} F_{\mu\nu}^{\rm clover}(x)
                      q(x)
                    \Big\} \,,
\label{action_a}
\end{eqnarray}  
where
\begin{eqnarray}
   \stackrel{\rightarrow}{\Delta^+_\mu} q (x)
             &=& {1\over a} \left[ U_\mu(x)q(x+a\hat{\mu}) - q(x) \right] \,,
                                                  \nonumber \\
   \stackrel{\rightarrow}{\Delta^-_\mu} q (x)
             &=& {1\over a} \left[ q(x) -  U^\dagger_\mu(x-a\hat{\mu})
                                                q(x-a\hat{\mu}) \right] \,,
\end{eqnarray}
so that (see eq.~(\ref{Ddefs}))
\begin{equation}
   \stackrel{\rightarrow}{D_\mu} = 
       \half \left( \stackrel{\rightarrow}{\Delta^+_\mu} + 
                    \stackrel{\rightarrow}{\Delta^-_\mu}  \right) \,.
\end{equation}
$am_q$ is defined in eq.~(\ref{bare_qm_def}) and
\begin{equation}
   am_{0c} = {1\over 2} \left( {1\over \kappa_c} - 8 \right) \,.
\end{equation}
In this latter form eq.~(\ref{action_a}) shows the additional
$O(a)$ operators most clearly.

In Table~\ref{table_run_params} we give our parameter values
\begin{table}[t]
\begin{small}
   \begin{center}
      \begin{tabular}{||l|l|l|l|c||l|l||}
         \hline
         \hline
\multicolumn{1}{||c}{$\beta$}  &
\multicolumn{1}{|c}{$c_{sw}$}  &
\multicolumn{1}{|c}{$\kappa$}  & 
\multicolumn{1}{|c}{Volume}    &
\multicolumn{1}{|c||}{$\#$ configs.} &
\multicolumn{1}{c|}{$am_{ps}$} &
\multicolumn{1}{c||}{$am_N$}  \\
         \hline
6.0 & 1.769 & 0.1320 & $16^3\times 32$ & $O(445)$  & 0.5412(9) & 0.9735(40) \\
6.0 & 1.769 & 0.1324 & $16^3\times 32$ & $O(560)$  & 0.5042(7) & 0.9353(25) \\
6.0 & 1.769 & 0.1333 & $16^3\times 32$ & $O(560)$  & 0.4122(9) & 0.8241(34) \\
6.0 & 1.769 & 0.1338 & $16^3\times 32$ & $O(520)$  & 0.3549(12)& 0.7400(85) \\
6.0 & 1.769 & 0.1342 & $16^3\times 32$ & $O(735)$  & 0.3012(10)& 0.7096(48) \\
         \hline
         \hline
6.2 & 1.614 & 0.1333 & $24^3\times 48$ & $O(300)$  & 0.4136(6) & 0.7374(21) \\
6.2 & 1.614 & 0.1339 & $24^3\times 48$ & $O(300)$  & 0.3565(7) & 0.6655(28) \\
6.2 & 1.614 & 0.1344 & $24^3\times 48$ & $O(300)$  & 0.3034(6) & 0.5963(29) \\
6.2 & 1.614 & 0.1349 & $24^3\times 48$ & $O(470)$  & 0.2431(6) & 0.5241(39) \\
         \hline
         \hline
6.4 & 1.526 & 0.1338 & $32^3\times 48$ & $O(220)$  & 0.3213(8) & 0.5718(28) \\
6.4 & 1.526 & 0.1342 & $32^3\times 48$ & $O(120)$  & 0.2836(9) & 0.5266(31) \\
6.4 & 1.526 & 0.1346 & $32^3\times 48$ & $O(220)$  & 0.2402(8) & 0.4680(37) \\
6.4 & 1.526 & 0.1350 & $32^3\times 48$ & $O(320)$  & 0.1923(9) & 0.4156(34) \\
6.4 & 1.526 & 0.1353 & $32^3\times 64$ & $O(260)$  & 0.1507(8) & 0.3580(47) \\ 
         \hline
         \hline
      \end{tabular}
   \end{center}
\caption{Parameter values used in the simulations, together with
         the measured pseudoscalar and nucleon masses. Note that
         the statistics refers to the number of (independent)
         configurations used for the $3$-point functions;
         the masses have sometimes been determined with
         a larger statistic.}
\label{table_run_params}
\end{small}
\end{table}
used in the (quenched) fermion simulations together with the 
pseudoscalar, $am_{ps}$, and nucleon mass, $am_N$.

We now give a series of tables tabulating the bare
matrix elements $v_n^{(q)}$ for $q = u$, $d$, the improvement
operator matrix elements $av_2^{(q;i)}$ for $i = 1$, $2$
and the mixing operators $v_3^{(q;m_i)}$ for $i = 1$, $2$ and
$v_4^{(q;m_i)}$ for $i = 1$, $2$ and $v_4^{(q;m_3)}/a$.

% ----------------------------------------------------------------------

\begin{table}[t]
\begin{small}
   \begin{center}
      \begin{tabular}{||l||r|r|r|r|r||}
         \hline
         \hline
\multicolumn{1}{||c}{$\kappa$}                                & 
\multicolumn{1}{|c}{0.1320}                                   &
\multicolumn{1}{|c}{0.1324}                                   &
\multicolumn{1}{|c}{0.1333}                                   &
\multicolumn{1}{|c}{0.1338}                                   &
\multicolumn{1}{|c||}{0.1342}                                   \\
         \hline
         \hline
\multicolumn{6}{||c||}{$\vec{p}=\vec{p}_1$}                       \\
         \hline
$v_{2a}^{(u)}$    & 0.359(19)   & 0.348(16)   & 0.335(21)   & 0.382(45)
                       & 0.312(35)   \\
$v_{2a}^{(d)}$    & 0.1696(89)  & 0.1557(82)  & 0.144(11)   & 0.163(19)
                       & 0.123(17)   \\
$v_{2a}$          & 0.189(11)   & 0.1925(95)  & 0.192(13)   & 0.219(31)    
                       & 0.188(25)   \\
         \hline
         \hline
$av_{2a}^{(u;1)}$ &-0.0254(17)  &-0.0261(15)  &-0.0288(40)  &-0.0288(40) 
                       &-0.0226(36)   \\
$av_{2a}^{(d;1)}$ &-0.01321(93) &-0.01372(85) &-0.0136(12)  &-0.0115(21)
                       &-0.0130(25)   \\
$av_{2a}^{(1)}$   &-0.01211(98) &-0.01238(91) &-0.0117(14)  &-0.0170(29)
                       &-0.00974(292) \\
         \hline
         \hline
$av_{2a}^{(u;2)}$ & 0.0190(14)  & 0.0162(13)  & 0.00676(165)& 0.00188(206)
                       &-0.00361(302) \\
$av_{2a}^{(d;2)}$ & 0.01025(82) & 0.00933(75) & 0.00592(104)&-0.00017(142)
                       & 0.00136(227) \\
$av_{2a}^{(2)}  $ & 0.00872(79) & 0.00693(82) & 0.00111(117) & 0.00204(200)
                       &-0.00448(260) \\
         \hline
         \hline
\multicolumn{6}{||c||}{$\vec{p}=\vec{0}$}                         \\
         \hline
$v_{2b}^{(u)}$   & 0.4066(32)  & 0.4108(27)  & 0.4130(48)  & 0.4196(78)  
                       & 0.414(11)   \\
$v_{2b}^{(d)}$   & 0.1894(17)  & 0.1886(16)  & 0.1851(26)  & 0.1845(41)
                       & 0.1798(55)  \\
$v_{2b}$         & 0.2171(18)  & 0.2221(16)  & 0.2278(29)  & 0.2350(52)       
                       & 0.2336(69)  \\
         \hline
         \hline
$av_{2b}^{(u;1)}$ &-0.02844(39) &-0.02895(40) &-0.03002(81) & -0.0328(13)
                       &-0.0311(25)  \\
$av_{2b}^{(d;1)}$ &-0.01490(26) &-0.01503(27) &-0.01585(51) & -0.0185(11)
                       &-0.0189(16)  \\
$av_{2b}^{(1)}$   &-0.01354(21) &-0.01390(25) &-0.01406(54) & -0.0141(11)
                       &-0.0117(16)  \\
         \hline
         \hline
$av_{2b}^{(u;2)}$ & 0.03954(52) & 0.03552(52) &0.0266(11)   & 0.0235(19)
                       & 0.0173(43)  \\
$av_{2b}^{(d;2)}$ & 0.02015(31) & 0.01822(35) & 0.01494(72) & 0.0173(43)
                       & 0.0152(30)  \\
$av_{2b}^{(2)}$   & 0.01936(29) & 0.01722(32) & 0.01135(69) & 0.00741(162)
                  & 0.00108(219)     \\
         \hline
         \hline
\multicolumn{6}{||c||}{$\vec{p}=\vec{p}_1$}                     \\
         \hline
$v_{2b}^{(u)}$    & 0.3918(54)  & 0.3959(49)  & 0.3908(88)  & 0.417(24)
                       & 0.370(19)   \\
$v_{2b}^{(d)}$    & 0.1842(30)  & 0.1807(28)  & 0.1727(49)  & 0.186(13)
                       & 0.155(11)   \\
$v_{2b}$          & 0.2076(31)  & 0.2152(32)  & 0.2180(59)  & 0.231(13)
                       & 0.215(14)   \\
         \hline
         \hline
$av_{2b}^{(u;1)}$ &-0.02642(60) &-0.01472(41) &-0.0288(12)  &-0.0266(32)
                       &-0.0268(41)  \\
$av_{2b}^{(d;1)}$ &-0.01420(37) &-0.01488(42) &-0.01563(85) &-0.0171(20)
                       &-0.0185(47)  \\
$av_{2b}^{(1)}$   &-0.01222(42) &-0.01320(39) &-0.01297(90) &-0.0102(24)
                       &-0.00898(402)\\
         \hline
         \hline
$av_{2b}^{(u;2)}$ & 0.03714(80) & 0.03357(78) & 0.0244(17)  & 0.0168(45)
                       & 0.0114(63) \\
$av_{2b}^{(d;2)}$ & 0.01944(48) & 0.01944(48) & 0.01433(11) & 0.0149(26)
                       & 0.0168(58) \\
$av_{2b}^{(2)}$   & 0.01777(52) & 0.01604(50) & 0.00988(109)& 0.00353(335)
                       &-0.00315(519)\\
         \hline
         \hline
      \end{tabular}
   \end{center}
\caption{The bare results for $v_2$
         from eq.~(\ref{R_def_practical}) for $\beta = 6.0$,
         $c_{sw}=1.769$.}
\label{table_run_v2_b6p00c1p769}
\end{small}
\end{table}

% ----------------------------------------------------------------------

\begin{table}[t]
\begin{small}
   \begin{center}
      \begin{tabular}{||l||r|r|r|r||}
         \hline
         \hline
\multicolumn{1}{||c}{$\kappa$}                                & 
\multicolumn{1}{|c}{0.1333}                                   &
\multicolumn{1}{|c}{0.1339}                                   &
\multicolumn{1}{|c}{0.1344}                                   &
\multicolumn{1}{|c||}{0.1349}                                   \\
         \hline
         \hline
\multicolumn{5}{||c||}{$\vec{p}=\vec{p}_1$}                       \\
         \hline
$v_{2a}^{(u)}$    & 0.403(24)   & 0.393(23)   & 0.413(38)   & 0.422(42)     \\
$v_{2a}^{(d)}$    & 0.187(11)   & 0.179(11)   & 0.182(17)   & 0.175(18)     \\
$v_{2a}$          & 0.217(14)   & 0.214(13)   & 0.232(23)   & 0.246(28)     \\
         \hline
         \hline
$av_{2a}^{(u;1)}$  &-0.0252(18)  &-0.0247(18)  &-0.0251(28)  &-0.0269(32)   \\
$av_{2a}^{(d;1)}$  &-0.01294(96) &-0.01224(97) &-0.0130(16)  &-0.0123(18)   \\
$av_{2a}^{(1)}$    &-0.0122(10)  &-0.0125(11)  &-0.0122(19)  &-0.0144(22)   \\
         \hline
         \hline
$av_{2a}^{(u;2)}$  & 0.0117(12)  & 0.00453(116)&-0.00110(179)&-0.00493(254) \\
$av_{2a}^{(d;2)}$  &0.00622(71)  & 0.00308(68) &-0.0067(110) &-0.00224(144) \\
$av_{2a}^{(2)}$    &0.00550(71)  & 0.00143(81) &-0.00140(146)&-0.00273(209) \\
         \hline
         \hline
\multicolumn{5}{||c||}{$\vec{p}=\vec{0}$}                         \\
         \hline
$v_{2b}^{(u)}$    & 0.4100(28)  & 0.4020(39)  & 0.4071(52)  & 0.4052(58)    \\
$v_{2b}^{(d)}$    & 0.1920(14)  & 0.1839(21)  & 0.1834(28)  & 0.1775(35)    \\
$v_{2b}$          & 0.2179(17)  & 0.2181(22)  & 0.2237(34)  & 0.2278(39)    \\
         \hline
         \hline
$av_{2b}^{(u;1)}$  &-0.02554(35) &-0.02494(48) &-0.02775(65) &-0.02883(96)  \\
$av_{2b}^{(d;1)}$  &-0.01316(21) &-0.01302(31) &-0.01444(42) &-0.01523(68)  \\
$av_{2b}^{(1)}$    &-0.01239(20) &-0.01191(28) &-0.01323(42) &-0.01357(68)  \\
         \hline
         \hline
$av_{2b}^{(u;2)}$  & 0.02801(41) & 0.02091(58) & 0.01874(79) & 0.0152(12)   \\
$av_{2b}^{(d;2)}$  & 0.01443(26) & 0.01141(38) & 0.01071(56) & 0.00959(87)  \\
$av_{2b}^{(2)}$    & 0.01360(24) &0.00949(35)  & 0.00777(58) & 0.00548(97)  \\
         \hline
         \hline
\multicolumn{5}{||c||}{$\vec{p}=\vec{p}_1$}                     \\
         \hline
$v_{2b}^{(u)}$    & 0.4153(56)  & 0.3987(58)  & 0.419(14)   & 0.406(19)    \\
$v_{2b}^{(d)}$    & 0.1966(29)  & 0.1840(32)  & 0.1938(77)  & 0.186(11)    \\
$v_{2b}$          & 0.2186(33)  & 0.2146(40)  & 0.2255(81)  & 0.220(10)    \\
         \hline
         \hline
$av_{2b}^{(u;1)}$  &-0.02536(49) &-0.02377(76) &-0.0269(11)  &-0.0288(19)  \\
$av_{2b}^{(d;1)}$  &-0.01316(30) &-0.01251(59) &-0.01438(63) &-0.0148(14)  \\
$av_{2b}^{(1)}$    &-0.01219(30) &-0.01144(42) &-0.01233(82) &-0.0136(15)  \\
         \hline
         \hline
$av_{2b}^{(u;2)}$  & 0.02761(67) & 0.01973(92) & 0.0174(15)  & 0.0151(25)  \\
$av_{2b}^{(d;2)}$  & 0.01437(42) & 0.01088(74) & 0.0105(10)  & 0.00970(179)\\
$av_{2b}^{(2)}$    & 0.01326(44) &0.00901(56)  & 0.00664(125)& 0.00454(215)\\
         \hline
         \hline
      \end{tabular}
   \end{center}
\caption{The bare results for $v_2$
         from eq.~(\ref{R_def_practical}) for $\beta = 6.2$,
         $c_{sw}=1.614$.}
\label{table_run_v2_b6p20c1p614}
\end{small}
\end{table}

% ----------------------------------------------------------------------

\begin{table}[t]
\begin{small}
   \begin{center}
      \begin{tabular}{||l||r|r|r|r|r||}
         \hline
         \hline
\multicolumn{1}{||c}{$\kappa$}                                & 
\multicolumn{1}{|c}{0.1338}                                   &
\multicolumn{1}{|c}{0.1342}                                   &
\multicolumn{1}{|c}{0.1346}                                   &
\multicolumn{1}{|c}{0.1350}                                   &
\multicolumn{1}{|c||}{0.1353}                                     \\
         \hline
         \hline
\multicolumn{6}{||c||}{$\vec{p}=\vec{p}_1$}                       \\
         \hline
$v_{2a}^{(u)}$    & 0.368(21)   & 0.348(30)   & 0.340(28)   & 0.339(34)
                       & 0.329(63)  \\
$v_{2a}^{(d)}$    & 0.173(11)   & 0.161(15)   & 0.150(12)   & 0.140(16)
                       & 0.122(24)  \\
$v_{2a}$          & 0.196(12)   & 0.185(17)   & 0.191(18)   & 0.199(23)    
                       & 0.205(44)  \\
         \hline
         \hline
$av_{2a}^{(u;1)}$  &-0.0210(13)  &-0.0181(17)  &-0.0208(18)  &-0.0214(24)
                       &-0.0242(55) \\
$av_{2a}^{(d;1)}$  &-0.01092(75) &-0.0092(10)  &-0.0104(11)  &-0.0103(14)
                       &-0.00981(291) \\
$av_{2a}^{(1)}$    &-0.01010(73) &-0.00853(99) &-0.0104(11)  &-0.0107(16)  
                       &-0.0141(39) \\
         \hline
         \hline
$av_{2a}^{(u;2)}$  & 0.00470(68) &-0.00203(101)&-0.00181(99) &-0.00525(152)
                       &-0.00456(307) \\
$av_{2a}^{(d;2)}$  & 0.00323(45) & 0.000043(690)&0.000384(646)&-0.000831(942)
                       &-0.00049(205) \\
$av_{2a}^{(2)}$    &0.00147(44)  &-0.00206(64) &-0.00229(75) &-0.00472(121)
                       &-0.00352(269) \\
         \hline
         \hline
\multicolumn{6}{||c||}{$\vec{p}=\vec{0}$}                         \\
         \hline
$v_{2b}^{(u)}$    & 0.4088(27)  & 0.4140(47)  & 0.4017(47)  & 0.3951(70)
                       & 0.397(13)    \\
$v_{2b}^{(d)}$    & 0.1914(14)  & 0.1881(23)  & 0.1842(24)  & 0.1781(39)
                       & 0.1674(71)   \\
$v_{2b}$          & 0.2174(17)  & 0.2259(32)  & 0.2174(31)  & 0.2169(47)       
                       & 0.2294(90)   \\
         \hline
         \hline
$av_{2b}^{(u;1)}$  &-0.02267(23) &-0.02191(58) &-0.02290(38) &-0.02295(75)
                       &-0.0213(22)  \\
$av_{2b}^{(d;1)}$  &-0.01161(15) &-0.01118(31) &-0.01190(28) &-0.01265(50)
                       &-0.0120(16)  \\
$av_{2b}^{(1)}$    &-0.01105(13) &-0.01076(34) &-0.01093(26) &-0.01028(50)
                       &-0.00988(133) \\
         \hline
         \hline
$av_{2b}^{(u;2)}$  & 0.01943(28) & 0.01406(68) & 0.01122(48) & 0.00728(90)
                       & 0.00206(265) \\
$av_{2b}^{(d;2)}$  & 0.01015(19) & 0.00797(33) & 0.00672(37) & 0.00611(66)
                       & 0.00485(201) \\
$av_{2b}^{(2)}$    & 0.00926(15) & 0.00624(44) & 0.00445(32) & 0.00115(64)
                       &-0.00261(160) \\
         \hline
         \hline
\multicolumn{6}{||c||}{$\vec{p}=\vec{p}_1$}                     \\
         \hline
$v_{2b}^{(u)}$    & 0.4097(42)  & 0.4059(97)  & 0.4012(91)  & 0.385(16)
                       & 0.390(34)  \\
$v_{2b}^{(d)}$    & 0.1906(23)  & 0.1828(46)  & 0.1813(46)  & 0.1697(72)
                       & 0.162(15)  \\
$v_{2b}$          & 0.2191(27)  & 0.2231(61)  & 0.2200(59)  & 0.215(12)
                       & 0.228(25)  \\
         \hline
         \hline
$av_{2b}^{(u;1)}$  &-0.02238(38) &-0.0212(11)  &-0.02219(70) &-0.0209(15)
                       &-0.0252(40) \\
$av_{2b}^{(d;1)}$  &-0.01156(24) &-0.01044(57) &-0.01189(48) &-0.01251(91)
                       &-0.0113(24) \\
$av_{2b}^{(1)}$    &-0.01079(21) &-0.01088(67) &-0.01022(46) &-0.00844(101)
                       &-0.0142(29) \\
         \hline
         \hline
$av_{2b}^{(u;2)}$  & 0.01898(44) & 0.0142(13) & 0.01042(78) & 0.00566(163)
                       & 0.00748(445) \\
$av_{2b}^{(d;2)}$  & 0.01003(28) & 0.00731(63)& 0.00675(56) & 0.00639(111)
                       & 0.00358(326) \\
$av_{2b}^{(2)}$    & 0.00895(23) & 0.00706(79)& 0.00366(55) &-0.00046(124)
                       & 0.00485(456) \\
         \hline
         \hline
      \end{tabular}
   \end{center}
\caption{The bare results for $v_2$
         from eq.~(\ref{R_def_practical}) for $\beta = 6.4$,
         $c_{sw}=1.526$.}
\label{table_run_v2_b6p40c1p526}
\end{small}
\end{table}

% ----------------------------------------------------------------------

\begin{table}[t]
\begin{small}
   \begin{center}
      \begin{tabular}{||l||r|r|r|r|r||}
         \hline
         \hline
\multicolumn{1}{||c}{$\kappa$}                                & 
\multicolumn{1}{|c}{0.1320}                                   &
\multicolumn{1}{|c}{0.1324}                                   &
\multicolumn{1}{|c}{0.1333}                                   &
\multicolumn{1}{|c}{0.1338}                                   &
\multicolumn{1}{|c||}{0.1342}                                     \\
         \hline
         \hline
\multicolumn{6}{||c||}{$\vec{p}=\vec{p}_1$}                       \\
         \hline
$v_{3}^{(u)}$     & 0.0981(55)  & 0.0973(52)  & 0.0961(72)  & 0.114(15)
                  & 0.0955(125)       \\
$v_{3}^{(d)}$     & 0.0435(26)  & 0.0408(25)  & 0.0378(36)  & 0.0407(61)
                  & 0.0342(62)        \\
$v_{3}$           & 0.0546(34)  & 0.0566(31)  & 0.0582(46)  & 0.0710(98)
                  & 0.0606(85)        \\
         \hline
         \hline
$v_{3}^{(u;m_1)}$ & 0.00051(215)& 0.00165(221)& 0.00193(373)
                  & 0.00694(638)&-0.00015(818)      \\
$v_{3}^{(d;m_1)}$ & 0.00007(129)&-0.00140(145)&-0.00208(232)
                  &-0.00158(379)&-0.00160(557)      \\
$v_{3}^{(m_1)}$   & 0.00045(150)& 0.00056(155)&-0.00009(298)
                  & 0.0103(57)  & 0.00113(687)      \\
         \hline
         \hline
$v_{3}^{(u;m_2)}$ &-0.00490(717)&-0.0102(73)  &-0.0070(132)
                  & 0.0350(269) &-0.0205(311)       \\
$v_{3}^{(d;m_2)}$ &-0.00342(459)&-0.0108(54)  &-0.0118(100)
                  & 0.0053(167)   & 0.0146(227)     \\
$v_{3}^{m_2}$     &-0.00134(487)  & 0.00002(522)& 0.0022(112)
                  & 0.0308(212)   & 0.0073(273)     \\
         \hline
         \hline
\multicolumn{6}{||c||}{$\vec{p}=\vec{p}_1$}                       \\
         \hline
$v_{4}^{(u)}$     & 0.0272(28)  & 0.0330(27)  & 0.0342(39)
                  & 0.0339(74)  & 0.0331(64)        \\
$v_{4}^{(d)}$     & 0.0128(15)  & 0.0128(16)  & 0.0112(24)
                  & 0.0141(42)  & 0.00780(485)      \\
$v_{4}$           & 0.0143(19)  & 0.0201(18)  & 0.0232(27)
                  & 0.0206(55)  & 0.0253(47)        \\
         \hline
         \hline
$v_{4}^{(u;m_1)}$ &-0.00588(626)&-0.00625(648)&-0.0048(104)
                  &-0.0109(217) & 0.0196(243)       \\
$v_{4}^{(d;m_1)}$ & 0.00355(394)&-0.00113(450)& 0.00144(708)
                  & 0.0086(129) & 0.0152(153)       \\
$v_{4}^{(m_1)}$   &-0.00905(399)&-0.00512(392)&-0.00612(799)
                  &-0.0202(144) & 0.0047(214)       \\
         \hline
         \hline
$v_{4}^{(u;m_2)}$ &-0.00680(598)&-0.00273(503)&-0.00053(841)
                  &-0.0052(198) & 0.0163(201)       \\

$v_{4}^{(d;m_2)}$ & 0.00180(298)&-0.00188(359)&-0.00142(591)
                  & 0.0066(100) &-0.0068(150)       \\
$v_{4}^{(m_2)}$   &-0.00803(382)&-0.00169(327)& 0.00038(692)
                  &-0.0085(150) & 0.0243(180)       \\
         \hline
         \hline
$v_{4}^{(u;m_3)}/a$ & 0.0178(75)& 0.00521(918)& 0.0082(157)
                  & 0.0028(330) & 0.0378(447)       \\
$v_{4}^{(d;m_3)}/a$ & 0.00244(508)&0.00523(630)& 0.0093(108)
                  &-0.0111(188) & 0.0434(496)       \\
$v_{4}^{(m_3)}/a$ &0.0156(48)   &-0.00127(587)&-0.0041(107)
                  & 0.0150(255) &-0.0062(351)       \\
         \hline
         \hline
      \end{tabular}
   \end{center}
\caption{The bare results for $v_3$ and $v_4$
         from eq.~(\ref{R_def_practical}) for $\beta = 6.0$,
         $c_{sw}=1.769$.}
\label{table_run_v3+v4_b6p00c1p769}
\end{small}
\end{table}

% ----------------------------------------------------------------------

\begin{table}[t]
\begin{small}
   \begin{center}
      \begin{tabular}{||l||r|r|r|r||}
         \hline
         \hline
\multicolumn{1}{||c}{$\kappa$}                                & 
\multicolumn{1}{|c}{0.1333}                                   &
\multicolumn{1}{|c}{0.1339}                                   &
\multicolumn{1}{|c}{0.1344}                                   &
\multicolumn{1}{|c||}{0.1349}                                     \\
         \hline
         \hline
\multicolumn{5}{||c||}{$\vec{p}=\vec{p}_1$}                       \\
         \hline
$v_{3}^{(u)}$     & 0.1087(72)  & 0.1075(64)  & 0.110(11)   & 0.115(12)    \\
$v_{3}^{(d)}$     & 0.0493(36)  & 0.0467(30)  & 0.0478(53)  & 0.0455(54)   \\
$v_{3}$           & 0.0594(41)  & 0.0608(40)  & 0.0628(71)  & 0.0697(88)   \\
         \hline
         \hline
$v_{3}^{(u;m_1)}$ &-0.00141(292)& 0.00278(316)& 0.00002(551)&-0.00474(849)\\
$v_{3}^{(d;m_1)}$ &-0.00233(195)&-0.00023(187)&-0.00364(328)&-0.00716(453)\\
$v_{3}^{(m_1)}$   & 0.00062(175)& 0.00294(225)& 0.00364(415)& 0.00319(699)\\
         \hline
         \hline
$v_{3}^{(u;m_2)}$ &-0.0011(101) &-0.0038(120) & 0.0201(228) & 0.0449(378)  \\
$v_{3}^{(d;m_2)}$ &-0.00088(704)&-0.00597(810)& 0.0136(159) & 0.0262(224)  \\
$v_{3}^{(m_2)}$   & 0.00191(505)& 0.00098(765)& 0.0135(143) & 0.0223(288)  \\
         \hline
         \hline
\multicolumn{5}{||c||}{$\vec{p}=\vec{p}_1$}                       \\
         \hline
$v_{4}^{(u)}$     & 0.0333(41)   & 0.0336(55) & 0.0331(69)  & 0.0442(85)   \\
$v_{4}^{(d)}$     & 0.0127(21)   & 0.0129(30) & 0.00757(388)& 0.00803(561) \\
$v_{4}$           & 0.0205(31)   & 0.0206(37) & 0.0238(62)  & 0.0344(89)   \\
         \hline
         \hline
$v_{4}^{(u;m_1)}$ & 0.0212(116)  &-0.0113(147)& 0.0339(256) & 0.0375(332)  \\
$v_{4}^{(d;m_1)}$ & 0.0102(69)   &-0.0117(90) & 0.0176(136) & 0.0088(189)  \\
$v_{4}^{(m_1)}$   & 0.0118(73)   & 0.0083(906)& 0.0172(180) & 0.0270(267)  \\
         \hline
         \hline
$v_{4}^{(u;m_2)}$ & 0.0074(101)  &-0.0007(130)  & 0.0157(213)& 0.0357(310) \\
$v_{4}^{(d;m_2)}$ & 0.00098(587) &-0.00384(687) & 0.0082(121)& 0.0092(168) \\
$v_{4}^{(m_2)}$   & 0.00756(606) & 0.00218(904) & 0.0100(150)& 0.0257(249) \\
         \hline
         \hline
$v_{4}^{(u;m_3)}/a$ & 0.0225(166)&-0.0201(192)  & 0.0324(364)&-0.0271(578) \\

$v_{4}^{(d;m_3)}/a$ & 0.0090(110)&-0.00002(1285)& 0.0198(235)& 0.0291(390) \\

$v_{4}^{(m_3)}/a$   & 0.0136(104)&-0.0134(129)  & 0.0097(275)&-0.0520(497) \\
         \hline
         \hline
      \end{tabular}
   \end{center}
\caption{The bare results for $v_3$ and $v_4$
         from eq.~(\ref{R_def_practical}) for $\beta = 6.2$,
         $c_{sw}=1.614$.}
\label{table_run_v3+v4_b6p20c1p614}
\end{small}
\end{table}

% ----------------------------------------------------------------------

\begin{table}[t]
\begin{small}
   \begin{center}
      \begin{tabular}{||l||r|r|r|r|r||}
         \hline
         \hline
\multicolumn{1}{||c}{$\kappa$}                                & 
\multicolumn{1}{|c}{0.1338}                                   &
\multicolumn{1}{|c}{0.1342}                                   &
\multicolumn{1}{|c}{0.1346}                                   &
\multicolumn{1}{|c}{0.1350}                                   &
\multicolumn{1}{|c||}{0.1353}                                     \\
         \hline
         \hline
\multicolumn{6}{||c||}{$\vec{p}=\vec{p}_1$}                       \\
         \hline
$v_{3}^{(u)}$     & 0.0996(62)  & 0.0940(88)  & 0.0963(88)  & 0.0946(104)
                       & 0.0898(179)   \\
$v_{3}^{(d)}$     & 0.0453(31)  & 0.0413(43)  & 0.0410(40)  & 0.0381(54)
                       & 0.0313(71)    \\
$v_{3}$           & 0.0543(37)  & 0.0516(53)  & 0.0554(59)  & 0.0564(73)
                       & 0.0594(148)   \\
         \hline
         \hline
$v_{3}^{(u;m_1)}$ &-0.00088(245)&-0.00250(353) &-0.00078(418)&-0.00659(559)
                       & 0.0180(117)   \\
$v_{3}^{(d;m_1)}$ &-0.00166(157)&-0.00301(268) &-0.00270(278)&-0.00400(400)
                       &-0.00394(813)  \\
$v_{3}^{(m_1)}$   &-0.00056(139)& 0.00018(258) & 0.00187(271)&-0.00341(430)
                       & 0.0213(109)   \\
         \hline
         \hline
$v_{3}^{(u;m_2)}$ &-0.0194(96) & 0.0135(151)  &-0.0240(180) &-0.0067(282)
                       & 0.0557(718)   \\
$v_{3}^{(d;m_2)}$ &-0.0121(58) &-0.0016(110)  &-0.0152(110) &-0.0039(183)
                       & 0.0212(400)   \\
$v_{3}^{(m_2)}$   &-0.00752(523)& 0.0131(103) &-0.0057(109) & 0.0014(191)
                       & 0.0383(582)   \\
         \hline
         \hline
\multicolumn{6}{||c||}{$\vec{p}=\vec{p}_1$}                       \\
         \hline
$v_{4}^{(u)}$     & 0.0355(49)  & 0.0382(78)  & 0.0440(75)  & 0.0496(104)
                       & 0.0403(227)   \\
$v_{4}^{(d)}$     & 0.0136(32)  & 0.0136(37)  & 0.0161(52)  & 0.0169(72)
                       & 0.0303(137)   \\
$v_{4}$           & 0.0216(27)  & 0.0249(51)  & 0.0280(46)  & 0.0310(69)
                       & 0.0123(166)   \\
         \hline
         \hline
$v_{4}^{(u;m_1)}$ &-0.0067(130)  & 0.0220(222)&-0.0004(217)
                  & 0.0194(296)  & 0.0878(792)\\
$v_{4}^{(d;m_1)}$ &-0.00681(859) & 0.0162(132)&-0.0117(151)
                  & 0.0046(199)  & 0.0106(445)\\  
$v_{4}^{(m_1)}$   &-0.00076(782) &-0.0006(131)& 0.0085(154)
                  & 0.0107(230)  & 0.0727(662)\\
         \hline
         \hline
$v_{4}^{(u;m_2)}$ &-0.0055(112)  & 0.0189(190)  & 0.0064(187)
                  & 0.0356(250)  & 0.0102(674) \\
$v_{4}^{(d;m_2)}$ &-0.00045(699) & 0.0163(141)  & 0.0050(117)
                  & 0.0264(164)  & 0.0029(425) \\
$v_{4}^{(m_2)}$   &-0.00455(654) & 0.00174(936) &-0.0001(127)
                  & 0.0075(208)  & 0.0038(508) \\
         \hline
         \hline

$v_{4}^{(u;m_3)}/a$ & 0.0306(223) &-0.0381(328) & 0.0570(402)
                    & 0.142(57)   &-0.069(127)  \\

$v_{4}^{(d;m_3)}/a$ & 0.0182(124) &-0.0155(202) & 0.0325(221)
                    & 0.0961(364) &-0.100(84)   \\

$v_{4}^{(m_3)}/a$   & 0.0072(124) &-0.0211(255) & 0.0100(259)
                    & 0.0488(449) & 0.028(108)  \\
         \hline
         \hline
      \end{tabular}
   \end{center}
\caption{The bare results for $v_3$ and $v_4$
         from eq.~(\ref{R_def_practical}) for $\beta = 6.4$,
         $c_{sw}=1.526$.}
\label{table_run_v3+v4_b6p40c1p526}
\end{small}
\end{table}

% ----------------------------------------------------------------------

\clearpage

% ----------------------------------------------------------------------

% ----------------------------------------------------------------------

\end{document}